\documentclass[a4paper,fleqn,usenatbib]{mnras}

\usepackage[T1]{fontenc}
\usepackage{ae,aecompl}

\usepackage{graphicx}	
\usepackage{amsmath}	
\usepackage{amssymb}	
\usepackage{multicol}

\title[Adiabatic vs impulsive core formation]{The nature of core formation in dark matter haloes: adiabatic or impulsive?}

\author[Burger \& Zavala]{
Jan D. Burger$^{1}$\thanks{E-mail: jdb5@hi.is (JB)} and
Jes\'us Zavala$^{1}$
\\
$^{1}$centre for Astrophysics and Cosmology, Science Institute, University of Iceland, Dunhagi 5, 107 Reykjavik, Iceland\\
}

\date{Accepted XXX. Received YYY; in original form ZZZ}

\pubyear{2018}

\begin{document}
\label{firstpage}
\pagerange{\pageref{firstpage}--\pageref{lastpage}}
\maketitle


\begin{abstract}
It is well established that the central deficit of dark matter (DM) observed in many dwarf galaxies disagrees with the cuspy DM haloes predicted in the collision-less and cold DM (CDM) model. Plausible solutions to this problem are based on an effective energy deposition into the central halo with an origin that is either based on baryonic physics (e.g. supernova-driven gas blowouts; SNF) or on new DM physics (e.g. self-interacting DM, SIDM). We argue that the fundamental difference between the two is whether the process is impulsive or adiabatic, and explore novel ways to distinguishing them by looking at the response of stellar orbits. We perform idealised simulations of tracers embedded in a $1.48\times 10^{10}{\rm M}_{\odot}$ spherical halo, and model the creation of a $\sim1$kpc DM matter core 
in SIDM with $\sigma_T/m_{\chi} = 2\,{\rm cm}^2{\rm g}^{-1}$ and injecting energy into the halo with 
a sudden mass removal equivalent to $\mathcal{O}(10\%)$ of the halo's potential energy.
Choosing idealised initial orbital configurations for the tracers, we find that radial actions are conserved (changed) in the SIDM (impulsive) case.
The adiabaticity of the SIDM case prevents tracers from changing their orbital family during core formation whereas SNF separates tracers of initially the same family to a variety of orbits. We show that these key features remain in a cosmological halo, albeit for a few dynamical timescales.
The number density and velocity dispersion profile of a Plummer sphere with $r_{1/2} =  500$ pc change only marginally under adiabatic core formation, whereas SNF causes a substantial expansion of the sphere driving it out of Jeans equilibrium. Our results point towards promising ways of differentiating adiabatic from impulsive core formation. 
\end{abstract}




\section{Introduction}\label{sec_intro}

The vast amount of evidence pointing towards the existence of dark matter (DM)  
remains purely gravitational and thus the nature of DM as a particle remains a mystery. Among the different avenues followed down to search for clues about the DM nature, a promising one is that of looking for the dynamical signature of non-gravitational DM physics in the physical properties of galaxies. In particular, dwarf galaxies are among the best targets to look for new DM physics since they are the most DM-dominated self-gravitating systems in the Universe. These galaxies have also been a constant point of contention over the last decades in the otherwise successful Cold Dark Matter (CDM) model of structure formation, where gravity is the only relevant DM interaction in the physics of galaxies (for a review see \citealt{Bullock2017}). It remains an open question whether the CDM model coupled with a complete description of gas and stellar physics, commonly referred to as {\it baryonic physics}, can fully explain the properties of dwarf galaxies. To this date, the possibility remains that non-gravitational DM physics plays a major role in galaxy formation and evolution. 

A prominent feature in dwarf galaxies that requires explanation is the DM deficit within their innermost regions relative to the predictions of the CDM model (without baryonic physics). Traditionally, two distinct issues are commonly associated with this deficit, both of which might share a common solution. The better known of these is that many dwarf galaxies most likely reside in cored DM haloes as opposed to the centrally cusped CDM haloes, which is the so-called core-cusp problem \citep[e.g.][]{Moore1994,deBlok2008,Kuzio2008,Walker2011}. The second, most recent 
issue is known as the too-big-to-fail-problem (\citep{Boylan-Kolchin2011,Papastergis2015}). It refers to an over-abundance of mass found within $\sim$1 kpc of a significant number of dwarf-size haloes in CDM, making their enclosed mass inconsistent with the internal kinematics of the dwarf galaxies they are naturally assumed to host. 

Among the most viable mechanisms to explain the observed inner DM deficit in dwarf galaxies 
are those which efficiently deposit energy into the centre of DM haloes. In particular, we focus on two distinct mechanisms that have been studied extensively in the past and that represent solutions based on either baryonic physics or new DM physics. In both, dwarf-size haloes are initially cuspy and dynamically cold at their centre and evolve to develop a hotter constant density core within the innermost regions. The first is a (gravitational) mechanism of energy transfer to DM particles in the centre of haloes following the rapid removal of gas during outflows caused by supernovae (supernova feedback; SNF) \citep[e.g.][]{Navarro1996,Gnedin2002,Pontzen:2011ty,DiCintio2014,Chan2015,Read2016,Tollet2016}. The second is a mechanism of energy redistribution in the centre of haloes from the outside-in caused by (strong) elastic self-scattering between DM particles \citep[SIDM;][]{Spergel2000,Yoshida2000,Dave2001,Colin2002,Vogelsberger2012,Rocha2013}. 

Both of these mechanisms of cusp-core transformation are currently plausible and efforts are being made to validate their key ingredients. In the case of SNF, although the evidence of gas outflows in galaxies is substantial \citep[e.g.][]{Martin2012}, the efficiency of this mechanism in dwarf galaxies remains unclear.
In particular, the efficiency depends on the bulk of the energy deposition occurring at much shorter time scales than the characteristic dynamical time of the galaxy \citep{Pontzen:2011ty}. There is evidence that bursts of star formation (and subsequent supernovae) happen in time scales at least comparable to the dynamical time scale of galaxies with $10^8$~M$_\odot$ in stellar mass \citep{Kauffmann2014}, but the supernova mechanism of cusp-core transformation requires the latter time scale to be considerably larger than the former. The situation is notably more uncertain for low-mass dwarf galaxies where the CDM problems mentioned above are more severe, and where the validation of the supernova mechanism requires a time resolution for bursts of star formation that cannot be achieved currently \citep{Weisz2014}.\\

On the other hand, the viability of DM self-scattering as a cusp-core transformation mechanism depends on whether or not the cosmological SIDM model is observationally allowed overall. Substantial effort has been put in constraining SIDM using different astrophysical probes, with the stringent (and more reliable) constraints being at the scale of elliptical galaxies \citep{Peter2013} and galaxy clusters \cite[e.g.][]{Robertson2017,Robertson2018} with the transfer cross section per unit mass limited to $\sigma_T/m_\chi\lesssim2$~cm$^2$ g$^{-1}$; for a recent review on SIDM constraints see \citet{Tulin2018}. To put this value into the context of the CDM challenges in dwarf galaxies, it has been shown that for $\sigma_T/m_\chi\lesssim0.1$~cm$^2$ g$^{-1}$, SIDM would be essentially indistinguishable from the CDM predictions \citep{Zavala2013}. Until very recently, the SIDM model remained essentially unconstrained at the scale of dwarf galaxies, but now a couple of works suggest that $\sigma_T/m_\chi\lesssim1$~cm$^2$ g$^{-1}$ given the presence of a cusp in the Milky-Way satellite Draco \citep{Read2018}, and the diversity of inner DM densities in the Milky-Way satellites (Zavala et al. in prep.). Both of these works, however, are based on SIDM simulations without taking into account baryonic physics. Although the effect the latter has on the DM mass distribution within the most massive Milky-Way subhaloes is expected to be small,
it remains to be seen if these stringent SIDM constraints remain after 
this has been taken into account. 
Even if these constraints are validated, reaching the lower limit for the cross section for
SIDM to be an alternative to CDM, $0.1$~cm$^2$ g$^{-1}$, 
might prove to be quite challenging since full galaxy formation and evolution models within CDM and SIDM are expected to have global predictions for the galaxy population that have degeneracies \citep[e.g.][]{Harvey2018}.
Given the current situation, it is crucial to identify ways in which the dark and baryonic mechanisms of energy transfer into the centre of dwarf-scale DM haloes can be differentiated in a definitive way. This objective has remained largely unexplored and we here take the first steps towards making progress in this direction by identifying distinct dynamical signatures in stellar orbits as they respond to these mechanisms. We focus on the fundamental difference that arises between a transfer of energy that occurs adiabatically (as in SIDM) and one that occurs impulsively (as in SNF). Using controlled simulations of an isolated halo with an embedded stellar population, we show that it is possible to form a DM core with similar global properties (size and central density) through both mechanisms. Yet, the responses in the kinematics of stars exhibit dramatic differences.

This paper is organized as follows. In Section \ref{basics} we provide the physical basis of our work with a simple one-dimensional toy model presented in appendix \ref{app1}, which illustrates the key differences between the adiabatic and impulsive cases.
In Section \ref{sec_simulations}, we describe the initial conditions of the DM halo and the modeling of the two cusp-core transformation scenarios in our simulations.
In Section \ref{tracersim} we construct three different initial configurations 
for tracers (stars) that are added to our simulations in order to explore the insights obtained in Section \ref{basics}. In Section \ref{results}, we discuss or main results. In Section \ref{section:cosmohalo} we discuss how abandoning the premise of an initially spherically symmetric isolated halo affects these results by analysing a cosmological halo. We then draw our conclusions in Section \ref{conclusion}. 

\section{Adiabatic vs impulsive cusp-core transformation in DM haloes}\label{basics}

In order for supernovae to efficiently transfer energy to the surrounding DM particles, the removal of gas following the explosion needs to occur impulsively, i.e., the gravitational potential needs to change rapidly relative to the orbital period of the DM particles. This phenomenon was
extensively investigated in \citet{Pontzen:2011ty} and we follow their results in detail in this paper. We start by considering their toy model of a one-dimensional harmonic oscillator which transitions between two different frequencies (modelling a change in gravitational potential) during a fixed transition time interval. If this time interval is much longer (shorter) than the oscillation period, then the change in the potential is adiabatic (impulsive).
This toy model is a nice illustration of the key physical differences between the adiabatic and impulsive cases and has the advantage of having solutions that are easily derived. We describe the model and its solution in both the impulsive and adiabatic cases in Appendix \ref{app1}, while in the following we discuss the main insights we obtain from this exercise.

In the case in which the potential changes adiabatically, the oscillation transitions smoothly from the initial to the final frequency. The ratio of the final to initial amplitudes and the shift in the phase of the oscillation are both independent of the phase of the oscillation prior to the change in potential.
They are completely determined by the frequency transition (see Eqs.~\ref{AOFT} and \ref{psi_2}), i.e., by the energy change in the potential. By contrast, if the potential changes impulsively, the transition is sudden and the change in amplitude and phase depends not only on the frequency (energy) change, but also explicitly on the initial phase (see Eqs.~\ref{amp} and \ref{phasetrafo}). This implies that for an ensemble of oscillators, the adiabiatic case will preserve the original distribution of amplitudes and phases, with just a constant re-scaling of the amplitude and a constant phase shift, whereas in the impulsive case, the original distributions are modified (see Eq. \ref{phase_red}).  

If we extend the implications of this toy model to particle orbits in a spherical potential, we expect different orbital responses to the reduction of the central potential in either the adiabatic or the impulsive case.
In both cases we expect a net expansion of the initial orbits of the tracer particles. In the adiabatic case, however, we expect the expansions to be fully determined by the integrals of motion associated with the trajectories of the tracers. In the impulsive case, on the other hand, we expect the phase of the radial oscillation to be of key importance, leading to a different response to the change in the potential depending on whether particles are, for instance, moving toward the halo centre or away from it.
We expect this behaviour as a consequence of the time averages theorem for stellar orbits \citep[see for instance][]{1987gady.book.....B}, which states that averaging physical quantities associated with the orbit of a star over a time longer than the orbital period is equivalent to calculating an ensemble average over the phase space volume which is accessible to the star. This theorem can only be applied if the accessible phase space volume does not change in a time scale of the order of an orbital period. 
Under this condition, the time averages theorem implies that the actions of the stars are conserved quantities, so-called adiabatic invariants. However, if the change in the potential is impulsive, the time averages theorem does not hold, and stars originally in a similar orbital family will split during the explosive event 
and end up on widely different orbits.

This fundamentally different nature of the adiabatic and the impulsive mechanisms of core formation offers a promising possibility to distinguish the SIDM and supernova-driven scenarios. The distribution of stellar orbits in a cored halo might contain signatures of the type of process that created the core. Our objective in this work is to model these core formation scenarios using simulations of isolated haloes, and study in detail the consequence of the adiabatic (SIDM) and impulsive (supernova-driven) cases for the orbits of collision-less tracers (stars).

\section{Initial conditions and core formation modelling}\label{sec_simulations}

We perform idealised (non-cosmological) simulations of isolated DM haloes using the {\scriptsize AREPO} code \citep{Springel:2009aa} with an added algorithm that implements
elastic DM self-scattering \citep[as described in detail in][]{Vogelsberger2012}.
For all our main results, the simulated halo has a total mass of $1.48\times 10^{10}$~M$_\odot$ (see Section~\ref{sec_sim_profiles} for more details). This value is chosen as it is a characteristic mass for the haloes of the progenitors of the most massive satellites of the Milky Way. The halo has a virial radius\footnote{Defined as the radius where the mean enclosed DM density is 200 times the critical density.} $r_{200}=52\,{\rm kpc}$ and a concentration $c_{200}=r_{200}/r_s=15$ where $r_s$ is the scale radius (see Eq.~\ref{HQrho} below). The concentration value we choose is within the $1\sigma$ range of the concentration-mass relation for haloes of this mass in a Planck cosmology \citep[e.g.][]{Pilipenko2017}. The Plummer equivalent softening length for the DM particles is chosen following 
\citet{Power:2002sw} (see also Appendix \ref{app2}): 
\begin{align}
	\epsilon = 4\frac{r_{200}}{\sqrt{N_{200}}}
	\label{powerradius}
\end{align}
where $N_{200}$ is the number of enclosed DM particles within $r_{200}$. 
After testing for stability and convergence (see appendix \ref{app2}), and given the fact that we want to observe effects related to the formation of a $\approx 1$kpc-sized core, we have chosen to run all of the simulations with $N=10^7$ equal-mass DM particles. The gravitational softening according to Eq. \ref{powerradius} is then 60  pc. 

\subsection{Initial conditions: equilibrium configuration}\label{setup}

All our haloes are spherically symmetric isotropic Hernquist haloes \citep{Hernquist:1990be} with an exponential cutoff beyond the virial radius that are initially set up in dynamical equilibrium following the algorithm described in
\citet{Kazantzidis:2003im}. Specifically, we generate a distribution of simulation particles 
with the following density profile:
\begin{align}
\rho(r) = 
\begin{cases}
\frac{\rho_{\mathrm{s}}r_{\mathrm{s}}^4}{r(r+r_{\mathrm{s}})^3} & r \le r_{200}\\
\frac{\rho_{\mathrm{s}}}{c_{200}(1+c_{200})^3}\left(\frac{r}{r_{200}}\right)^{\epsilon_d}\exp\left(-\frac{r-r_{200}}{r_{\mathrm{d
ecay}}}\right) & r > r_{200}.
\end{cases}
\label{HQrho}
\end{align}
Here, $\rho_{\mathrm{s}}$ is the scale density and 
$r_{\mathrm{decay}}$ is a parameter that determines how sharp the exponential cutoff is. The choice of the latter parameter is somewhat arbitrary, although a sensible choice is for it to be smaller than $r_{200}$ in order to guarantee a reasonable convergence of the equilibrium configuration, and to avoid having large additional mass outside $r_{200}$. Choosing the exponent $\epsilon_d$ to be: 
\begin{align}
	\epsilon_d = \frac{-1-4c_{200}}{1+c_{200}} - \frac{r_{200}}{r_{\mathrm{decay}}}
\end{align}
renders a continuous density profile with an equally continuous logarithmic slope. 

To achieve (isotropic) dynamical equilibrium we first calculate the phase space distribution using Eddington's formula \citep{Eddington1916}:
\begin{align}
	f(\mathcal{E}) = \frac{1}{\sqrt{8}\pi^2}\int_{0}^{\sqrt{\mathcal{E}}}du\frac{d^2\rho}{d\Psi^2}(r(\Psi(u))),
\end{align}
where $u = \sqrt{\mathcal{E}-\Psi}$ and  $\mathcal{E}$ and $\Psi(r)$ are the (negative) specific energy and gravitational potential, respectively, both shifted by the zero point of the gravitational potential (set at $r\rightarrow\infty$). We then assign velocity vectors to all the DM particles in the halo by sampling their magnitude from the phase space distribution calculated above using a rejection sampling algorithm and then randomly drawing directions from the unit sphere. See Appendix \ref{app2} for our tests on the stability of the equilibrium configuration.

\subsection{Modelling supernova-induced core formation}\label{pesec}

In order to transform an initially cuspy density profile into a cored one, a certain amount of energy needs to be deposited into the centre of the halo. An estimate of the required energy 
is obtained by comparing the potential energy of the halo before and after the cusp-core transformation
(see \citealt{Penarrubia:2012bb}): 
\begin{align}
	\Delta E = \frac{W_{\mathrm{core}}-W_{\mathrm{cusp}}}{2}.
	\label{energyrequirement}
\end{align}
The potential energy of the halo is 
\begin{align}
	W = -4\pi G \int_{0}^{\infty}\rho(r)M(<r)\,r\,dr,
\end{align}
where $M(<r)$ is the enclosed mass. Neglecting the exponential cutoff of the profile in Eq.~\ref{HQrho}, we can take the potential energy of the Hernquist halo as an approximation for $W_{\rm cusp}$: 
\begin{align}
	W_{\rm cusp}\sim W_{\mathrm{HQ}} = -\frac{GM^2_{\infty}}{6r_s}.
\end{align}  
where $M_{\infty}$ is the total mass of the halo  $M_\infty = (16/15)^2\times M_{200}$. In order to have an estimate of $W_{\rm core}$ we can model the final cored halo with the following density profile: 
\begin{align}
	\rho_{\mathrm{cored}} = \frac{\rho_sr_s^4}{(r+r_c)(r+r_s)^3},
\end{align}
where $r_c$ is the core radius. 
With these considerations we estimate that the required energy to form a $r_c=1$~kpc core in a $1.48\times 10^{10}{\rm M}_{\odot}$ halo with a concentration $c_{200} = 15$ is roughly 11\% of the halo's initial potential energy. 
We take this energy estimate into consideration in the modelling of supernova gas blowouts below to form a core of the relevant size.

We follow the procedure described in \citet{2013MNRAS.433.3539G} 
to implement a simple model of an episode of gas accretion followed by a sudden gas blowout. This is done by adding an external time-dependent Hernquist potential centered at the halo centre, which represents an effective baryonic mass distribution $M_{\rm gal}(t)$ with scale radius $r_{\rm s, gal}$, 
and induces an additional acceleration of the particles in the simulation:
\begin{align}
	\mathbf{a}_{\mathrm{ext}} = -\frac{GM_{\mathrm{gal}}(t)}{(r+r_{\mathrm{s,gal}})^2}\frac{\mathbf{r}}{r},
	\label{aext}
\end{align}
where $\mathbf{r}$ is the position vector relative to the centre of mass of the halo. 
Contrary to \citet{2013MNRAS.433.3539G}, we do not introduce an extra gravitational softening since the acceleration does not diverge at $r\rightarrow0$, but rather reaches a constant value. 
We explore two simple models for the behaviour of $M_{\rm gal}(t)$:
\begin{itemize}
	\item {\it single explosion:} starting from a halo in equilibrium (set up as described in Section~\ref{setup}), we add a constant external potential $M_{\rm gal}(t)=M_{\rm single}$ at $t=0$ and let the halo contract adiabatically into the centre until it reaches a new equilibrium. Once the configuration is stable, we set the mass to zero instantaneously, mimicking a single large mass blowout.
	\item {\it multiple explosions:} starting from a halo in equilibrium (set up as described in Section~\ref{setup}) at $t=0$, we grow the mass distribution linearly with time until a maximum value $M_{\rm mult}$ is reached at $t=\tau$ ($M_{\mathrm{gal}}(t) = M_{\mathrm{mult}}\times t/\tau$), afterwards we set the mass to zero instantaneously. This procedure is repeated periodically, which mimics several consecutive mass blowouts.
\end{itemize} 
Each time the mass is set to zero, a total amount of energy: 
\begin{align}
	E = \frac{GM_{\mathrm{gal}}^2}{6r_{\mathrm{s,gal}}}
\end{align}
is deposited into the halo, which implies that if we keep $r_{\rm s,gal}$ fixed, the transferred energy scales as the square of the baryonic mass distribution. This means that in order to achieve the same effect with $N$ smaller explosions as with a single explosion we require
\begin{align}
	M_{\mathrm{mult}} = \frac{M_{\mathrm{single}}}{\sqrt{N}}.
	\label{rescale}
\end{align} 

\subsection{Core formation in SIDM}\label{sec_core_SIDM}
\begin{figure*}
		\includegraphics[height=6.5cm,width=8.5cm,trim=0.5cm 0.5cm 0cm 0.0cm, clip=true]{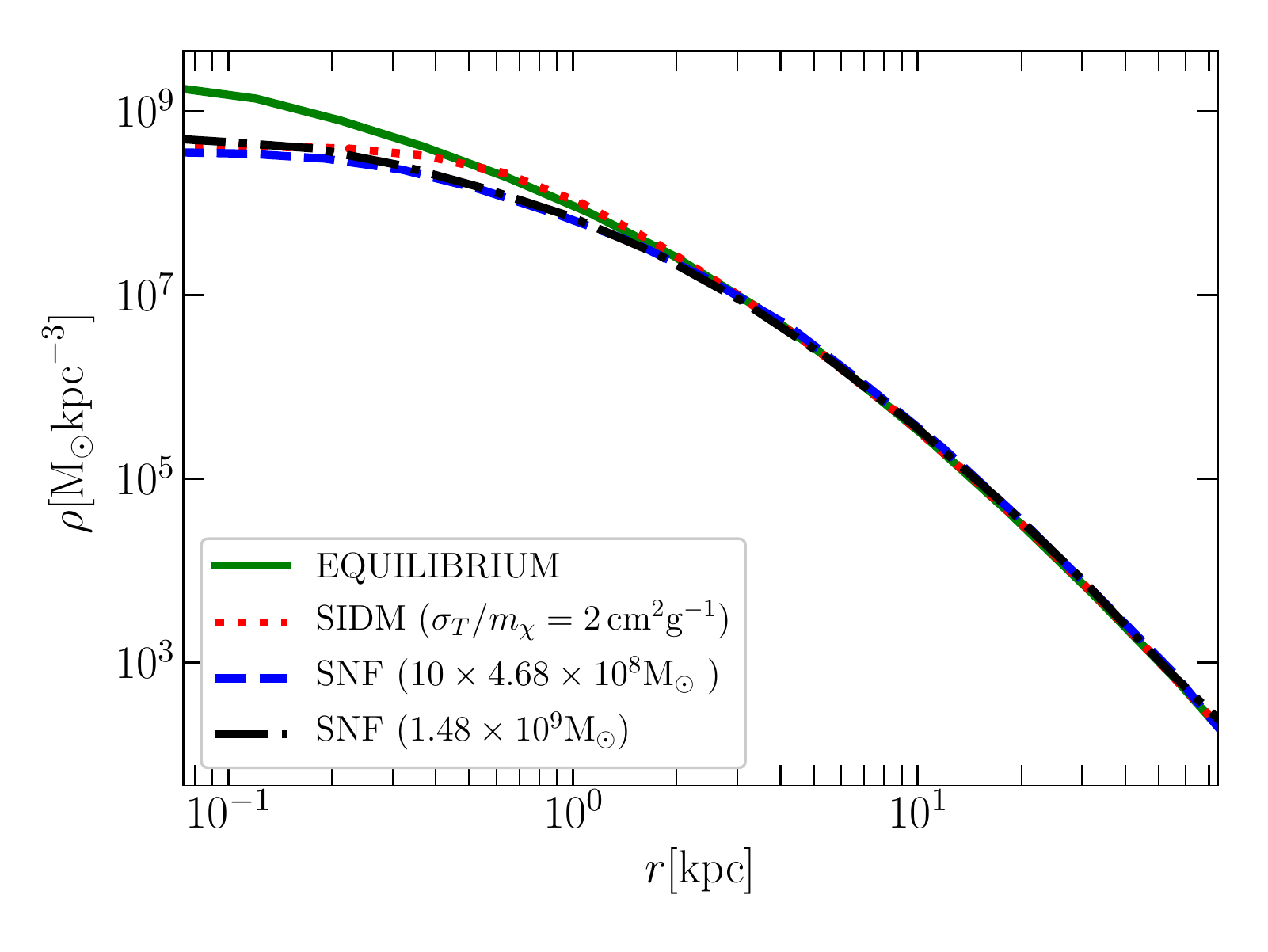}
		\includegraphics[height=6.5cm,width=8.5cm,trim=0cm 0.5cm 0.5cm 0.0cm, clip=true]{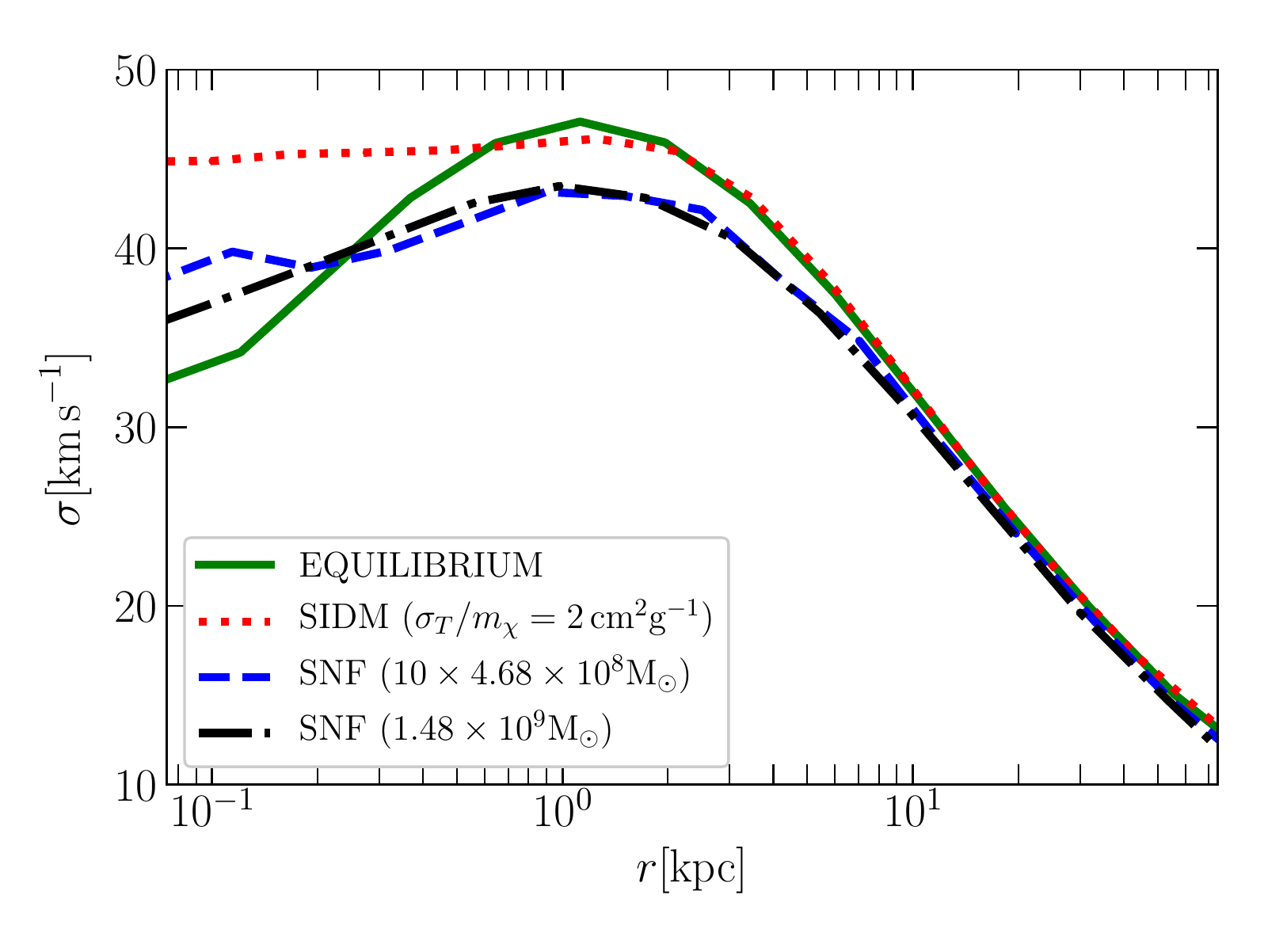}
		\caption{DM core formation in the impulsive (SNF) and adiabatic (SIDM) scenarios: density (left) and velocity dispersion profiles (right) for a $1.48\times 10^{10}$~M$_\odot$ halo. The initial condition is that of a Hernquist halo with a cutoff at large radii (see Eq.~\ref{HQrho}) set in dynamical equilibrium. 
		Different lines belong to the different simulations: the green solid line denotes the equilibrium case in which we evolve the halo profile set up as in section \ref{setup} including only self-gravity between the DM particles; the dashed black line refers to the SNF model with a single large explosion removing a mass of $1.48\times 10^{9}\,{\rm M}_{\odot}$, the dashed blue line refers to the SNF model with ten consecutive explosions depositing a total amount of energy equivalent to the single explosion case; the red dotted line refers to the case where DM is self-interacting with $\sigma_T/m_\chi=2~\mathrm{cm}^2\mathrm{g}^{-1}$. Except for the equilibrium case, the profiles are shown in their stable configuration after the cusp-core transformation has finalized.
		}
		\label{profileevolution}
\end{figure*}
DM haloes naturally develop cores with a size that is comparable to the scale radius of the halo if the cross section for self-scattering is sufficiently large to thermalize the inner halo. In particular, collisions redistribute energy among the DM particles from the dynamically hotter intermediate regions near the scale radius (where the velocity dispersion profile approximately peaks) to the central colder regions \citep[see e.g. Fig. 2 of][]{Colin2002}.
The time scale for full thermalisation of the halo is approximately given by the condition of having on average one collision per particle in the lifetime of the halo centre ($t_{\rm age}$):
\begin{equation}
    \rho_{\rm p}(\sigma_T/m_\chi) v_{\rm p} t_{\rm age}\approx 1
\end{equation}
where $\rho_p$ and $v_{\rm p}$ are the density and average relative velocities between DM particles, respectively, at the peak of the velocity dispersion profile. If the average number of collisions is less than one, then the cusp core transformation is incomplete. In fact, for dwarf-sized SIDM haloes, if $\sigma_T/m_\chi\sim0.1~\mathrm{cm}^2\mathrm{g}^{-1}$, then the resulting density profile is only slightly less cuspy than the corresponding CDM halo \citep{Zavala2013}. We can consider this as a lower limit of the cross section for SIDM to be a distinct alternative to CDM. On the other hand, if the average number of collisions is much larger than one, then the unavoidable gravothermal collapse is triggered and the SIDM halo eventually becomes even cuspier than its CDM counterpart \citep[e.g.][]{Koda2011,Pollack:2014rja}. In the case of dwarf-sized haloes, the threshold for collapse is reached when $\sigma_T/m_\chi$ is of $\mathcal{O}(10)~\mathrm{cm}^2\mathrm{g}^{-1}$ (\citealt{Elbert2015} and Zavala et al. in prep.). Thus, for dwarf-sized haloes, a cross section of $\mathcal{O}(1)~\mathrm{cm}^2\mathrm{g}^{-1}$ provides the right amplitude for fully thermalized SIDM haloes. In our SIDM simulations we pick a value of $2~\mathrm{cm}^2\mathrm{g}^{-1}$.

As mentioned at the beginning of Section~\ref{sec_simulations}, we model the effect of DM self-interactions by using the probabilistic algorithm for elastic self-scattering in {\scriptsize AREPO} described in detail in \citet{Vogelsberger2012}. 

\subsection{Cored haloes in SIDM and CDM}\label{sec_sim_profiles}

Fig.~\ref{profileevolution} shows a comparison of the density profile (left panel) and velocity dispersion profile (right panel) of a halo that develops a core impulsively (SNF) and adiabatically (SIDM; dotted red line) starting from the same initial condition. For the former case we show the single explosion case (dot-dashed black line) and a case with 10 explosions with an injected energy equivalent to the single explosion case (dashed blue line); see Section \ref{pesec}. We also show the case where there is no energy deposition or redistribution (solid green line). This simulation serves as a reference for the degree to which the equilibrium method described in Section~\ref{setup} is stable across the simulation time. It is also a baseline for comparison with the cuspy Hernquist profile, which is the initial condition for all the simulations.
Except for this reference case, all the other profiles are shown after the halo has reached a new equilibrium configuration after the cusp-core transformation.
We remark that tn the SIDM case, the equilibrium is only quasi-stable. Thermalization is fully complete after 1.2 Gyrs have passed from the start of the simulation. Afterwards, the central density increases due to the beginning of the gravothermal collapse phase, which is a stage that we want to avoid since our goal is to compare DM cores formed due to self-interactions with cores formed due to supernovae.
For the cases of impulsive core formation (SNF), the scale radius of the external baryonic distribution is fixed to 300~pc (a typical half-light radius among bright Milky-Way satellites), while its mass is chosen to be larger than the mass of typical dwarf galaxies in order to create a core with an approximate size of 1 kpc (comparable to the SIDM case explored) 
according to the energy requirement outlined in Section~\ref{pesec}.
In the single explosion explosion case, this corresponds to $M_{\mathrm{gal}}=1.48\times 10^9{\rm M}_{\odot}$.
As mentioned in Section~\ref{pesec}, this baryonic mass is added as an external Hernquist potential, and thus, the halo responds to it by adiabatically contracting substantially given the large baryonic mass. Once the halo is fully relaxed, the explosion is triggered and we follow the evolution of the halo until it reaches its final equilibrium state, which is the one shown in Fig.~\ref{profileevolution}.
In the multiple explosion case, the baryonic mass distribution is added periodically as a linear function of time with a maximum at $M_{\rm mult}= \sqrt{10}\times 1.48\times 10^8{\rm M}_{\odot}$ (see Eq.~\ref{rescale}) and a period of 0.3 Gyrs followed by an explosion. A total of 10 explosions are simulated in this way to achieve the same energy deposition as in the single explosion case. After the last explosion, we let the halo evolve into its final equilibrium configuration.
Although this episodic model of accretion and mass blowout also results in adiabatic contraction of the halo during the accretion phase, it is naturally milder than the single explosion case.

Fig.~\ref{profileevolution} shows that both of the impulsive (SNF) cases result in a cored halo that looks very much alike, with both of them having very similar density and velocity dispersion profiles. This shows that the energy requirement is fulfilled regardless of whether the explosion occurs only once, or is episodic as long as the energy transferred during the explosion(s) is equivalent. This result is a validation of the scaling of the deposited energy on the square of the baryonic mass mentioned in Section~\ref{pesec} \citep[based on][]{Penarrubia:2012bb}, which supports our modelling of explosions as an accurate implementation. We note that this scaling with mass is apparently in disagreement with the results in Fig. 4 of \citet{2013MNRAS.433.3539G}. There, the authors compare the single and multiple explosion cases; and although they mention that the single explosion case is more efficient than the multiple explosion case (for the same amount of blowout mass), it is clear from their Fig. 4 that the difference is not very strong. We have found, however, that if the total blowout mass is set to the same value as in the single explosion case ($10\times1.48\times10^8$~M$_\odot$), then the size of the formed core reduces significantly.
Given the energetic arguments in Section~\ref{pesec}, we are confident that our implementation is accurate.  

Looking at the adiabatic (SIDM) case in Fig.~\ref{profileevolution}, we can see that the adiabatic core is fully isothermal within the radius where the velocity dispersion peaks ($\sim 1$~kpc). As we mentioned in Section~\ref{sec_core_SIDM} this is a signature of full thermalization of the core. We note that the isothermal region extends further outwards from the halo centre than the constant density plateau, which agrees with results found previously \citep[e.g.][]{Koda2011,Vogelsberger2012}. 
Comparing the SIDM and SNF cases, it is apparent that although we have achieved our goal of creating a DM core with the same asymptotic central density and similar core size, there are clear differences in the density and velocity dispersion profiles. The density core is flatter in SIDM and the impulsive case is not fully isothermal. Instead, it retains a similar slope - albeit shallower - as in the initial cuspy halo. Within the central region ($\lesssim15$~kpc), both the average density and velocity dispersion of DM particles are larger in the SIDM case than in the SNF case. This is a signature of the difference between the central energy redistribution taking place in SIDM, which bounds the expansion of orbits within the central regions, 
versus the energy transfer in the SNF case, which causes the orbits to expand to much larger radii. We study this difference in detail below where we explore the impact of these two mechanisms of core formation on the orbits of tracer particles.

\section{Simulations with tracer particles}\label{tracersim}

In order to study the effect of a time-dependent potential on the orbits of stars in detail, we need to introduce (collision-less) tracer particles. The reason for this is twofold: (i) since we need to track the orbits accurately in time, it will be computationally prohibitive (and unnecessary) to use all the $N=10^7$ DM particles as tracers; (ii) for the SIDM case, the DM particles do not act as tracers since they are collisional. 
We require that the population of tracer particles fulfill the following requirements:
\begin{enumerate}
	\item The combined mass of all the tracer particles should be smaller than the mass of one DM particle. In this way, they have a negligible impact on the motion of DM particles. We choose $1.48 \times 10^{-1}{\rm M}_{\odot}$ as the mass of a single tracer.
	\item The orbits of the particles are computed exactly from the sum of the gravitational forces exerted by all DM particles\footnote{This is a direct summation approach that requires $N$ partial forces per tracer particle, while for DM particles, gravitational forces are computed using a tree algorithm that requires $\mathcal{O}({\rm log}~N)$ calculations per DM particle.}.
	\item The softening length of the tracer particles is small (we pick $\epsilon_{\rm tracer}=15$~pc), and we neglect self-gravity to ensure detailed time integration. 
\end{enumerate}  
An exact calculation of the forces exerted on the tracer particles is computationally expensive and thus we cannot afford to include a very large number of tracers, still, the sample needs to be large enough to obtain statistically significant results. We have found that a sample of 2000 particles is sufficient for our purposes.  

Following the arguments in Section \ref{basics} we expect that the actions of tracer particles will be conserved when the change in the potential is adiabatic (SIDM case), whereas we do not expect that in the impulsive (SNF) case. In particular, the radial action of bound orbits will be an adiabatic invariant:
\begin{align}
J_r &= \frac{1}{\pi}\int_{r_{\mathrm{peri}}}^{r_{\mathrm{apo}}}dr\sqrt{2E-2\Phi(r)-\frac{L^2}{r^2}}
\label{radialaction}
\end{align}
where $E$ is the energy of the tracer particle, $L$ the magnitude of its angular momentum, $\Phi$ is the gravitational potential, and $r_{\rm apo}$ and $r_{\rm peri}$ are the apocentre and pericentre of the orbit of the tracer.

Our strategy to investigate the differences between an impulsive and an adiabatic change in potential is to put the tracer particles in configurations that probe quantities related to the radial action. In the rest of this Section we describe the setup for three different configurations we have chosen to accomplish this objective.

\subsection{A single orbital family}\label{orbitalfamily}

Following \citet{2018ApJ...855...87M}, we expect different orbital families to respond differently to a change in the gravitational potential, as well as to effects of diffusion and mixing. For this reason, we decided to investigate how one particular orbital family responds to the impulsive and adiabatic core formation processes. An orbital family is defined by having similar values of $E$ and $L$.
Given the spherical symmetry of our problem, these two requirements are sufficient and it is not necessary that the orientation of the angular momentum vector be the same for orbits to be of the same family. Since we want to compare the adiabatic and impulsive cases of core formation, we need to pick an orbital family with ($E,L$) values which corresponds to particles in orbits around the region where the core will form.
In the case of spherically symmetric problems there is a natural way to make a sensible choice by connecting ($E,L$) to the apocentre and pericentre of the orbit. In the stable case we can write the energy for a central potential as: 
\begin{align}
E = \frac{1}{2}\left(\frac{dr}{dt}\right)^2 + \frac{1}{2}\frac{L^2}{r^2} +\Phi(r)
\end{align}
In a bound orbit, $dr/dt = 0$ at the apocentre and pericentre of the orbit. Therefore, 
\begin{align}
0 = 2\left[E-\Phi(r_{\mathrm{apo(peri)}})\right]-\frac{L^2}{r^2_{\mathrm{apo(peri)}}}
\end{align}
which can be solved to yield $E$ and $L$ in terms of the apocentre and pericentre radii:
\begin{align}
L &= r_{\mathrm{apo}}r_{\mathrm{peri}}\sqrt{\frac{2\left[\Phi(r_{\mathrm{apo}})-\Phi(r_{\mathrm{peri}})\right]}{r^2_{\mathrm{apo}}-r^2_{\mathrm{peri}}}}\\
E &= \Phi(r_{\mathrm{apo}})+\frac{1}{2}\frac{L^2}{r^2_{\mathrm{apo}}}
\end{align}
In this way, we can chose the orbital family we wish to inspect by simply choosing an interesting radial range, spanning from radii that are heavily affected by the change in potential to radii that are barely affected. 
Specifically, we draw pericentre and apocentre radii for all the 2000 tracers from Gaussian distributions with a spread of 0.075 kpc centered at 0.75 kpc and 3 kpc, respectively. We then calculate the associated angular momentum and energy pairs and subsequently populate the radial range with orbits at different radial positions (phases), sampling from a flat distribution between pericentre and apocentre. The direction of rotation is randomly chosen. 

\subsection{Gaussian distribution in radial action}\label{sec_Jr_setup}

To directly investigate the degree to which radial actions (Eq.~\ref{radialaction}) are (not) conserved in the (impulsive) adiabatic case, we construct a Gaussian distribution of radial actions $p(J_r)$ (with a mean of $45~\mathrm{km}\,\mathrm{s}^{-1}\mathrm{kpc}$ and a standard deviation of $3.75~\mathrm{km}\,\mathrm{s}^{-1}\mathrm{kpc}$) and assign $J_r$ values to each of the 2000 tracers using a rejection sampling algorithm \citep{von_1951} as follows:
\begin{enumerate}
	\item draw a random radius $r^\ast$ between two reference radii (0.15 kpc and 3 kpc).
	\item draw a random energy $E^\ast$ value from a flat prior between the potential $r^\ast$ and $E^\ast = 0$, the energy at which the tracer is no longer gravitationally bound.
	\item draw a random $L^\ast$ value from a flat prior bounded by the maximum possible value given by the radius and energy drawn in (i) and (ii). 
	\item calculate the value of the radial action $J_r^\ast$ for the set ($r^\ast,E^\ast,L^\ast$) using Eq.~\ref{radialaction}.
	\item accept $J_r^\ast$ if $p(J_r^\ast)$ is larger than a random variable drawn in the interval $[0,{\rm max}(p(J_r))]$, reject otherwise
\end{enumerate}
In case $J_r^\ast$ is accepted, we assign a velocity vector to the tracers which is in agreement with the pair ($E^\ast,L^\ast$). 

\subsection{A Plummer sphere}

Finally, in order to model a stellar population with a more realistic distribution, we also sample the positions of the 2000 tracers from a radial Plummer profile  with a half-light radius of 
$500\mathrm{pc}$, which is a typical value for the bright Milky Way satellites.
This is the only free parameter of the Plummer distribution since by construction the stellar population is a set of tracers that does not contributes to the gravitational potential. We set the total mass of the Plummer profile to be equal to $N_{\star}m_{\star}$, where $N_\star=2000$ and $m_\star=1.48 \times 10^{-11}{\rm M}_{\odot}$.
The total mass needs to be consistent in order to be able to sample the tracers' positions from a Plummer density profile with the same method we used to sample the DM particles' positions from a Hernquist profile.
Since in this setting the gravitational potential is fully determined by the DM distribution, we use the spherical Jeans equation (assuming isotropic orbits for the tracers) to compute the equilibrium velocity dispersion profile of the Plummer sphere. The velocities of the tracers are then sampled from a Maxwellian distribution with a radially dependent velocity dispersion. 

\section{Results}\label{results}

\subsection{Numerical errors in circular orbits}\label{errordisc}

\begin{figure}
	\includegraphics[width=\linewidth, trim=0.5cm 0.5cm 0cm 0.0cm, clip=true]{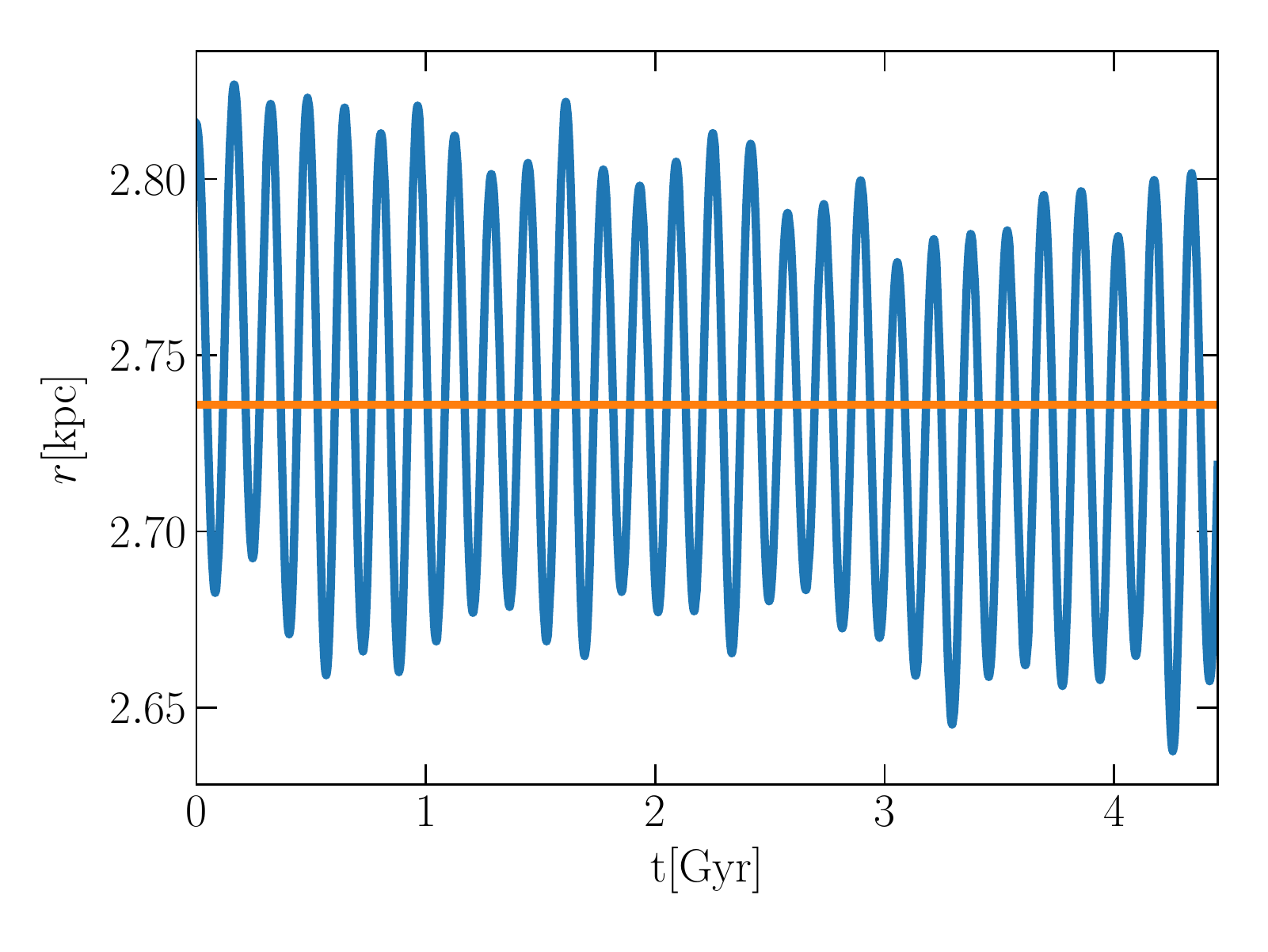}
	\caption{Numerical errors from circular orbits. The radial distance of one particular tracer particle in a Hernquist halo relative to the centre of mass of the halo as a function of time (the red line is the time average). Despite the tracer initially being  on a circular orbit, the radial distance oscillates in time, which also implies a non-zero radial velocity. We use these oscillations to quantify the baseline errors (in the simulation where the halo is in equilibrium) in the time integration scheme we use.}
	\label{orbit8}
\end{figure}

\begin{figure*}
	\includegraphics[height=6.0cm,width=8.5cm,trim=0.5cm 0.5cm 0cm 0.5cm, clip=true]{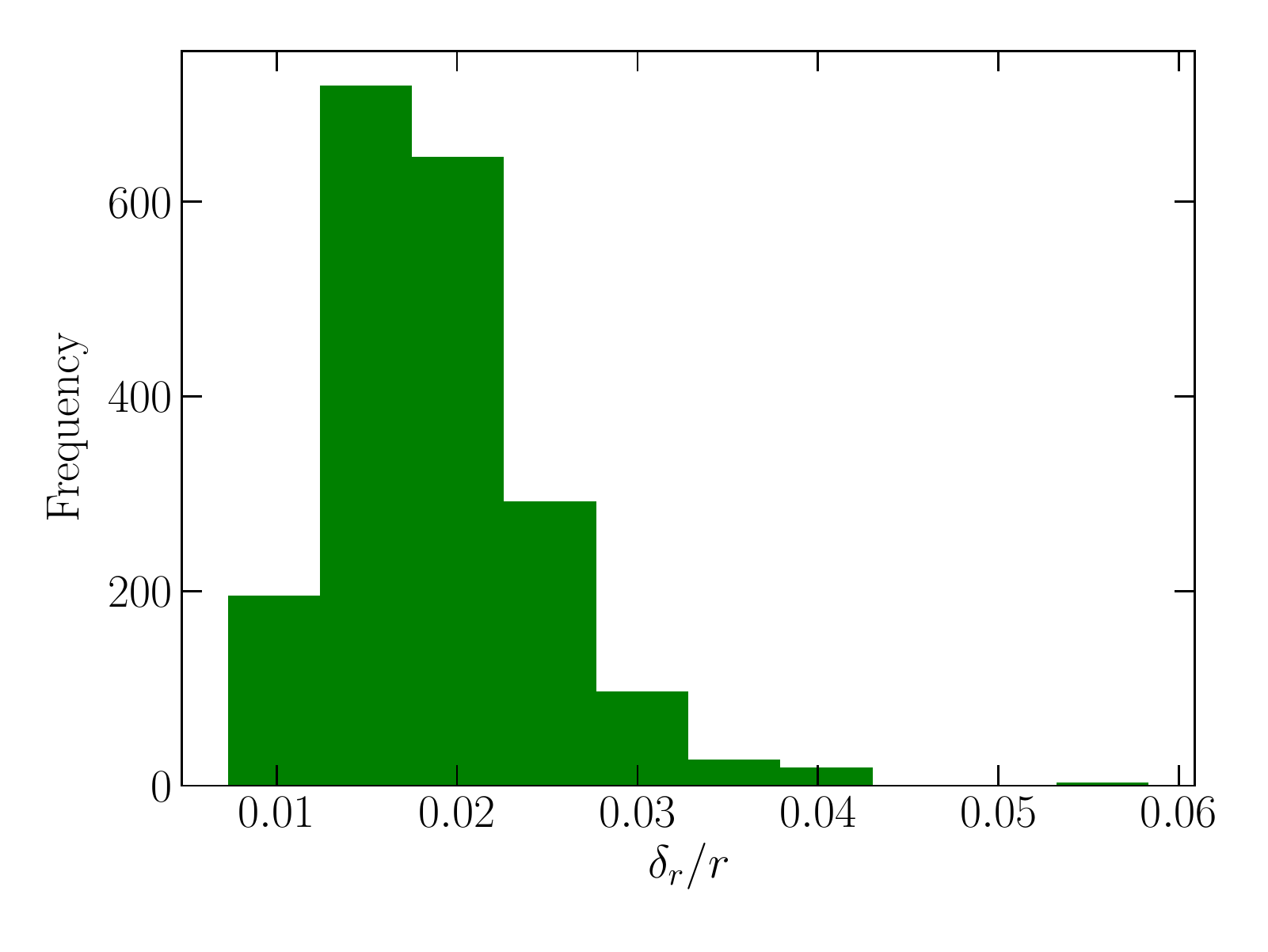}
	\includegraphics[height=6.0cm,width=8.5cm,trim=0.5cm 0.5cm 0cm 0.5cm, clip=true]{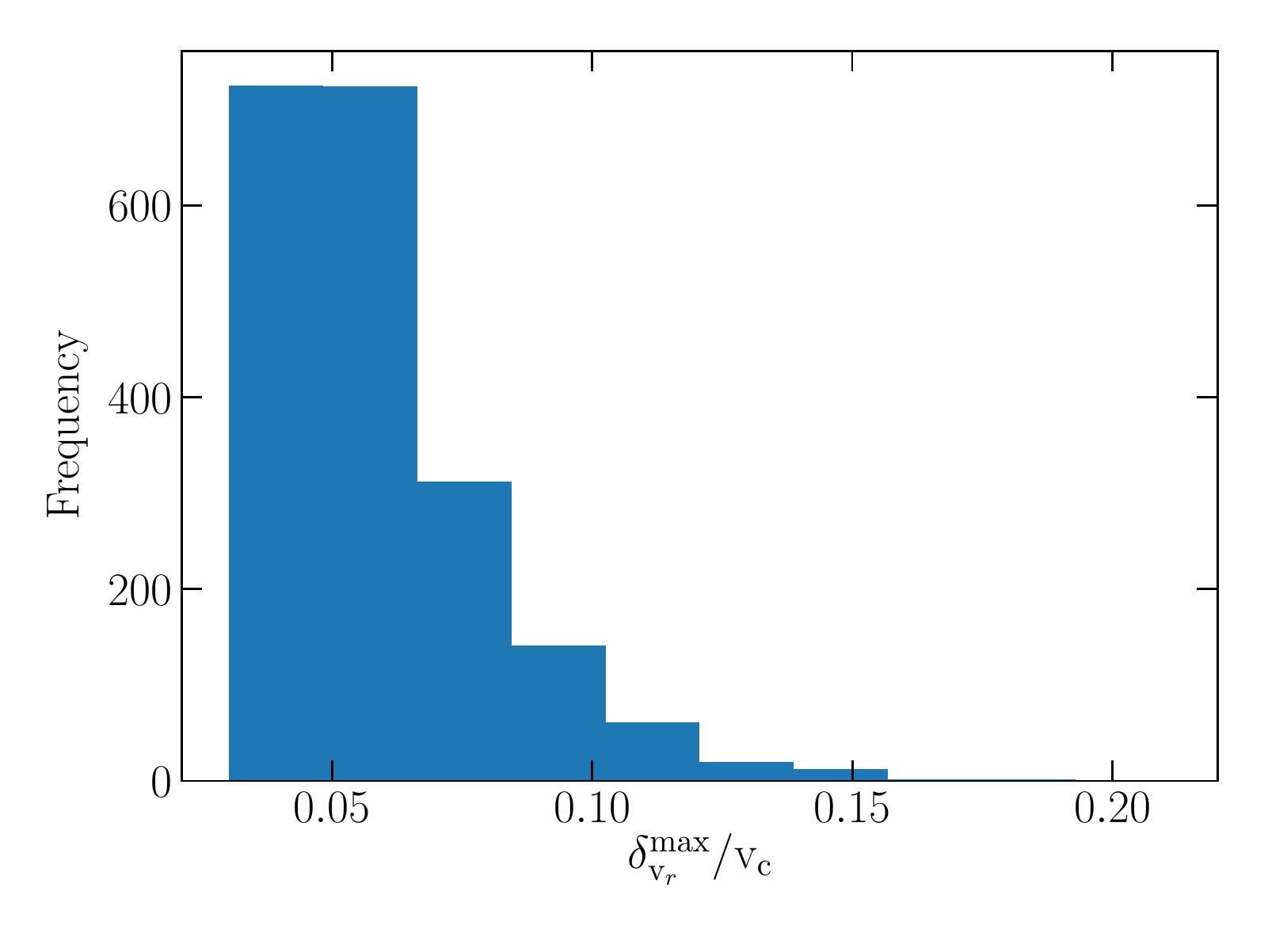}\\
	\includegraphics[height=6.0cm,width=8.5cm,trim=0.5cm 0.5cm 0.0cm 0.5cm, clip=true]{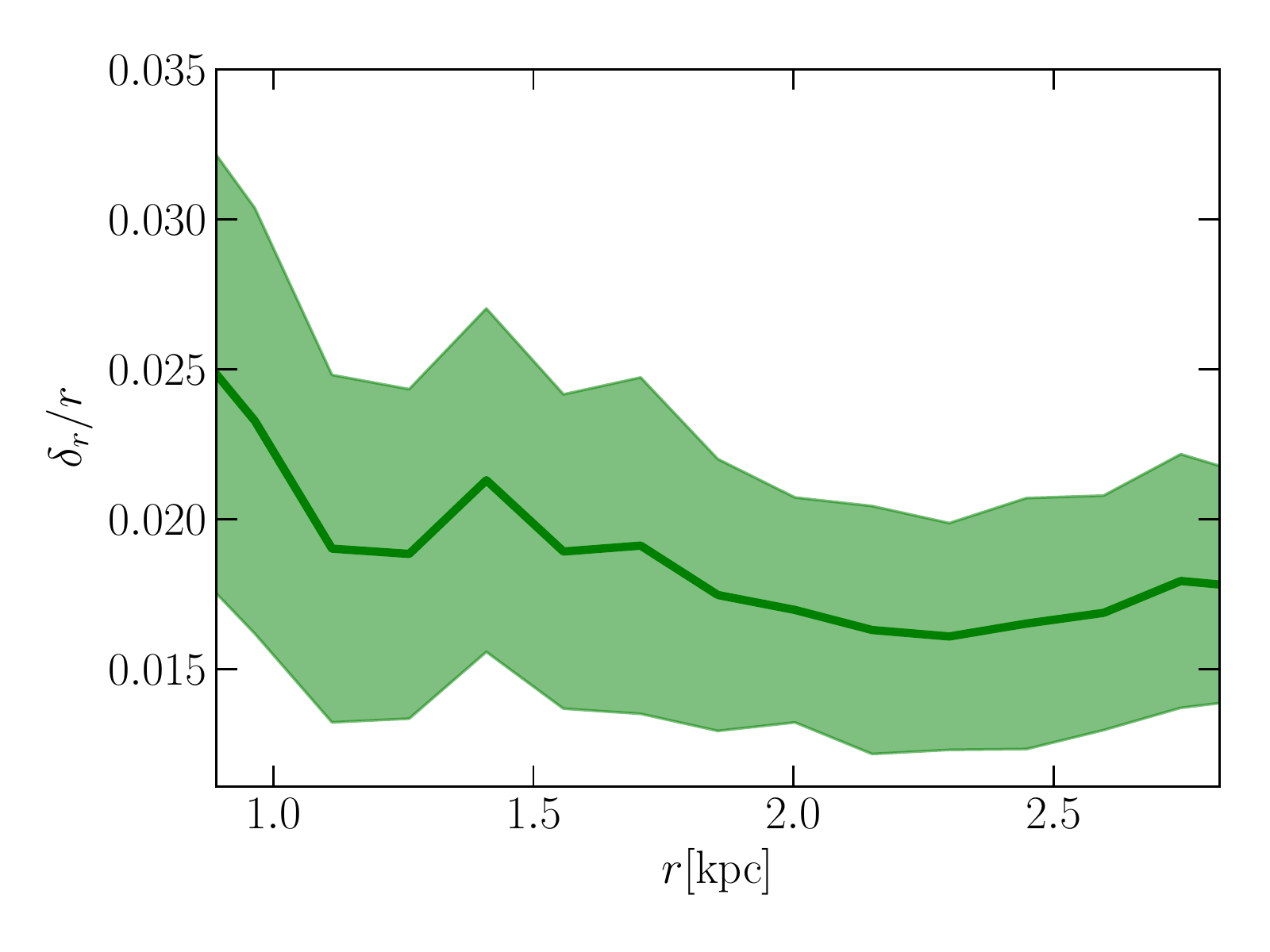}
	\includegraphics[height=6.0cm,width=8.5cm,trim=0.5cm 0.5cm 0cm 0.5cm, clip=true]{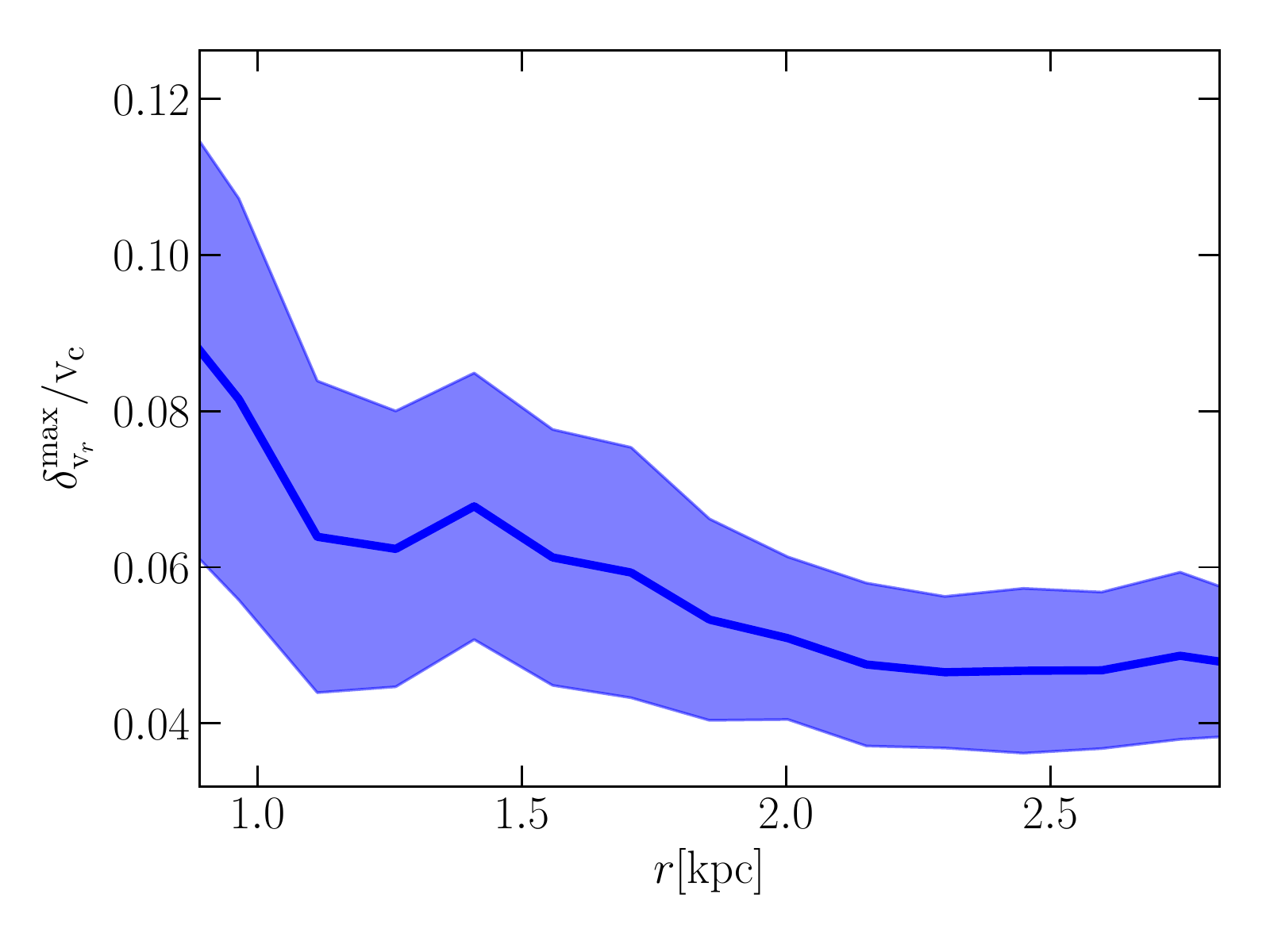}
	\caption{Baseline integration errors for circular orbits. On the left panels, histogram and radial profile of the relative errors in the orbital radius, and on the right panels, histogram and radial profile of the maximum errors in the radial velocity divided by the circular velocity at the radius where the maximum error occurs. The relative error in the orbital radius is centered around 2 per cent, whereas the absolute error in radial velocity is centered around 4~$\mathrm{km\,s}^{-1}$, which translates to a relative error of around 7 per cent.
	The bottom panels show a clear dependence of the orbital errors on the orbital radius, which can be mainly attributed to the number of enclosed particles being smaller at smaller radii and hence the sampling of the gravitational potential being poorer towards the halo centre. The Poissonian sample variance is included in the lower panels and it exhibits no dependence on radius.}
	\label{errorsstable}
\end{figure*}

Before looking at the evolution of the tracer particles as they respond to the two mechanisms of cusp-core transformation, we will first study the amplitude of the time integration errors, force errors, and discreteness effects which are inherent to the code and setup we are using to track the orbits of tracers.
Looking at simulated circular orbits in the case where the potential is fixed (equilibrium) is a simple way to estimate the magnitude of these numerical errors. In particular, we look at deviations from a true circular orbit where the radius is constant and the radial velocity is zero. This will serve as a benchmark when studying the orbital response in a time-varying potential. To construct this benchmark, we set up the 2000 tracers within the halo in equilibrium on circular orbits whose radii are sampled from a flat distribution between 0.75 and 3 kpc, and evolve the simulation over 4.5 Gyrs.

Fig. \ref{orbit8} shows one typical circular orbit, albeit at a rather large radius within the range we probe. We can see that the orbit is not completely circular, with the orbital radius oscillating in time by $\sim\pm2\%$ around an average value. 
We look at the statistical behaviour of the ensemble of circular orbits to estimate the baseline integration errors. To give a measure to the errors in the radius of a single orbit, we calculate the mean and variance as: 
\begin{align}
	\langle r\rangle &= \frac{1}{N_{\mathrm{steps}}}\sum_{i=0}^{N_{\mathrm{steps}}}r_i\\
	\delta^2_r &= \frac{1}{N_{\mathrm{steps}}-1}\sum_{i=0}^{N_{\mathrm{steps}}}\left(r_i-\langle r\rangle\right)^2,
\end{align}
where $r_i$ is the orbital radius at each timestep in the simulation and $N_{\rm steps}$ is the total number of timesteps. For the velocity, we provide
a worst-case estimate of the errors by recording the maximal radial velocity encountered throughout the whole simulation time for each tracer.

Fig.~\ref{errorsstable} shows the orbital errors of all the 2000 tracer particles. The upper panels show the histograms of the relative error in radius on the left ($\delta_r/\langle r\rangle$) and the worst-case error in the radial velocity divided by the circular velocity at the radius where the maximum error occurs ($\delta_{\rm v_r}/{\rm v_c}$); the latter has been normalised to the circular velocity, which, representing the characteristic magnitude of the tracers' velocities at a given radius, is the relevant velocity to compare the errors to. We see that on average, the relative error in the radius is 2 per cent, with a distribution tail extending to 4 per cent. On the other hand, the mean maximum value of $\delta_{\rm v_r}/{\rm v_c}$ is around 7 per cent with a distribution tail extending to 15 per cent. We emphasize that in the case of velocities, these are worst-case errors over the entire simulation time.

\begin{figure*}
	\includegraphics[height=6.5cm,width=8.5cm,trim=2.0cm 0.5cm 0.75cm 0.5cm, clip=true]{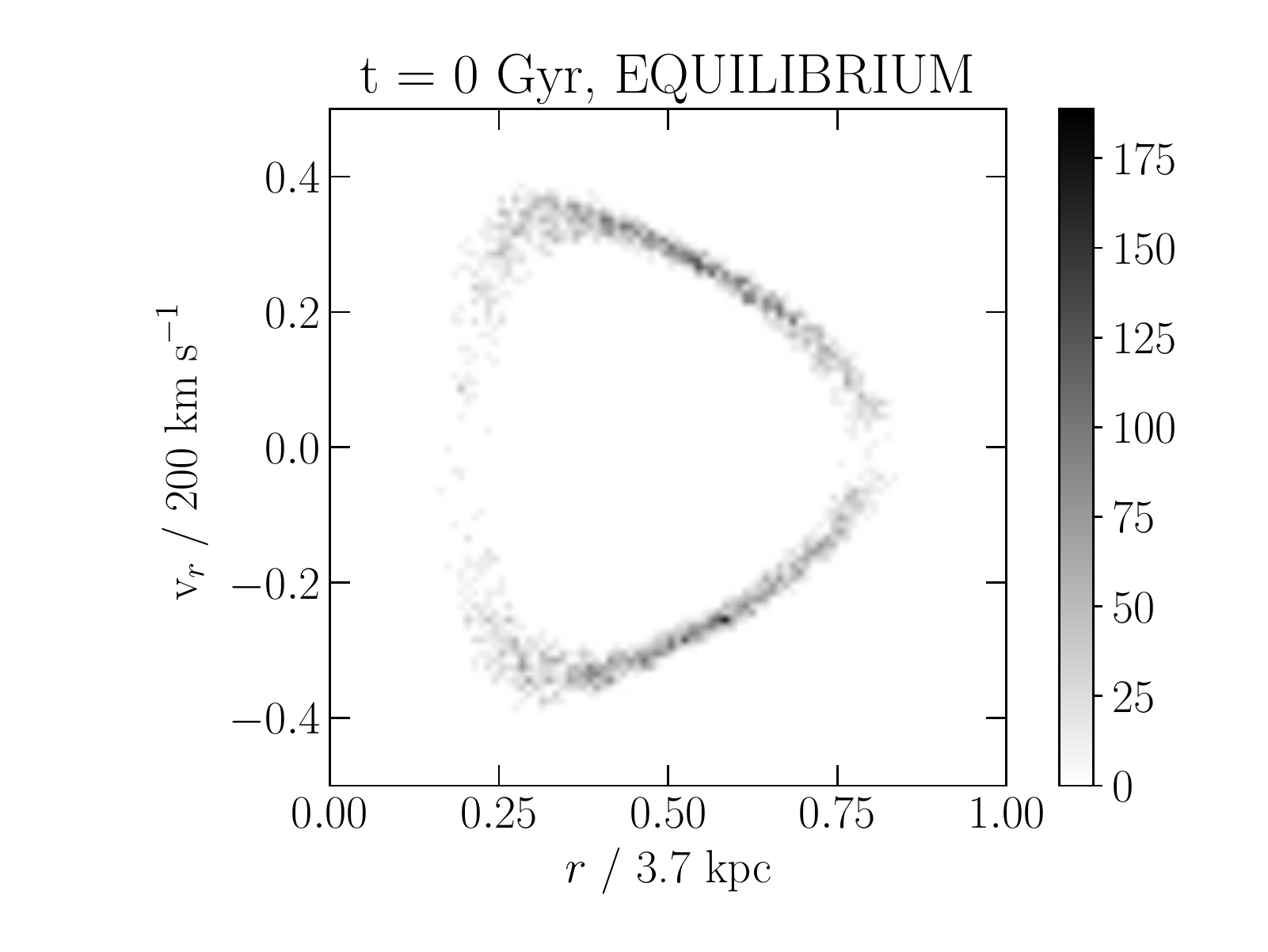}
	\includegraphics[height=6.5cm,width=8.5cm,trim=2.0cm 0.5cm 0.75cm 0.5cm, clip=true]{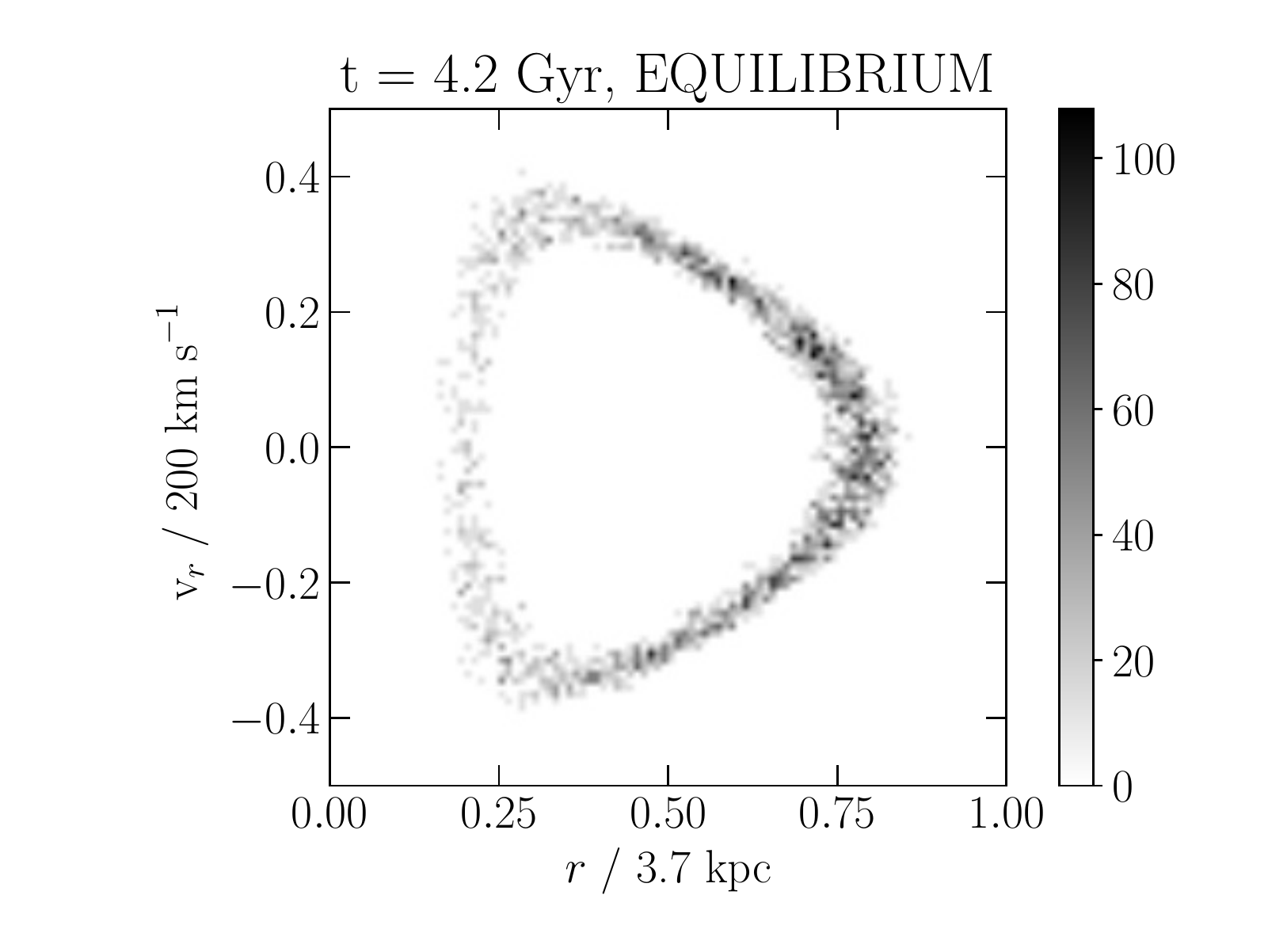}
	\caption{Initial and final configuration (after 4.2 Gyrs) of the radial 2D phase space density (sampled with 2000 tracers) of the orbital family described in \ref{orbitalfamily} in a purely gravitational simulation of a dwarf-size halo set in equilibrium. The radial coordinate is in units of 3.7 kpc, whereas the radial velocity is in units of 200 km $\mathrm{s}^{-1}$. The colour bar refers to the 2D phase space density in units of $[\mathrm{s\,km}^{-1}\mathrm{kpc}^{-1}]$. 
	In the long term, the tracer particles are preferentially located near the apocentre of their orbits, which is in general agreement with expectations from collision-less kinematics. Furthermore, the spread of the distribution near the apocentre increases, which can be attributed to the numerical integration errors discussed in section \ref{errordisc} (see fig. \ref{errorsstable}). Despite this, the overall evolution in phase space is minimal, which allows us to use the final distribution as a reasonable baseline of comparison with the simulations in the adiabatic and impulsive core formation scenarios in Figs.~\ref{comprvr} and \ref{singleexp}.}
	\label{baselinervr}
\end{figure*}

The lower panels of Fig.~\ref{errorsstable} show the dependence of the orbital errors on the (time averaged) radius of the orbit. The sample of tracers is split in radial bins according to the individual (time averaged) radius with the solid line in the left (right) panel showing the mean value of $\delta_r/\langle r\rangle$ ($\delta^{\rm max}_{\rm v_r}/{\rm v_c}$) over the ensemble of tracers, while the shaded area shows the Poisson sample variance. We find that the orbital errors are larger at smaller radii, where the impact of particle discreteness is larger. However, over the radial range shown in the plots, the mean of the orbital errors changes by less than a factor of 2. 
Thus, we conclude that the
central value of the upper left histogram in fig. \ref{errorsstable} suffices as the leading order estimate for the relative baseline error in the orbital radius. In the case of errors in orbital velocities, we can see that relative to circular velocities, the median of the errors is around 5-7 per cent. We note that in absolute terms the distribution of errors is centered at around 4~$\mathrm{km\,s}^{-1}$. The approximate magnitudes of these errors will help us to quantify the significance of any detected discrepancy between the behaviour of tracer particles when undergoing adiabatic or impulsive core formation.

\subsection{A single orbital family}\label{sec_res_orbit}

To study the evolution of the orbital family defined in Section~\ref{orbitalfamily}, we will analyse the radial 2-dimensional phase space density of orbits in the $r-v_r$ plane. Although this is only a projection of the six-dimensional phase space density, it contains the relevant information to describe the orbits of the tracers since the equation of motion in a spherically symmetric potential is two-dimensional.
Fig.~\ref{baselinervr} shows the initial (left) and final (right) radial 2-dimensional phase space density of the 2000 tracers embedded in the halo in equilibrium after 4.2 Gyrs.  
As we can see in the left panel, the phase space of the orbital family is not sampled completely even initially, particularly in the vicinity of the apocentre and the pericentre of the orbits where the sampling is sparse. This is due to a non-perfect random number generator for the uniform distribution, which undersamples the extrema. Nevertheless, the distribution of points essentially fills a closed area in phase space, which was our intention in Section~\ref{orbitalfamily} when introducing the orbital family configuration.
After 4.2 Gyrs of evolution within a halo that essentially remains in equilibrium (see Appendix~\ref{app2}), we find that the orbital family remains within essentially the same phase space area.
There is some minor evolution, however, most noticeably the phase space area is more evenly distributed and has now spread slightly compared to the initial configuration. This is a consequence of the baseline errors in the orbits described in Section~\ref{errordisc}.
Another difference is that there are noticeably more stars near the apocentre than near the pericentre, which is due to the tracers spending a longer period of time in the vicinity of the apocentre where the radial velocities are lower \citep[see also][]{2013MNRAS.433.2576P}.
We take this baseline numerical diffusion in phase space as the benchmark for the extent of the irreducible artificial evolution that occurs in our simulations, and move on to discussing the evolution of the orbital family under the core formation scenarios.

\begin{figure*}
	\includegraphics[height=6.5cm,width=8.5cm,trim=2.0cm 0.5cm 0.75cm 0.5cm, clip=true]{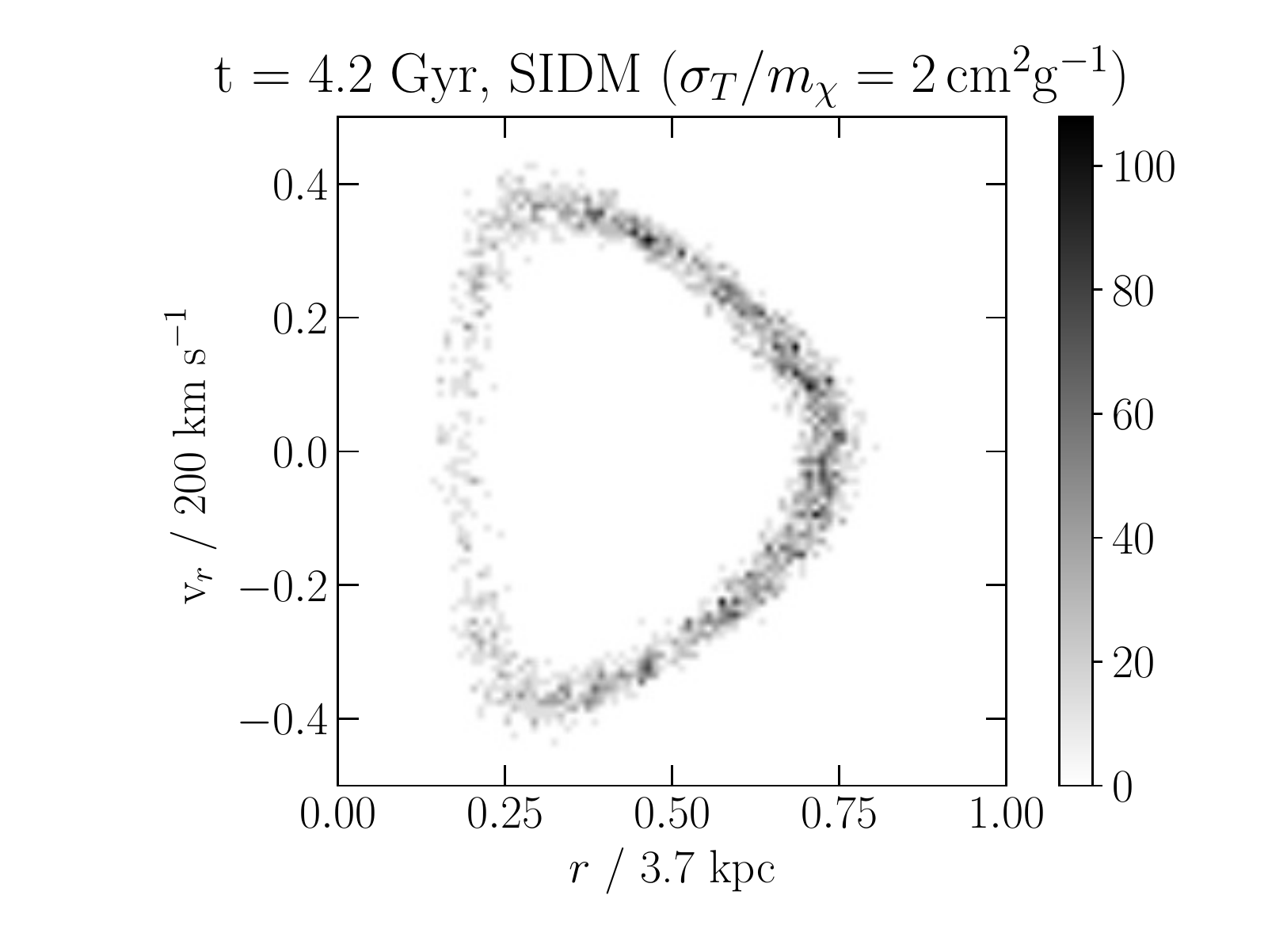}
	\includegraphics[height=6.5cm,width=8.5cm,trim=2.0cm 0.5cm 0.75cm 0.5cm, clip=true]{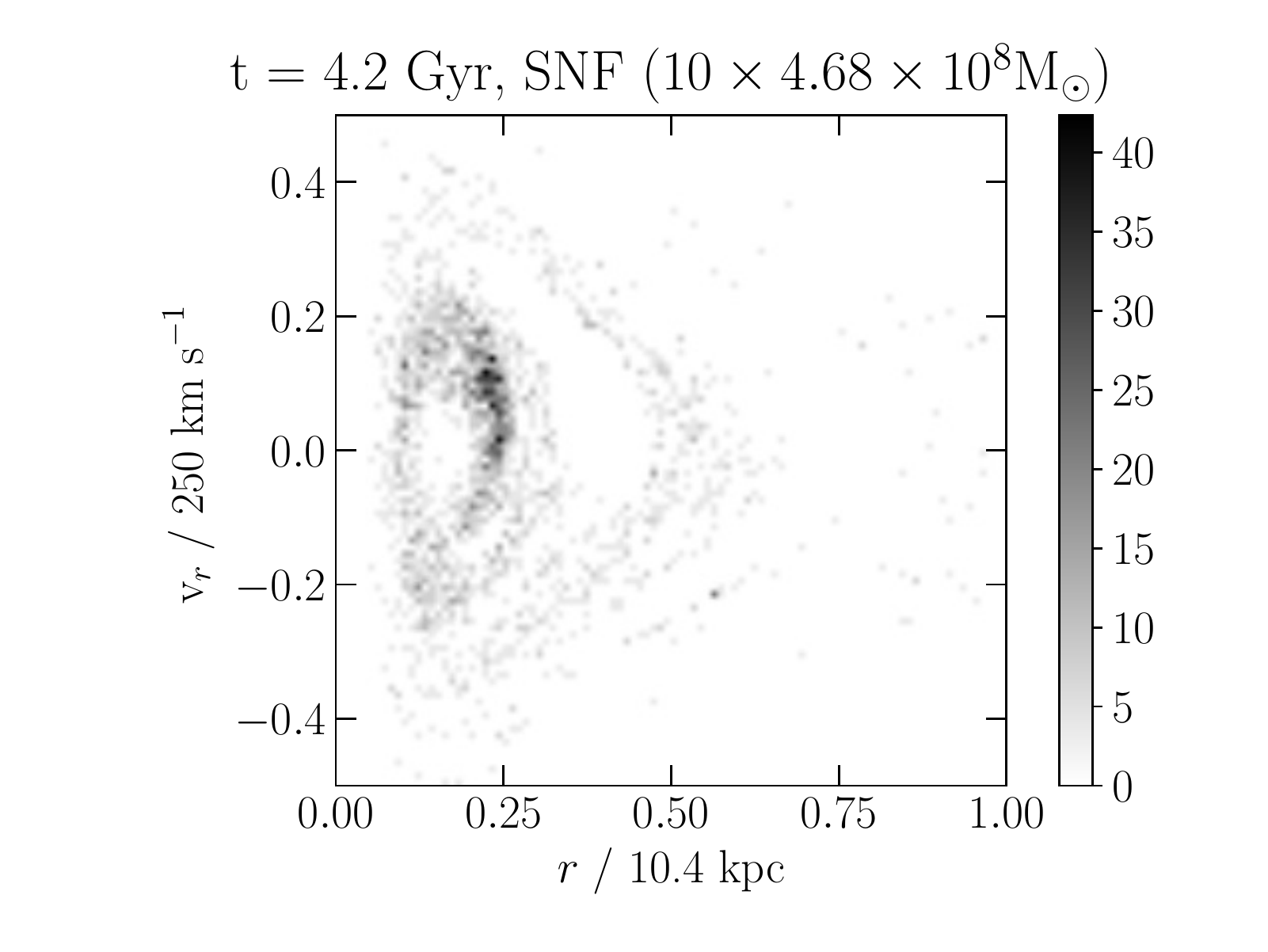}
	\caption{The same as in fig. \ref{baselinervr} but showing the final configuration (at 4.2 Gyrs) of the adiabatic (SIDM) and impulsive (SNF with 10 explosions) simulations to the left and right, respectively.  
	Notice that in the SNF case (right) the axes have been scaled differently to the SIDM (left) and equilibrium (Fig.~\ref{baselinervr}) cases. This is because in this case the tracer orbits expand to considerably larger radii in response to the impulsive mass removal, with some even expanding to radii beyond the range shown in the figure, up to 4 times as large. We choose to plot this radial range in order to focus on the most densely populated region of the phase space.  
	Although both mechanisms of core formation differ from the equilibrium case depicted in Fig. \ref{baselinervr}, the evolution in the SIDM (adiabatic) case is not very far from it, while the SNF (impulsive) case shows striking differences.}
	\label{comprvr}
\end{figure*}
Fig. \ref{comprvr} shows the 2-dimensional phase space distributions of the adiabatic (SIDM, left panel) and impulsive (SNF with 10 explosions, right panel) cases after 4.2 Gyrs. 
The final configuration in the SIDM case is remarkably similar to the equilibrium case shown in the right panel of Fig. \ref{baselinervr}, both in the area occupied by the orbital family and in the phase space density values across this area (notice that the grey scale is the same in both).
The main difference is found in the mean apocentre of the orbits, which has contracted sligthly by $\sim10\%$.
This is a result of the conservation of radial actions as the central potential is reduced adiabatically. 
This result is a nice illustration of the time averages theorem (see Section~\ref{basics}) being valid for the SIDM case, as we can directly see in the left panel of Fig.~\ref{comprvr} that tracers (stars) that are part of the same orbital family end up being part of the same orbital family after the adiabatic cusp-core transformation.

Looking at the impulsive case of core formation, shown in the right panel of Fig.~\ref{comprvr}, we find an entirely different picture. The tracers are no longer in a common orbital family, but instead they have separated into different families occupying a wider range of radial velocities and orbital radii. Some of the tracers are in bound orbital families with apocentres near $5-6$~kpc, twice as large as in the original orbital family. Other particles have moved to even larger radii out to almost $45~\,\mathrm{kpc}$, near the virial radius of the halo.
This clear separation of the orbits is a result of the time averages theorem not being valid in the impulsive case, and can be explained from the 
results of the 1-dimensional harmonic oscillator toy model (see Section~\ref{basics} and Appendix~\ref{app1}). From this model, we expected that
the expansion of an orbit (amplitude of an oscillator) after an explosion is determined by the magnitude of the change in potential (energy deposited) as well as the particular location of the tracer along its orbit (phase of the oscillator). It is due to this dependence on the phase of the orbit that the tracers group into different orbital families as the explosions occur.
We notice that a substantial fraction of the tracers is found in new orbits with a smaller apocentre than the initial orbital family (shown in the left panel of Fig.~\ref{baselinervr}).
This is likely due to the periodic cycles of growth of the baryonic mass distribution in the multiple explosion case (see Section~\ref{pesec}), which cause an adiabatic contraction of the collision-less tracers and DM particles.

\begin{figure}
	\includegraphics[height=6.5cm,width=8.5cm,trim=2.0cm 0.5cm 0.75cm 0.5cm, clip=true]{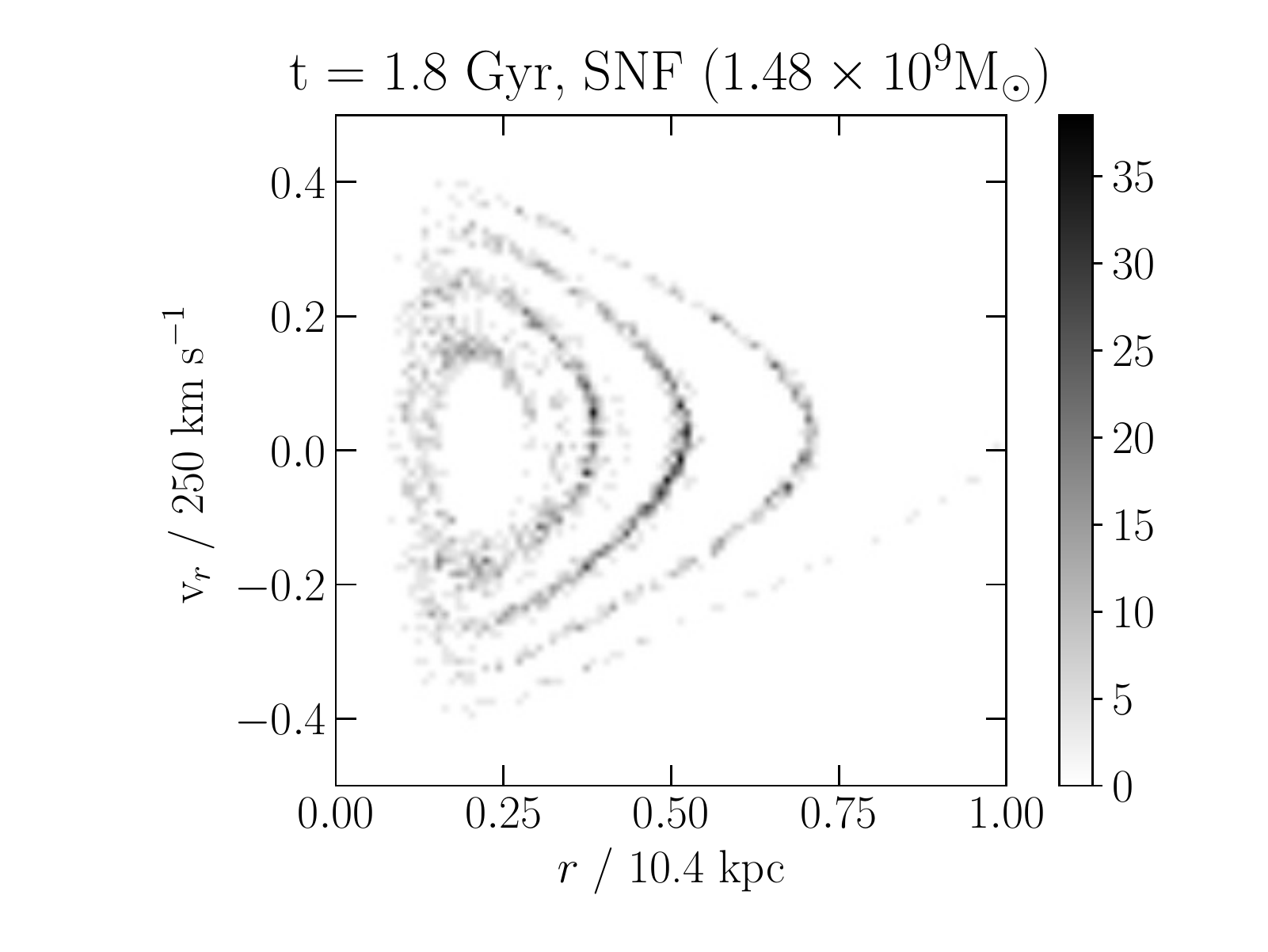}
	\caption{The same as the right panel of Fig.\ref{comprvr} but for the simulation with one single explosion. The time of the simulation is 1~Gyr after the explosion occurred (1.8~Gyrs from the simulation start). We can see that the initial orbital family (left panel of Fig.~\ref{baselinervr}) has been split into at least 3 distinct shells (orbital families), and a more diffuse area closer to the initially occupied phase space area. 
	Furthermore, we can identify a clear infall orbit 
	starting at the largest radii occupied by tracers.} 
	\label{singleexp}
\end{figure}
\noindent

\begin{figure*}
	\includegraphics[height=10cm,width=14cm,trim=0.5cm 0.5cm 0.5cm 0.5cm, clip=true]{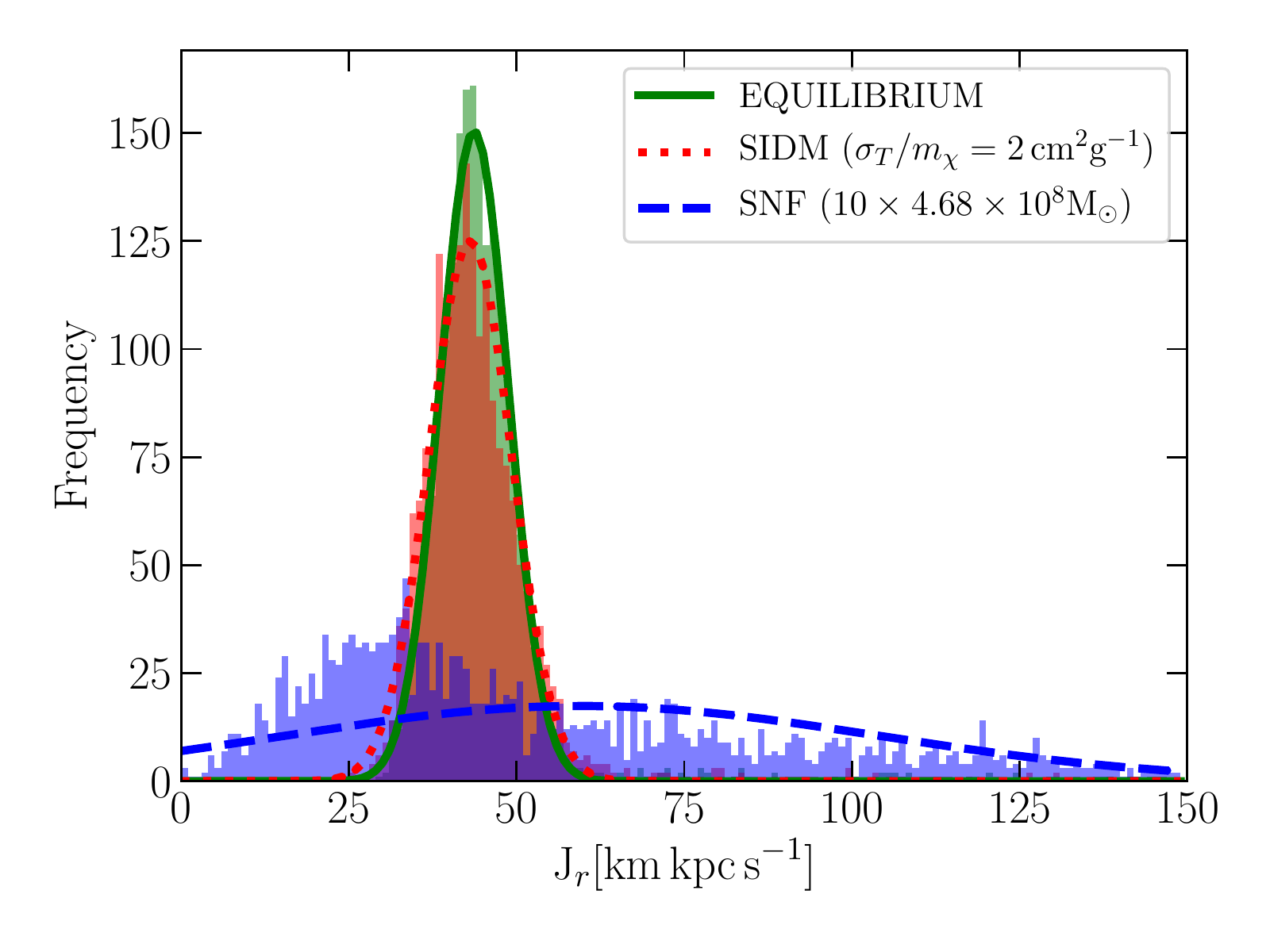}
	\caption{Distributions of radial actions after 4.2 Gyrs for the 2000 tracers originally set up following a Gaussian distribution in $J_r$ (see Section~\ref{sec_Jr_setup}). In green, red, and blue we show the equilibrium, adiabatic (SIDM), and impulsive (SNF with 10 explosions) cases respectively.
	The bin size for $J_r$ is 1~km kpc s$^{-1}$. The dashed lines are Gaussian functions with the mean and standard deviation of each distribution (see Table~\ref{tab_simulations}) as their parameters).}
	\label{raddists}
\end{figure*}

To look at the impact of the impulsive case on the orbital family in a more clear way, we show in Fig.~\ref{singleexp} the single explosion model of core formation. The 2-dimensional phase space density is shown $1.05\,\mathrm{Gyrs}$ after the explosion. Here the segregation of orbits into distinct orbital families (shells in phase space) is more clear than in the right panel of Fig.~\ref{comprvr}, where the combined effect of multiple explosions has diluted the effect.
We note, however, that as time progresses, more shells appear in this single explosion case as well, reducing the contrast between the different orbital families and thus diluting the overall effect.
We have chosen the time shown in Fig.~\ref{singleexp} to highlight the appearance of the shell-like structure of orbits.
Finally, we also identify a clear infall signature towards the halo centre, starting beyond $10$~kpc, which is also visible, albeit barely, in the multiple explosion case. The tracers in this orbit are those that were in an initial orbital phase that maximised the orbital expansion when the potential suddenly changed. They were thus expelled to very large radii and are now falling back into the halo. This resonant amplification of the energy transferred to the orbit is a distinct feature of the impulsive case. Along with the segregation of the original orbital family, it causes the final configuration of the SNF (single explosion) case to differ significantly from the adiabatic case.
These effects are, however, not as striking in the more realistic episodic explosion core formation scenario. Nevertheless, the preservation of the original orbital family in the adiabatic (SIDM) is very clear in general, and exhibits a dramatic contrast to the destructive nature of the impulsive (SNF) case.

\subsection{Gaussian distribution in radial action}\label{sec_res_Jr}

To analyse the degree to which the conservation of radial action is satisfied in the adiabatic (SIDM) case and violated in the impulsive (SNF) case, we examine the evolution of the tracers originally sampled from a Gaussian distribution in radial action (see Section~\ref{sec_Jr_setup}).
Fig. \ref{raddists} shows the final distributions of radial actions for the tracers in the 
benchmark equilibrium case (green), SIDM (red) and SNF (10 explosions, blue) cases.

To quantify the differences between the distributions, we calculate the mean and standard deviations of each distribution. We then interpret the final distribution as a Gaussian with parameters $\mu$ and $\sigma$ given by the calculated mean and standard deviation. We also quantify how much the actual final distribution deviates from a Gaussian with the calculated mean and spread by calculating $\chi^2/{\rm dof}$  (see Table~\ref{tab_simulations}).
We also investigate the impact of the baseline numerical errors presented in Section~\ref{errordisc} (see Fig.~\ref{errorsstable}) on the numerical diffusion of the initially seeded Gaussian distribution.
We do this by calculating the mean of the relative error in radius $\langle\delta_r/\langle r\rangle\rangle$ and the mean of the absolute error in radial velocity $\langle\delta^{\rm max}_{v_r}\rangle$. We then assume that the errors are Gaussian and use the mean errors to re-sample the tracers' phase space coordinates in the following way:
\noindent
\begin{enumerate}
	\item start with the initial Gaussian distribution $p(J_r)$ for the tracers.
	\item replace the orbital radius $r$ of the tracers by a randomly sampled value from a Gaussian distribution with a mean equal to $r$, and a standard deviation equal to the $\langle\delta_r/\langle r\rangle\rangle\times r$.
	\item replace the radial velocities $v_r$ of the tracers by a randomly sampled value from a Gaussian distribution with a mean equal to $v_r$, and a standard deviation equal to the $\langle\delta^{\rm max}_{v_r}\rangle$.
	\item re-calculate the action with the new coordinates and velocity vector components.
	\item repeat for every tracer and construct the new radial action distribution.
\end{enumerate}
Since we are including the (average) maximum radial velocity baseline errors\footnote{We recall that these are the maximum values over the tracers' orbits in the baseline benchmark, averaged over all tracers; see Section~\ref{errordisc}.}, we expect that this estimate will overestimate the numerical diffusion, if there are indeed no other sources of diffusion.
\begin{table*}
	\begin{center}
		\begin{tabular}{cccccccccc}
			\hline
			Simulation    &  $\sigma_T/m_\chi$ & Exploding mass & mean $J_r$ & spread in $J_r$           & Number of tracers              & $\chi^2$/dof   \\
			& 	[cm$^2$g$^{-1}$]	&	[${\rm M}_{\odot}$] &  $[\mathrm{km\,kpc\,s}^{-1}]$          & $[\mathrm{km\,kpc\,s}^{-1}]$    &  &                \\
			\hline     
			\hline 
			Equilibrium      &$-$  & $-$ & 43.7   & 5.3  & 1928     & 1.2 \\
			SIDM     & $2.0$ & $-$ & 43.1 & 5.9   & 1928  & 1.7   \\
			SNF     &$-$  & $10\times 4.68\times 10^{8}$ & 59.7 & \bf{44.2}          &  1928  & \bf{5.7}     \\
			\hline
			Initial     & $-$ & $-$& 44.9 & 3.5    & 1928   &  1.8 \\
			Re-sampled & $-$ & $-$ & 50.0 & 7.2    & 1901   &  1.8 \\
			\hline
		\end{tabular}
	\end{center}
	\caption{The columns list the main properties of the final radial action distributions for the tracers in each of the simulations shown in Fig.~\ref{radialaction}, as well as the initial and re-sampled Gaussian distributions. Column(1): simulation case; 
	Column (2): self-interaction transfer cross-section per unit mass for the SIDM case; Column(3): the total mass blowout in the SNF case (in 10 explosions); Columns (4-5) the mean and standard deviation of the radial action distribution; and Column (6): the amount of tracers taken into account in these distributions. The reason for not taking all of the 2000 particles into account each time lies in inaccuracies of the initial rejection sampling algorithm (see text for details). 
	Finally, Column (7) shows the $\chi^2$/dof with respect to a Gaussian with the calculated mean and spread in each case. Notice the large deviation of the SNF over the Gaussian distribution (highlighted in bold font).} 
	\label{tab_simulations}
\end{table*}

In Table~\ref{tab_simulations} we give an overview of the main characteristics of the radial action distributions in the different cases.
Since it is difficult to achieve a full convergence of the iterative rejection sampling algorithm in the tails of the Gaussian distribution, we note that we have removed $\lesssim5\%$ of the tracers that could not be assigned a sensible value of $J_r$.
In particular, we limit our analysis to those tracers that are initially at least within 5$\sigma$ of the mean of the initial Gaussian distribution. 
We then tag these particles and follow their evolution throughout the various simulations. 

By looking at the results in Fig.~\ref{radialaction} and Table~\ref{tab_simulations} we draw several interesting conclusions.
Firstly, by comparing the result of the simulation that evolves in equilibrium (green in Fig.~\ref{radialaction}) with the re-sampled distribution explained above (bottom row of Table~\ref{tab_simulations}), we can be certain that the numerical diffusion of the radial action distribution is fully contained within the expectations of the baseline orbital errors described in Section~\ref{errordisc} (notice that the spread of the re-sampled distribution is larger than that of the final configuration of the equilibrium simulation). Thus, the growth of the spread in the equilibrium case by the end of the simulation represents the degree to which the conservation of radial action is violated purely from numerical errors. This is quantified by looking at the difference in the standard deviations of the equilibrium and initial distributions shown in Table~\ref{tab_simulations}, which amounts to $\sim50\%$. 
Looking at the SIDM case, we find that the result is consistent with our expections based on the adiabaticity of the change in potential in this case. The radial action distribution is still well-fitted by a Gaussian that is quite close to the equilibrium case, albeit with slight distortions and a larger spread. The degree to which the radial actions are conserved in the SIDM case is remarkable.

The key result in this Section is the comparison between the final distributions of the impulsive (blue line) and adiabatic (red line) cases of core formation. 
It is clear that 
the initial Gaussian distribution is entirely dissolved in the impulsive case, with a final distribution considerably broader and skewed towards low $J_r$ values, relative to the mean of the original distribution, and with a long tail of very large $J_r$ values. The spread is in fact larger than in the SIDM case by a factor of $\sim7.5$ and the interpretation of the final distribution as a Gaussian is significantly worse than in the SIDM case (see the values of $\chi^2/$dof in table \ref{tab_simulations}).
The striking difference between the SIDM and SNF case in the degree to which radial actions are conserved in the former, and violated in the latter offers a promising guideline to look for observational quantities in cored dwarf galaxies that are related to radial actions, which could probe the adiabatic or impulsive nature of the core formation mechanism.

\subsection{A Plummer sphere}

\begin{figure*}
	\includegraphics[height=6.5cm,width=8.5cm,trim=0.5cm 0.5cm 0cm 0.0cm, clip=true]{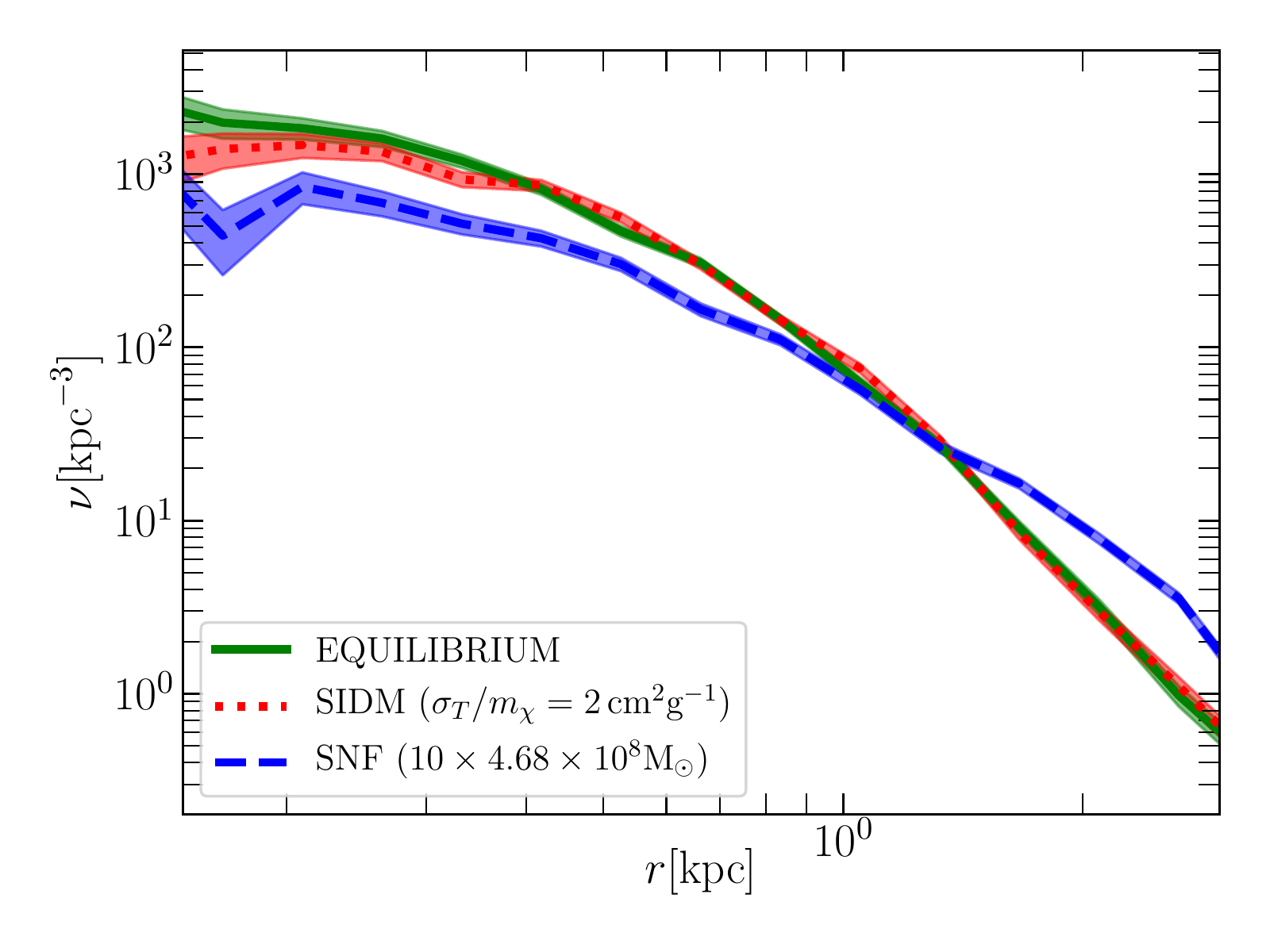}
	\includegraphics[height=6.5cm,width=8.5cm,trim=0.5cm 0.5cm 0cm 0.0cm, clip=true]{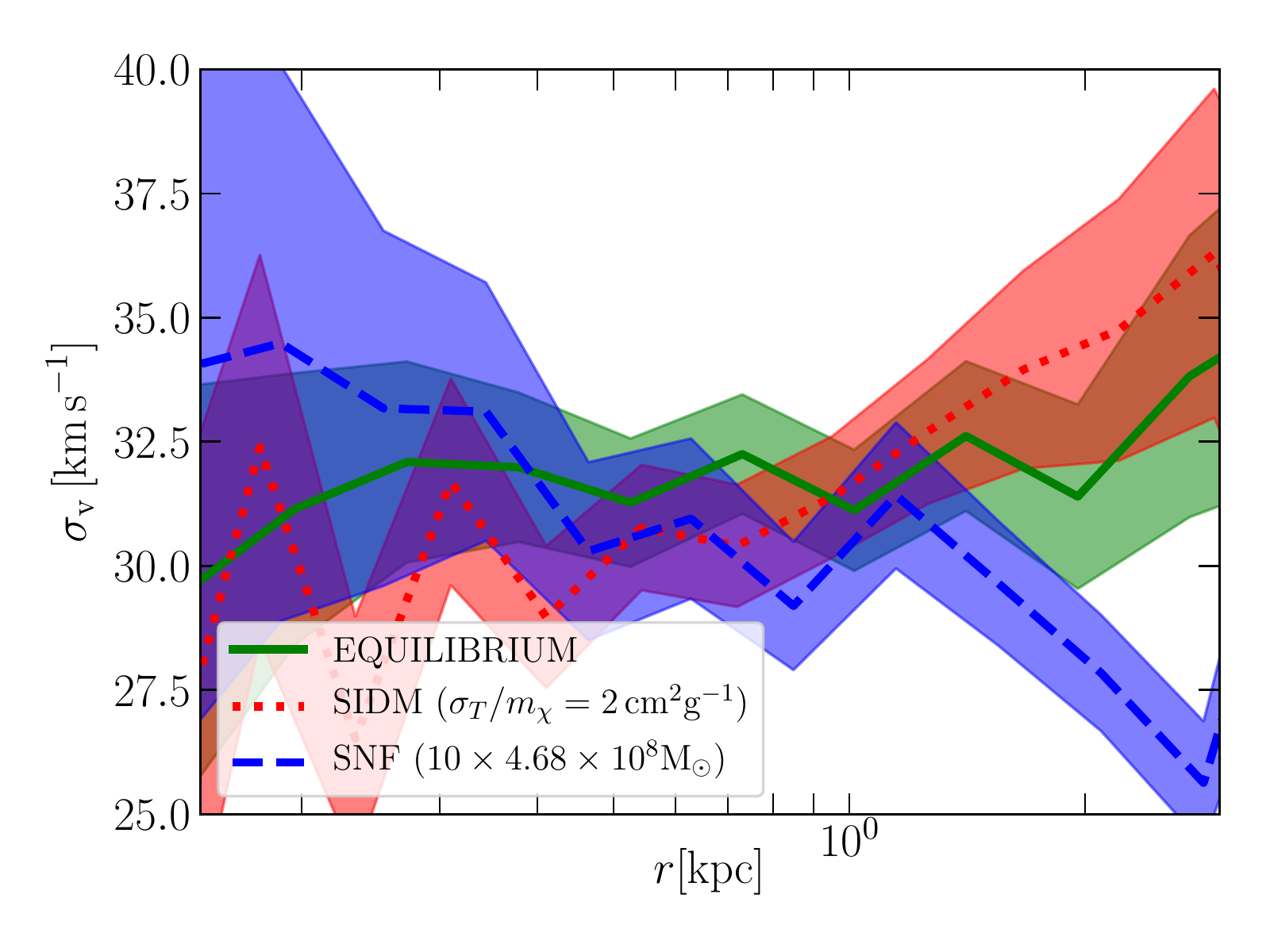}
	\caption{Impact on the stellar distribution (Plummer sphere) due to cusp-core transformation in the impulsive (SNF) and adiabatic (SIDM) scenarios. We show the number density and velocity dispersion profiles of the stars in the left and right panels, respectively. The stars are modelled as tracers initially following a Plummer density profile with a half light radius of 500 pc. The initial setup as an equilibrium configuration works as in the case of the Hernquist halo. 
	The SIDM case is shown after the isothermal core is fully formed, while the SNF and equilibrium cases are shown after 4.2 Gyrs, which is the time where the SNF DM core has the same central density as in the SIDM case (see Fig.~\ref{profileevolution}). The different cases are colour-coded as in previous figures. 
	The error bands correspond to Poisson counting errors, which are larger than the (systematic) numerical integration errors (see Fig.~\ref{errorsstable}). The number density in the impulsive (SNF) case is modified significantly relative to the equilibrium and adiabatic (SIDM) cases. The stellar kinematics (velocity dispersion profile) imply that the final SNF configuration is likely not in equilibrium.} 
	\label{plummer}
\end{figure*}

For the final configuration of tracers, we
investigate how a Plummer sphere with a half-light radius of 500 pc (typical of the bright Mikly Way satellites) evolves in the three scenarios. To obtain an adequate comparison to the behavior of the DM particles, we show the profiles at the same times as in Fig.~\ref{profileevolution} (after 4.2 Gyrs for SNF and equilibrium, after 1.2 Gyrs for SIDM). 
Fig. \ref{plummer} shows the number density (left panel) and the velocity dispersion (right panel) profiles. The solid lines show the calculated mean of the radial bins while the shaded area in the left panel shows the (Poisson) counting variance. 
The shaded area on the right panel shows the sampling errors on the velocity dispersion calculated from the biased estimator of the variance when assuming an underlying Gaussian distribution (see, for instance \citealt{LehmCase98}),
\begin{align}
	\delta_{\sigma_v} = \frac{\sigma_v}{\sqrt{2(N_{\rm shell}-1)}},
\end{align} 
where $N_{\rm shell}$ is the number of stars within the radial shell in which we calculate the velocity dispersion. 

As in the other two configurations of tracers, we find that there is a significant difference in the final tracer distribution depending on whether core formation proceeds adiabatically or impulsively.
In the adiabatic (SIDM) case, the radial distribution of the Plummer sphere remains close to the equilibrium case, showing only a slight reduction in central density (comparing the red and green lines in the left panel of Fig.~\ref{plummer}).
The velocity dispersion profile also responds adiabatically, moving from an almost flat profile (in the radial range shown in Fig.~\ref{plummer}) at the beginning (green line) to a slghtly steeper profile with positive slope.
The adiabatic formation of the DM core drives the slow evolution of the tracers towards a new state of Jeans equilibrium. These results roughly agree with previous ones from SIDM cosmological simulations (see Figs. 6 and 8 of \citealt{Vogelsberger2014}). 
Comparing both the number density profiles and the velocity dispersion profiles of the tracers in the SIDM and the equilibrium cases, it is evident that the Plummer sphere evolution in the SIDM case is quite mild relative to the evolution of the DM profile (see Fig.~\ref{profileevolution}). In fact, just as we have found in the previous Sections \ref{sec_res_orbit} and \ref{sec_res_Jr}, there is no clear observational signature of the formation of the core by just looking at the spatial and kinematic distribution of tracers (stars).

The impulsive case of core formation (SNF with 10 explosions), on the other hand, triggers a strong evolution of the Plummer sphere. 
Particularly, the number densities are considerably lower than in the equilibrium case within the half light radius of the sphere with the tracers being pushed out towards radii beyond $\sim1$~kpc, which would correspond to a more extended galaxy with a lower central surface brightness.
The velocity dispersion profile of the SNF case has similar central values as in the other cases. However, it has a negative slope with radius, which is particularly apparent beyond 1~kpc. 
Although the sampling errors are relatively large, the evidence for this trend seems sufficient.
The behaviour of the velocity dispersion profile seems to indicate that the tracers are not in isotropic equilibrium. This is most clearly seen within $\sim400$~pc, where the DM density profiles are very similar in all simulations (see left panel of Fig.~\ref{profileevolution}), while the profile of the tracers has essentially the same shape, but with considerable deficit in the SNF case relative to the other simulations (left panel of Fig.~\ref{plummer}). Under the assumption of isotropic Jeans equilibrium, these conditions would imply that the central velocity dispersion in the SNF case should be lower than in the other simulations. Although the sampling errors are large in the centre, this is not what is seen in the right panel of Fig.~\ref{plummer}, implying that the tracers are not in isotropic equilibrium. This circumstance could be a transient state resulting from the strong fluctuations caused by the multiple accretion and explosion events in this SNF configuration.

Overall, we can conclude that the response of a Plummer sphere to the formation of a DM core {\it of the same central density} is considerably different in the adiabatic and impulsive cases. For the former, the properties of the sphere remain remarkably unchanged, while the destructive nature of the latter one causes the centre of the sphere to expand considerably. This expansion proceeds beyond the extent of the central DM core, thus putting the orbits of the tracers out of equilibrium (at least within the time scales we simulated).

\section{Behaviour of a cosmological halo}\label{section:cosmohalo}

The results so far are based on the assumption of an isolated, spherically symmetric DM halo. This was done in order to test the theoretical predictions of a direct link between 
conserved quantities derived from the initial symmetry of the halo (angular momentum and radial action) and the nature of the process responsible for triggering the formation of the DM core. However, it is evident that realistic DM haloes are better described as being triaxial rather than spherically symmetric, and that they are clearly not isolated but rather assembled by a hierarchical merging process throughout their cosmological 
formation history.
When looking at real haloes we thus do not expect all of the predictions made in section \ref{results} to remain the same. In particular, one can argue that the results shown in Fig.~\ref{raddists} will be quite different in realistic haloes because 
radial actions will no longer be conserved in haloes with a significant degree of triaxiality, and thus the differences between the adiabatic and impulsive core formation scenarios that are obvious in Fig.~\ref{raddists} will be diluted.
This lack of conservation of radial actions for orbits in non-spherical potentials was for instance shown explicitly in \citet{Pontzen:2015ova} for a 3D anisotropic harmonic oscillatior, and for a cosmological dwarf-size halo (see in particular their figures A1 and A2).

\begin{figure*}
    \includegraphics[height=6.5cm,width=8.5cm,trim=2.0cm 0.5cm 0.75cm 0.5cm, clip=true]{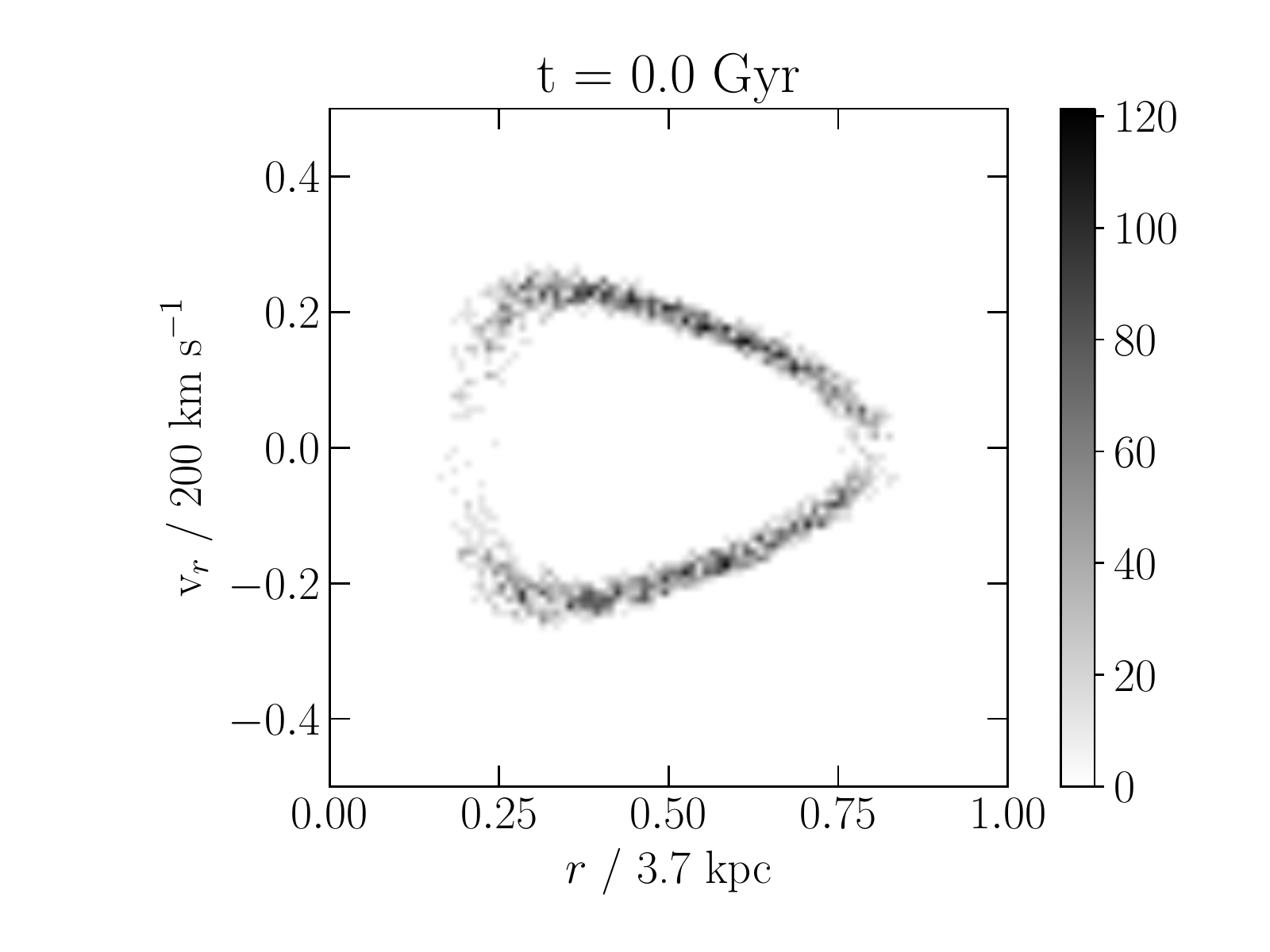}
    \includegraphics[height=6.5cm,width=8.5cm,trim=2.0cm 0.5cm 0.75cm 0.5cm, clip=true]{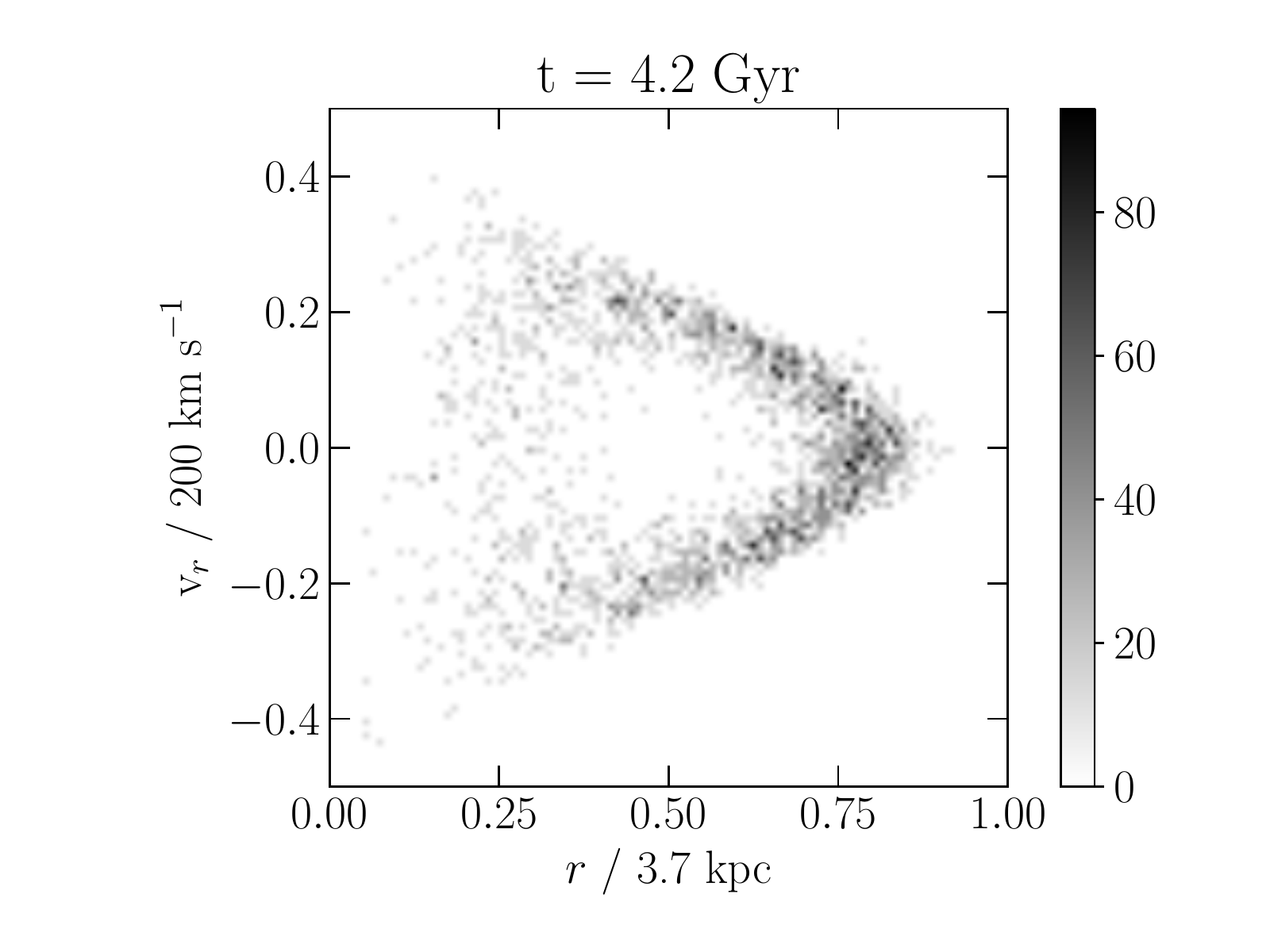}
    \caption{Evolution of the radial 2D phase space density of the orbital family described in Section \ref{sec_res_orbit}, but set up within the $z=0$ cosmological dwarf-size halo taken from \citet{Vogelsberger2014} and without any core formation mechanism.
    The left panel shows the initial configuration whereas the right panel shows the result after 4.2 Gyrs. As in figure \ref{baselinervr}, the scale on the x-axis is unchanged and all of the tracers remain within our initially chosen patch of phase space. However, there is a striking evolution in the phase space occupation which we attribute mainly to the triaxiality of the halo and the resulting non-conserved angular momentum. In particular, the final result exhibits a heavily broadened and scarce density in the pericentre vicinity relative to the original distribution of the orbital family.}
    \label{fig:cosmo_equi}
\end{figure*}

Regarding the predictions of Fig.~\ref{plummer}, we argue that they will remain qualitatively the same, as long as orbits expand significantly in the SNF case and stay relatively localized in the case of SIDM.  The most striking result we presented in the spherically symmetric case is that of the appearance of distinct features (shells) in phase space (discussed in Section \ref{sec_res_orbit} and shown in Figs.~\ref{comprvr} and \ref{singleexp}) due to the impulsive episodes of energy injection that lead to core formation in the SNF case. To test whether or not these features survive in a realistic halo, we took the zoom-in dark-matter-only simulation of a dwarf-size halo presented in \citet{Vogelsberger2014}. This halo (labelled dA-CDM-B-hi) is similar in mass and size to the one we analysed here and it also has a similar numerical resolution; its main properties are  
given in tables 2 and 3 of \citet{Vogelsberger2014}. The relevant ones for our purposes are: $M_{200}=1.07\times 10^{10}{\rm M}_{\odot}$, $r_{200} = 45.8 \,{\rm kpc}$, $\epsilon  = 34.2 \,{\rm pc}$, and a particle mass $m_{\rm DM} = 9.7 \times 10^2 {\rm M}_{\odot}$ (equivalent to $\sim1.1\times10^7$ particles within $r_{200}$). 

To perform our analysis, we only consider the particles in the original simulation at $z=0$ within a $(500 {\rm kpc})^3$-sized box around the centre of potential of the halo which will be the origin of the reference frame used in the subsequent analysis. 
We calculate the centre of the potential iteratively considering all particles within concentric shells with radii of 75, 7.5 and 0.75 kpc using a weighted mean with the potential at the position of each particle as a weight. This is done during the initial setup of the tracer particles as well as during each timestep of the orbital evolution. By using the centre of the potential instead of the center of mass as the origin of the reference frame we are taking into account the asymmetric distribution of mass in the halo. We note that the halo chosen has had most of its major merger activity in the past and is relatively quiescent by $z=0$.

\subsection{Orbital family in a cosmological halo}\label{sec_orb_cosmo}

\begin{figure*}
    \includegraphics[height=6.5cm,width=8.5cm,trim=2.0cm 0.5cm 0.75cm 0.5cm, clip=true]{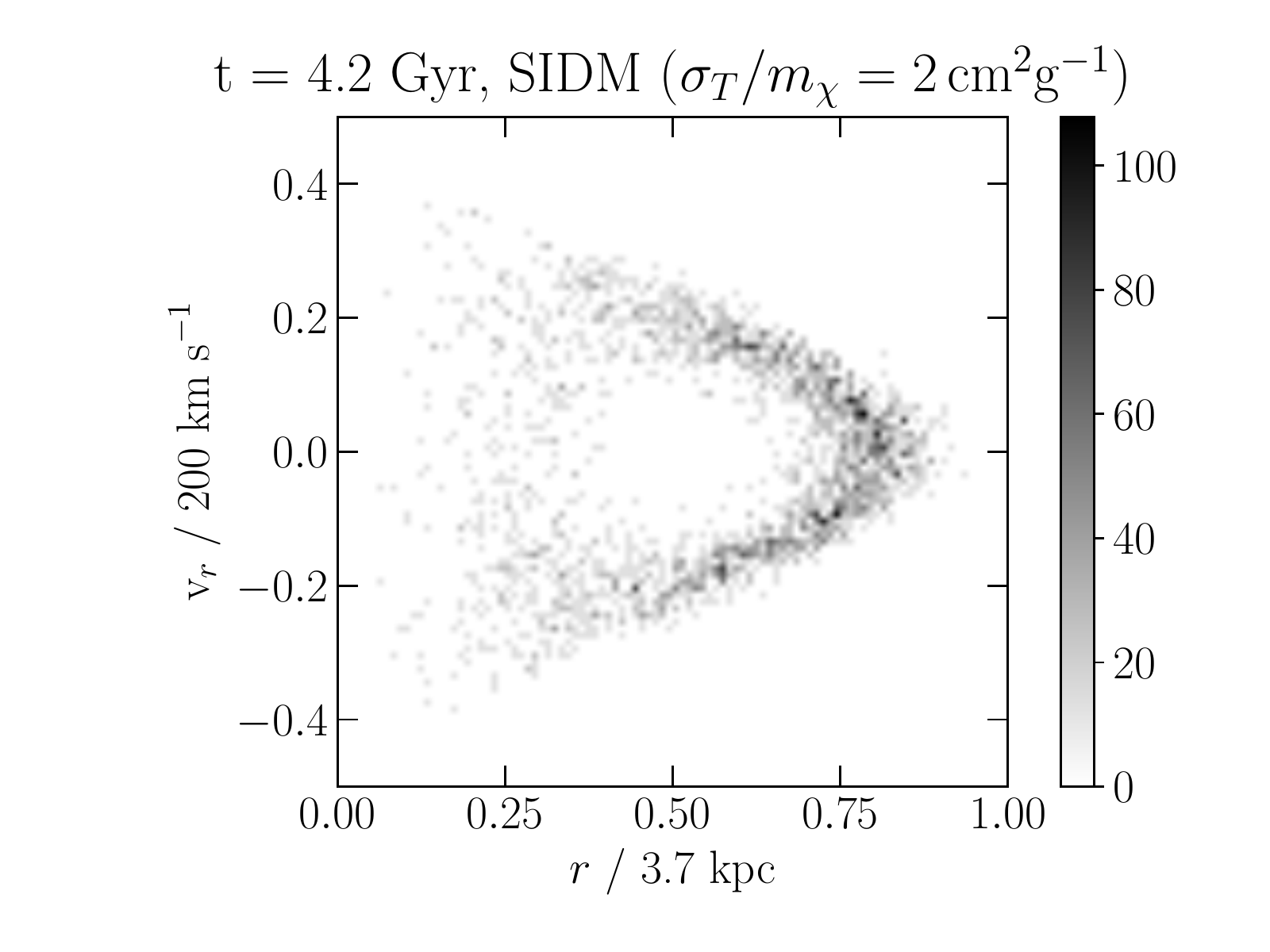}
    \includegraphics[height=6.5cm,width=8.5cm,trim=2.0cm 0.5cm 0.75cm 0.5cm, clip=true]{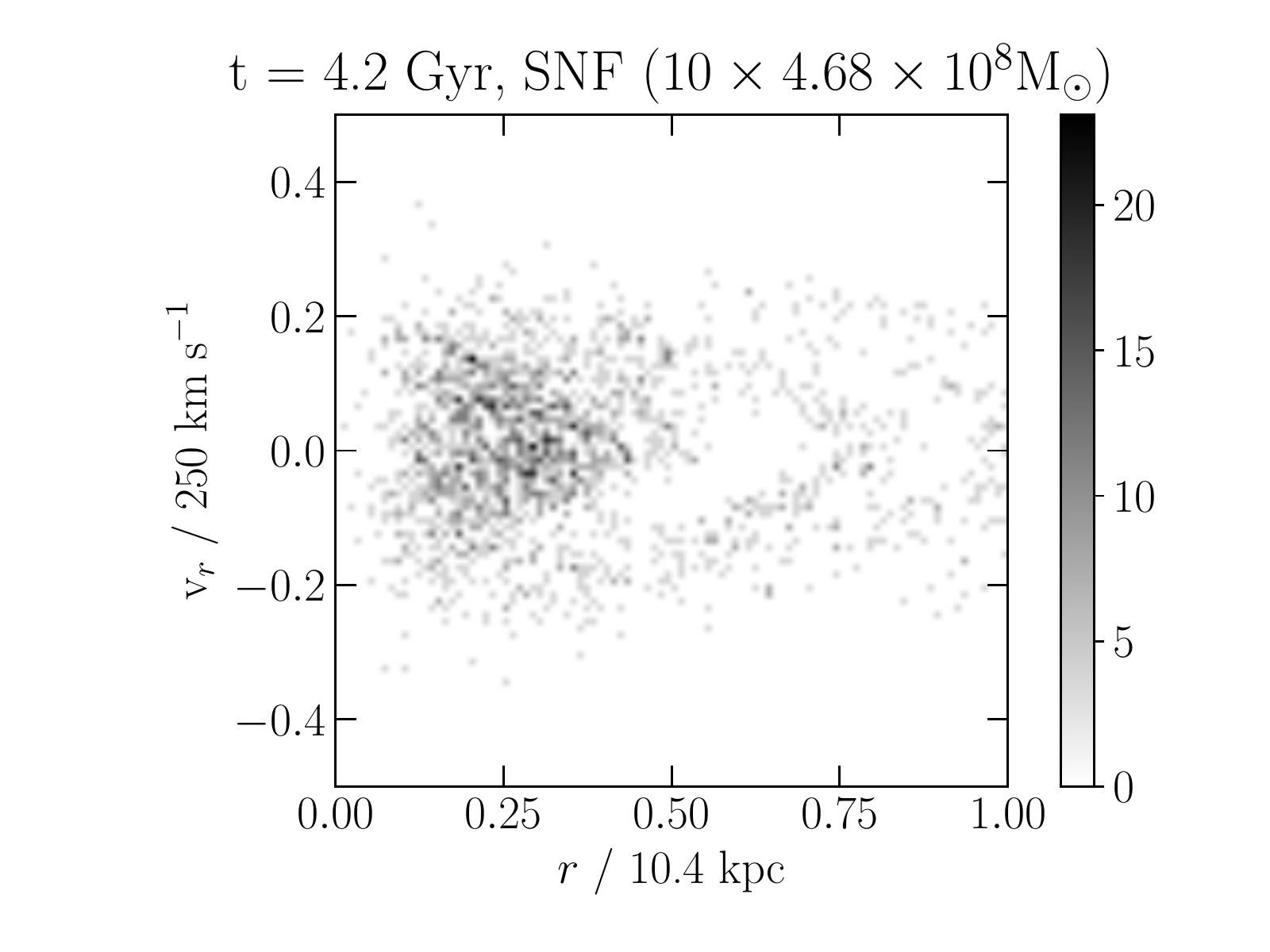}
    \caption{
    As Fig.~\ref{fig:cosmo_equi} but in the case of core formation driven by SIDM (adiabatic, left panel) and by SNF (impulsive multiple explosions, right panel) in a cosmological halo. The final profile in the case of adiabatic core formation can hardly be distinguished from the final phase space density in the baseline 
    case shown in the right panel of Fig.~\ref{fig:cosmo_equi}. The SNF case, however, clearly shows a strong average radial expansion compared to the SIDM case. Notice the change of scales on the x-axis in the right panel. The shell-like structures, clearly visible in Fig.~\ref{comprvr} are no longer as distinct when the host halo is no longer idealised (spherical and smooth).} 
    \label{fig:cosmo_core}
\end{figure*}

To analyse the impact of having a cosmological halo in our results,
we take the halo described above at $z=0$ and calculate its spherically averaged enclosed mass profile and radially dependent potential and set up an orbital family of tracer particles as described in section \ref{orbitalfamily}. We then evolve the system for 4.2 Gyrs in time 
in three configurations: no core formation mechanism (baseline), SIDM and SNF. The first case will serve as our new baseline in which we can measure the impact on the evolution of the orbital family caused purely by the lack of spherical symmetry and having a subhalo distribution rather than just a smooth halo.
We show the radial 2D phase space density evolution in this baseline 
case in Fig.~\ref{fig:cosmo_equi}: the left panel
where we set the orbital family at $t=0$, while on the right
we show the configuration after $t=4.2$~Gyr. Comparing this to Fig.~\ref{baselinervr} (the spherically symmetric smooth halo), we find that the final distribution is quite different in the two cases. In particular, we find a much stronger evolution - and dilution - of the initial phase space density in the cosmological halo than was the case in the idealized spherical halo. This is especially evident in the vicinity of the pericentres of the initial orbital family, where the initially narrow distribution is heavily broadened. Qualitatively however, we still observe that 
the area around the orbital apocentre of the orbits is more populated than other segments along the trajectories. 
Furthermore, even after 4.2 Gyrs, the occupied region in phase space still exhibits key features of the initial distribution. The most notable feature is that there is still a gap in phase space between the largest pericentre and the smallest apocentre radius. Moreover, we see no considerable net expansion of the radial range occupied by the orbital family.

Most likely, the main reason for the observed discrepancy between the results in the idealized case and in the case of the cosmological halo is the triaxiality of the latter. Since spherical symmetry is broken,
angular momenta are no longer conserved, and thus,
a configuration based on the premise of such conservation 
will not remain stable while evolving in a triaxial potential. 
We notice however, that
for tracers orbiting in a potential which does not strongly violate spherical symmetry, a quantity closely related to the angular momentum is an integral of motion \citep{1987gady.book.....B}. This is particularly relevant in our case since we are interested in dwarf-size haloes in our analysis, which have been shown to be closer to spherical symmetry than larger haloes (see e.g. Fig. 13 of \citealt{Bonamigo:2014rba}). This is likely the reason why the radial phase space density of the orbital family still remains confined to a relatively narrow region, which retains a similar shape to the original despite evolving significantly within a couple of dynamical times. This picture is supported by the fact that 
once we start the simulation, most of the evolution takes place within the first 300 Myrs, indicating that the orbital family 
quickly relaxes to a new steady state set by the actual shape of the cosmological halo and then remains relatively unchanged until the simulation ends. 

Fig.~\ref{fig:cosmo_core} shows the same comparison between the adiabatic and impulsive (multiple explosions) core formation scenarios in the case of a cosmological host halo as Fig.~\ref{comprvr} did in the case of an idealized spherically symmetric halo. Just as in the idealized case, we identify strong differences between adiabatic and impulsive core formation in terms of the response of tracer particles: (i) the expansion of the orbits to considerably larger radii in response to impulsive mass removal remains a clear SNF signature; (ii) the final orbital distribution in the SIDM case
can hardly be distinguished from that of the baseline case. 

On the other hand, tn the case of SNF we find that, contrary to the idealised case, the pronounced shell structures that were clearly visible in Fig.~\ref{comprvr} have disappeared almost entirely, with the final radial phase space density in the right panel of Fig.~\ref{fig:cosmo_core} being a mixture of several families overlapping and occupying most of the available phase space.
Keeping in mind that in the multiple explosion case, the last of the 10 explosions happens at 
3 Gyrs, after which the 
system relaxes into a new stable configuration, the question arises 
on which time-scales the shell-like structures dilute and how this compares to the dynamical time-scale of the tracers. As a benchmark, we estimate the dynamical time as the time one radial oscillation takes under the assumption of spherical symmetry, which is 
\begin{align}
   t_{\rm dyn} = 2\int_{r_{\rm peri}}^{r_{\rm apo}}{\rm d}r \frac{1}{\sqrt{2\left[E-\Phi(r)\right]-\frac{L^2}{r^2}}}, 
\end{align}
where all quantities are calculated from the initial peri- and apocentre radii and the spherically averaged potential. For our cosmological halo, we calculate a dynamical time of 
$t_{\rm dyn} \approx 170 {\rm Myrs}$. 

To check how fast the shell-like features dilute, we looked at several outputs from different simulation times and assessed the respective phase space densities in the SNF case.
\begin{figure*}
    \includegraphics[height=6.5cm,width=8.5cm,trim=2.0cm 0.5cm 0.75cm 0.5cm, clip=true]{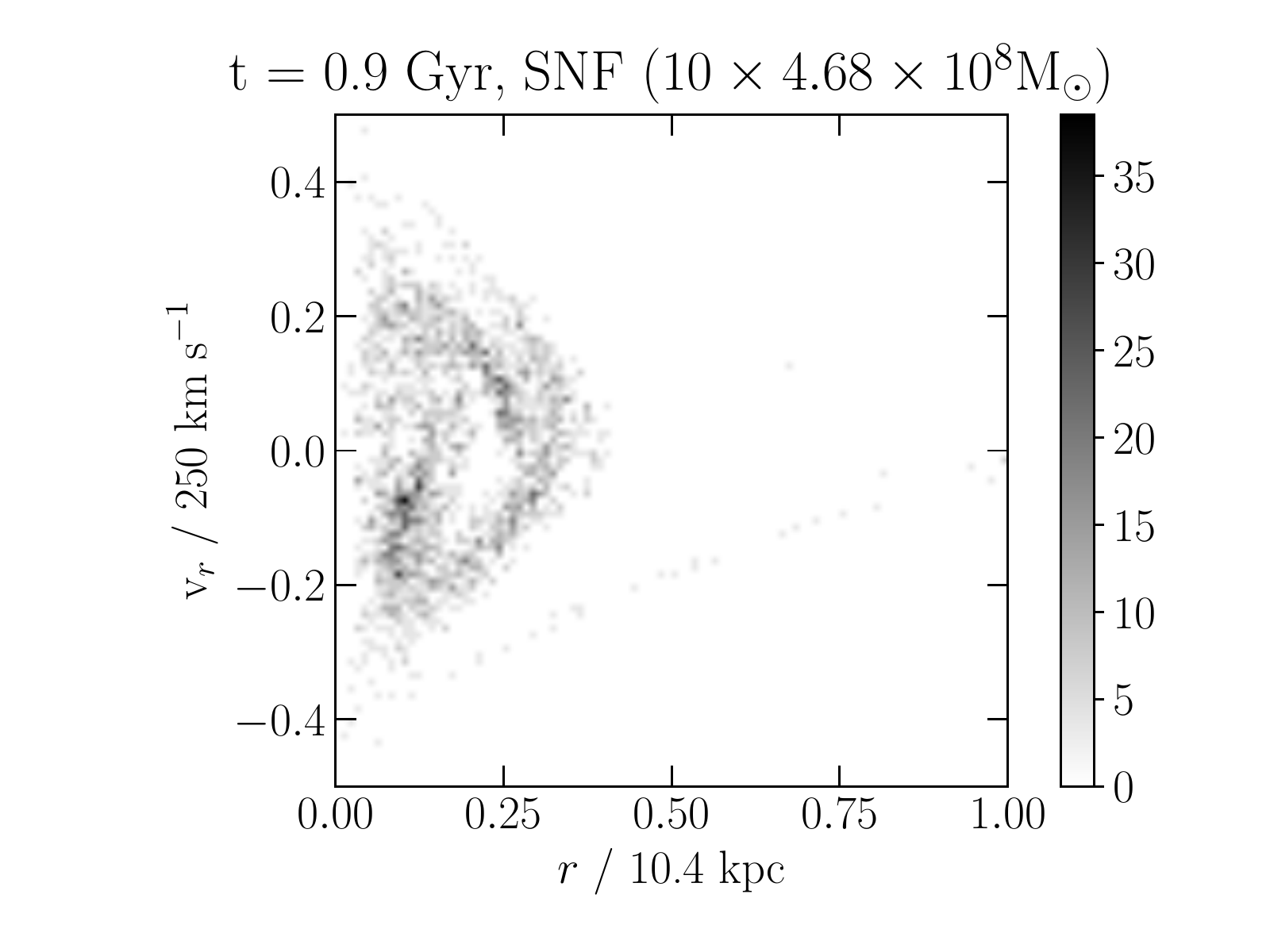}
    \includegraphics[height=6.5cm,width=8.5cm,trim=2.0cm 0.5cm 0.75cm 0.5cm, clip=true]{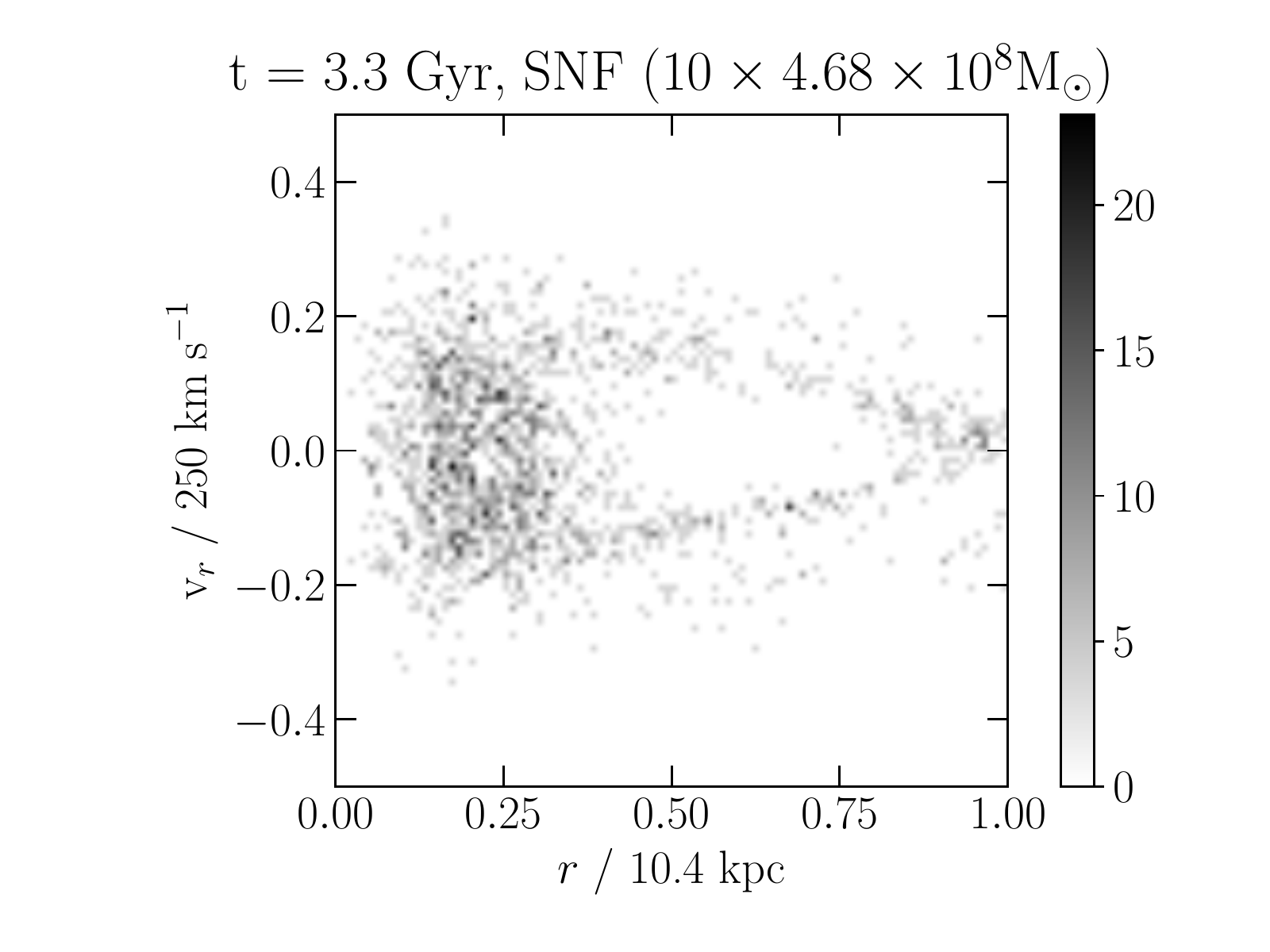}
    \caption{ 2D radial phase space density of the orbital family at two different times in the evolution of the SNF (10 explosions) core formation case in a cosmological halo (prior stages to that shown in the right panel of Fig.~\ref{fig:cosmo_core}). In the left panel,
    the time is $t=0.9$~Gyr (around the time of the third explosion), whereas in the right panel
    $t=3.3$~Gyr which is 300 Myr after the last explosion. In the left panel we can see shell-like features and a clear infall signature, which are the characteristic signatures of an impulsive event, as we pointed out in Fig.~\ref{comprvr} for the idealised halo. 
    The right panel shows that after letting the system relax for 300 Myrs after the last explosion, the phase mixing that appears in Fig.~\ref{fig:cosmo_core} (1.2 Gyr after the last explosion) is less advanced and impulsive features still remain.}
    \label{fig:cosmo_snf}
\end{figure*}
%
%
We show the 2D radial phase space density of the orbital family 
at two representative times in Fig.~\ref{fig:cosmo_snf}. In the left panel, we see 
that at an early stage of the simulation (after two explosions and right around the time of the third) the shell-like structures are indeed there and -moreover- are very pronounced, akin to what we have seen in the case of the idealized halo at the end of the simulation in Fig.~\ref{comprvr}. Furthermore, we clearly identify an infall branch in the phase space density.
It is thus the process of letting the tracer stars settle onto new orbits after a last sudden energy injection  that dilutes these features in a non-spherically symmetric potential. To investigate how long it takes for these features to disappear, we looked
at the evolution of the  orbital family
at several times after the explosion occurred. As an example we show the configuration at $t=3.3$~Gyr (0.3 Gyr after the last explosion) in the right panel of Fig.~\ref{fig:cosmo_snf}. At this time, there are at least two distinct shells that can be identified, one with apocentre at around 10 kpc and a very faint one containing only the most energetic tracers and extending out of the radial range shown in the plot.  
At the same time, however, it is apparent that orbits have already mixed and features which emerged as a result of the impulsive events have been diluted, particularly within the phase space  
region close to the halo centre. When letting the system relax further, we find that this trend quickly continues outwards, with most of the features being unidentifiable after 3.6 Gyrs. We thus estimate that the time during which effects of (impulsive) supernova-driven core formation could potentially be observed is $\lesssim0.5$ Gyrs, which equals roughly 3 orbital revolutions of the tracer particles, i.e. three dynamical times.

We have shown that the key differences between the impulsive (SNF) and adiabatic (SIDM) core formation scenarios remain even when
we abandon the assumption of a spherically symmetric isolated halo, and take instead a halo assembled in a cosmological setting, despite the fact that radial actions are no longer conserved in a non-spherical potential, an assumption that is at the core of most of our analysis. The fundamental complication in the more realistic setting of a cosmological halo is that phase mixing occurs within a few orbital time scales after an (impulsive) mass blowout episode in the SNF case, which tends to erase the remarkable shell-like features in phase space that are the clearest signatures of the impulsive scenario. Despite this complication, our results suggest that the differences in the stellar orbits between the adiabatic and impulsive core formation scenarios are relevant enough to warrant further investigation in a more realistic modelling (cosmological simulations with a full treatment of baryonic physics) and with actual stellar kinematic data in dwarf galaxies. We plan to move forward along both of these avenues in future work.

\section{Conclusion}\label{conclusion}

The surprising deficit of dark matter (DM) in the innermost regions of dwarf galaxies relative to the naive expectations of the Cold Dark Matter (CDM) model is a notable observational feature that remains poorly understood.  
Perhaps the most likely possibility to explain this DM mass deficit is to have a mechanism that efficiently deposits energy into the inner halo, effectively transforming the CDM cusp into a central core.  
Among the various possibilities, we focus on two which have been invoked extensively: supernovae and DM self-interactions (SIDM). In those two cases, the energy required to transform a cusp into a core originates from energy sources of a completely different nature, yet both are viable causes of core formation as they are capable of lowering the inner densities of CDM haloes to the observed levels (for references see Section~\ref{sec_intro}). 

In the case of supernovae, the viability of this process to explain the ubiquity of DM cores in dwarf galaxies has been discussed in terms of whether there is sufficient energy to achieve this goal \citep[e.g.][]{Penarrubia:2012bb} and whether the energy injection is efficient enough (e.g. \citealt{Tollet2016} versus \citealt{Bose2018}). On the other hand, in the case of SIDM, the discussion has focused on whether the SIDM transfer cross section per unit mass $\sigma_T/m_\chi$ needed to generate DM cores in dwarf galaxies, which is $\mathcal{O}(1)\,{\rm cm}^2{\rm g}^{-1}$, is consistent with a wide range of other astrophysical observations (see Table 1 of \citealt{Tulin2018} for a recent compilation of SIDM constraints). 

In this paper we take a different perspective and focus on the problem of how to distinguish these two mechanisms of cusp-core transformation. We approach this question in a fundamental way by considering the time scales of energy deposition relative to the dynamical time scales of DM orbits in the inner halo. If the former is much larger than the latter, as in SIDM, then the process is adiabatic, while in the opposite regime the process is impulsive and irreversible. This is the regime where supernova-driven mass outflows are effective in forming DM cores \citep{Pontzen:2011ty}.  
We argue that by looking for signatures of the adiabatic or impulsive nature of observed DM cores in dwarf galaxies, we can provide evidence for or against the necessity of new DM physics. This paper is a first exercise towards accomplishing this goal in which we analyse in detail the different evolution of stellar orbits in an idealized (spherical) potential that changes in an adiabatic or impulsive way. Furthermore, we also investigate the implications of having a realistic cosmological dwarf-sized halo providing the background potential instead of an idealised spherically symmetric halo. 

Our idealised halo model consists of a truncated Hernquist sphere with a virial mass of  
${\rm M} = 1.48\times 10^{10}{\rm M}_{\odot}$, a virial radius of $52$~kpc and a scale radius of $3.5$~kpc. These parameters are typical of the haloes corresponding to the dwarf galaxies where DM cores of $\mathcal{O}(1)$~kpc are common.
The halo is discretised with $N=10^7$ simulation particles and set up initially in equilibrium in a self-consistent way (see Section~\ref{setup} and Appendix~\ref{app2}).
To analyse what happens in a more realistic setting we also take the DM-only zoom-in simulation of a dwarf-size halo described in \citet{Vogelsberger2014}, which has a virial mass of $1.1\times 10^{10}{\rm M}_{\odot}$, and a mass and spatial resolution similar to ours ($1.1\times 10^7$ particles within the virial radius, and $\epsilon = 34$ pc.)
We then add the two different mechanisms of core formation into either one of the haloes and follow the evolution using the {\scriptsize AREPO} code \citep{Springel:2009aa}: 
\begin{itemize}
    \item {\it supernova-driven mass blowouts} are modeled by adding an external centrally concentrated Hernquist potential, which is then suddenly (impulsively) removed. This is either done in a single event that removes $10\%$ of the idealised halo mass, or in 10 consecutive explosions, each removing $\sqrt{10}\%$ of the halo mass. These masses are chosen to satisfy the energy requirements to form a core of $\sim1$~kpc in either the single or multiple explosion cases (see Section~\ref{pesec}). 
    \item {\it DM self-interactions} are added by letting the simulation particles collide elastically in a probabilistic way following the algorithm described in \citet{Vogelsberger2012}. The transfer cross section is fixed to $\sigma_T/m_{\chi} = 2\,{\rm cm}^2{\rm g}^{-1}$. This cross section is chosen to form a DM core of the same central density and similar size as in the supernova case (see Section~\ref{sec_sim_profiles} and Fig.~\ref{profileevolution}).
\end{itemize}

To study the impact of the cusp-core transformation on stellar orbits, we add 2000 tracer particles into the simulations in three different idealised configurations. The first two are constructed to study key fundamental differences between the adiabatic and impulsive transformations and follow from simple theoretical expectations of the behaviour of the stellar orbits (see Section~\ref{basics} and Appendix \ref{app1}). We test for the conservation of radial actions under an adiabatic change of potential in the idealized halo which is predicted by the time averages theorem. Furthermore, we look at the segregation of stars that were initially in a similar orbit (but with distinct orbital phase) into distinct orbits after the impulsive transformation (i.e. violation of the time averages theorem; see e.g. \citealt{1987gady.book.....B}). We first test this in the idealised halo and then check by how much the results change in the cosmological halo. We find strong evidence in support of these theoretical expectations.
The setup for the three configurations and our main results are as follows. 

\noindent
\begin{enumerate}
    \item {\it Orbital family.-} The tracers are originally set up with a similar energy and (magnitude of) angular momentum, which is equivalent to sampling similar orbits (similar apo- and pericentre but different phases).
    This leads to a small region in $r-v_r$ space being populated at a rather large density. This region remains the same, within the numerical errors in our simulations, if there is no change in the potential (see Fig.~\ref{baselinervr}). 
    The response of the orbits (and the evolution of the populated area in phase space) is dramatically different in the SIDM and supernova-driven core formation scenarios.
    In the SIDM case, we find that the region in phase space which is occupied by the tracer particles remains remarkably compact,
     proving that the SIDM case is fully in the adiabatic regime (see left panel of Fig.~\ref{comprvr}).
    In the case of supernova-driven mass blowouts, the situation is drastically different: the orbital family is split into distinct orbital families, depending on the location of the tracer on its orbit (phase) when the explosion(s) happen (a peculiar consequence of the violation of the time averages theorem). In the idealised spherically symmetric halo, the orbital families form a distinctive structure of concentric shells in the 2-dimensional phase space (see right panel of Fig.~\ref{comprvr} and Fig.~\ref{singleexp}). Although these shell-like structures are more evident in the single explosion case, traces of them clearly remain even in the more chaotic multiple explosion case. When looking at the evolution in the cosmological halo, we find that there is a significant amount of phase mixing after roughly three dynamical times. This phase mixing leads to a broadening of the originally occupied region in phase space (see Fig.~\ref{fig:cosmo_equi}) and a relatively quick dilution of the shell-like structures in the impulsive case (see Figs.~\ref{fig:cosmo_core} and \ref{fig:cosmo_snf}). We attribute this to the triaxiality of the host halo, which implies that angular momentum is not conserved. Nonetheless, the final occupied region in phase space remains remarkably compact in the case of SIDM and the aforementioned shell-like signatures of an impulsive event such as a supernova are present, albeit only for about 0.5 Gyrs.
    \item {\it Gaussian distribution in radial action.-} 
    Since the radial action of tracer orbits is only conserved in a spherically symmetric host potential, we perform this test only in the idealised halo to confirm our underlying theoretical expectations.
    The orbits of the tracers are initially sampled to have a narrow Gaussian distribution in radial action.
    If the potential does not change, then this distribution should remain the same, and indeed it does in our simulation with a small level of numerical diffusion.
    The evolution of the radial action distribution is drastically different in the two mechanisms of core formation. 
    In the SIDM case, radial actions are remarkably close to being conserved, with the final distribution remaining very close to the initial one (see Fig.~\ref{raddists}). There is, however, a small distortion in this adiabatic case towards a broader distribution, which is worth studying in the future (e.g. along the lines presented in \citealt{2013MNRAS.433.2576P}). 
    In the supernova-driven core formation scenario, however, there is a substantial evolution of the distribution of radial actions, which emphasizes the degree to which actions are not conserved in the impulsive regime
    The final distribution is considerably broader with a long tail of high radial action values, and  no longer consistent with a Gaussian distribution (see Fig.~\ref{raddists}). 
    This confirms the original expectation that the fundamental difference between adiabatic and impulsive core formation in a spherically symmetric host potential is whether or not the radial actions of tracer (star) particles are conserved.
    \item {\it Plummer sphere.-} To test how a more realistic stellar distribution would react to the different mechanisms of core formation, we initially arrange the tracers in an isotropic Plummer sphere with a half-light radius of 500~pc set up in Jeans equilibrium within the idealised halo (the sphere remains a collection of tracers, i.e., it exerts no gravity).
    We find relevant differences between the two core formation scenarios. In the SIDM case, the sphere responds by (slightly) expanding adiabatically towards a new state of Jeans equilibrium, with only mild deviations from the original density and velocity dispersion profiles (see Fig.~\ref{plummer}). This agrees with previous results based on SIDM cosmological simulations \citep{Vogelsberger2014}.
    In the supernova-driven case, on the other hand, the Plummer sphere expands considerably in response to the formation of the DM core, which we emphasize has the same central density as in the SIDM case. This eventually results in a more extended sphere with a lower central density (see Fig.~\ref{plummer}), corresponding to a galaxy with lower central surface brightness. Furthermore, we find that the episodic supernova events tend to drive the stellar configuration out of Jeans equilibrium for a rather long time (our simulation stops 1.2 Gyr after the last explosion and the sphere remains out of equilibrium). 
    This implies that, contrary to SIDM, repeated supernovae (in the impulsive regime) have a lasting impact on the spatial distributions, as well as the Jeans equilibrium, of stars in dwarf galaxies.  
\end{enumerate}

In summary, we have quantified in detail up to which degree the process of redistributing energy can be considered adiabatic in the SIDM case of core formation. Moreover, we have quantified the amount of energy (mass removal) which is required to form a core with similar characteristics in the impulsive regime. We have found that with the two processes constituting extremes of the adiabatic-impulsive spectrum (SIDM being very close to adiabatic and SNF being entirely impulsive), the consequences for stellar orbits are severe.  
In the idealised cases we have studied, we find that the impulsive case preserves information about the initial locations and velocities (phases) of the tracers before the cusp-core transformation. This information proves to have a sizeable impact on the final stellar orbit.
On the other hand, radial actions are not conserved in realistic cosmological haloes, which are not spherical and have a DM distribution that is not smooth. Moreover, the distinctive phase space signatures in the impulsive (SNF) case
are diluted due to phase mixing within a few dynamical times in a cosmological halo.
We nevertheless have shown that the impact of different core formation mechanisms on the orbits of tracer orbits remains substantial. We argue that these
signatures in the stellar kinematics
constitute a promising avenue towards
distinguishing the different types of core formation mechanisms.

The interplay between different physical processes and the core formation mechanisms is expected to be quite complex, particularly for the SIDM case \citep[for dwarf galaxies e.g.][]{Vogelsberger2014,Fry2015,Robles2017}. This interplay remains essentially unexplored when it comes to its effect on stellar orbits, and the type of signatures we have found here.  We aim to comprehensibly explore this interplay in the near future 
using cosmological hydrodynamical simulations with full baryonic physics.

\section*{Acknowledgements}

We thank Volker Springel for giving us access to the {\scriptsize AREPO} code, and thank Chaichalit Srisawat for useful discussions. We thank Andrew Pontzen for insightful comments that encouraged us to significantly improve our work. This work was supported by a Grant of Excellence from the Icelandic Center for Research (Rann\'is; grant number 173929$-$051).  We acknowledge that the results of this research have in part been achieved using the PRACE Research Infrastructure resource {\scriptsize CURIE} based in France at CEA. The simulations in this paper were carried out
on the Garpur supercomputer, a joint project between the University of Iceland and University of Reykjav\'ik with funding from Rann\'is.



\bibliographystyle{mnras}

\appendix

\section{Impulsive vs adiabatic change in the potential: 1D harmonic oscillator toy model}\label{app1}

To develop an intuition about
the conceptual differences between impulsive and adiabatic core formation, we use the 1-dimensional harmonic oscillator toy model introduced in \citet{Pontzen:2011ty}. For the impulsive regime, we simply reproduce the key results and equations derived by the authors, while we develop a detailed description of the adiabatic regime, which was only mentioned vaguely in \citet{Pontzen:2011ty}. 

The 1-dimensional model consists of a massless test particle responding to an external potential:
\begin{align}
	V(x;t) = V_0(t)x^2.
\end{align}
Here only the normalization of the potential $V_0$ is explicitly time-dependent while the functional form remains the same. Depending on whether the change in potential is impulsive or adiabatic, the solution to the equation of motion has to be obtained differently. 

\subsection{The impulsive regime}\label{impulsive}

To model an instantaneous change in the potential, we impose a sudden frequency change from $\omega_0$ to $\omega_1$ at time $t=0$. Before and after the frequency change, the test particle's equation of motion is that of an ordinary harmonic oscillator, which has the general solution 
\begin{align}
    x(t) = A\cos(\omega t + \psi),
\end{align}
where the phase $\psi$ and amplitude $A$ have different values before and after the frequency change. The final amplitude $A_1$ and phase $\psi_1$ of the oscillator can then be computed from the initial ones by requiring that $x(t)$ and $\dot{x}(t)$ be continuous at $t=0$, which results in \citep{Pontzen:2011ty}:
\begin{align}
&A_1^2 = A_0^2\left[1+\frac{\omega_0^2-\omega_1^2}{\omega_1^2}\sin^2\psi_0\right] \label{amp}\\
&\tan\psi_1 = \frac{\omega_0}{\omega_1}\tan\psi_0 \label{phasetrafo}
\end{align}
where $A_0$ and $\psi_0$ are the initial amplitude and phase, respectively.
The key feature of the solution is that there is an 
explicit dependence of the final amplitude and phase on the initial conditions. In particular, if we imagine two oscillators moving with the same amplitude $A_0$ initially but with different initial phases, they will end up oscillating with different amplitudes and phases after an impulsive change in the potential.
This is in essence a consequence of the time averages theorem not being valid for the impulsive case.

For 3-dimensional stellar orbits, this result suggests that stars that are initially on a common orbit (but with different phases) will in general end up on very different orbits after an impulsive gas outflow produced by supernovae. 
If we assume a flat (random) distribution of radial phases in the beginning and a subsequent substantial and explosive change in the potential ($\omega_1\ll\omega_0$ in Eq.~\ref{phasetrafo}), then after the 
explosion, as the orbits expand, most of the particles will be significantly closer to the pericentre of their new orbit than to the pericentre of their previous orbit. This implies that the initially flat distribution of radial phases changes to a non-flat distribution with a bias towards phases close to $\pi$. In the case of the harmonic oscillator, starting from a flat prior $2\pi p(\psi_0)=1$ results in a final distribution of phases given by \citep{Pontzen:2011ty}:
\begin{align}\label{phase_red}
    2\pi p(\psi_1) = \left( \frac{\omega_0}{\omega_1}\cos^2\psi_1 + \frac{\omega_1}{\omega_0}\sin^2 \psi_1\right)^{-1}.
\end{align}

\subsection{The adiabatic regime}

Although a numerical solution for the adiabatic case is shown in Fig. 3 of \citet{Pontzen:2011ty}, no details are given on how this solution was obtained. Thus, in the following we provide a full description of this regime.

To model an adiabatic change in the harmonic oscillator potential we use a smooth function for the time dependence of the frequency $\omega(t)$ which, contrary to the impulsive case, changes gradually from an initial value $\omega_0$ to $\omega_1$ over several oscillation periods. To be precise, the solution is subject to the conditions: 
\begin{align}
\lim\limits_{t\to -\infty}\omega(t) &= \omega_0\\
\lim\limits_{t\to\infty}\omega(t) &= \omega_1
\end{align}
and must obey the equation of motion:
\begin{align}
\ddot{x}+\omega(t)^2x = 0
\label{harmonic}
\end{align}
We propose as an ansatz that the solution has the general form of the harmonic oscillator at all times: 
\begin{align}
	x(t) = A(t)\cos(\omega(t) t +\psi(t)),
	\label{hansatz}
\end{align}
where $\omega^2(t)=2V_0(t)$, $A(t)$, and $\psi(t)$ are the frequency, amplitude and phase of the oscillation.
In addition, we define a new variable:
\begin{align}
\zeta = \omega + \dot{\omega}t +\dot{\psi},
\label{zetadef}
\end{align}
which is the total time derivative of the argument of the cosine function.
Demanding that Eq.~\ref{harmonic} be fulfilled leads to the following set of two equations: 
\begin{align}
\frac{\ddot{A}}{A} &= \left(\zeta^2-\omega^2\right),\label{eqa1}\\
\frac{\dot{A}}{A} &= -\frac{1}{2}\frac{\dot{\zeta}}{\zeta}.\label{eqa2}	
\end{align}
Now, since 
\begin{align}
\frac{d}{dt}\left(\frac{\dot{A}}{A}\right) = \frac{\ddot{A}}{A}-\left(\frac{\dot{A}}{A}\right)^2,
\end{align}
we can combine Eqs.~\ref{eqa1} and \ref{eqa2} to find the following second order non-linear differential equation for the variable $\zeta$:
\begin{align}
\frac{\ddot{\zeta}}{\zeta}-\frac{3}{2}\left(\frac{\dot{\zeta}}{\zeta}\right)^2 = -2\left(\zeta^2-\omega^2\right),
\label{deqz}
\end{align}
which can be solved numerically for a given functional form of $\omega(t)$. 
We can then solve for the amplitude and phase of the oscilator: 
\begin{align}
A(t) &= A_0\exp\left[-\frac{1}{2}\int_{-\infty}^{t}\frac{\dot{\zeta}}{\zeta}dt'\right],
\label{aoft}\\
\psi(t) &= \psi_0+\int_{-\infty}^{t}\left(\zeta-\omega-\dot{\omega}t'\right)dt'.
\label{psioft}
\end{align}
Assuming now that the frequency transition is fully adiabatic, we can theoretically halt it at any point and continue it later on. This implies that  if the ansatz in equation \ref{hansatz} is correct, 
the motion of our test particle is described by an ordinary harmonic oscillator at all times during the transition with an (instantaneous) period given approximately by $T\sim2\pi/\omega(t)$. As the variable $\zeta$ defined in equation \ref{zetadef} is the time derivative of the cosine's argument, adiabaticity thus demands that at any point in time 
\begin{align}
    \zeta(t) \sim \omega(t).
    \label{fullyadiabatic}
\end{align}
This immediately implies that 
\begin{align}
\nonumber A(t) &\sim A_0\exp\left[-\frac{1}{2}\int_{-\infty}^{t}\frac{\dot{\omega}}{\omega}dt'\right] = A_0\exp\left[-\frac{1}{2}\int_{\omega_0}^{\omega(t)}\frac{d\omega}{\omega}\right]\\
&= A_0\sqrt{\frac{\omega_0}{\omega(t)}}
\label{AOFT}
\end{align} 
which has the asymptotic behaviour $A_{t\rightarrow\infty}=A_0\sqrt{(\omega_0/\omega_1)}$, and agrees with the results shown in Fig. 3 of \citet{Pontzen:2011ty} (specifically, for the numerical values used there,  $A_{t\rightarrow\infty}\approx 1.78$). Using Eq.\ref{fullyadiabatic} we can also approximate Eq.~\ref{psioft} for the phase $\psi(t)$ as: 
\begin{align}
\psi(t) \sim\psi_0 -\int_{-\infty}^{t}\dot{\omega}t'\,dt'
\label{psi_2}
\end{align}
and see that the change in phase depends explicitly on the time coordinate, as different transitioning times demand a different phase change to ensure a smooth behaviour of the oscillator. Note also that, contrary to the impulsive case, the phase shift in no way depends on the initial phase.

Eqs. \ref{AOFT} and \ref{psi_2} are exact if the change in potential occurs fully adiabatically, i.e., in the limit of the frequency transition taking an infinite amount of time. In practice, this is never exactly the case. However, the adiabatic approximation in general works very well if the change in energy (or equivalently frequency) occurs slowly compared to the dynamical time of the system, which in this case is given by the instantaneous frequency of oscillation.
For adiabaticity, we thus demand that 
\begin{align}
    \left|\frac{\Delta E}{E_{{\rm min}}(\Delta t)}\right| \ll \frac{\omega_{{\rm min}}}{2\pi},\label{ad1}
\end{align}
where we choose the smallest characteristic time-scale and energy to obtain the most conservative bound. On the other hand, we have that the initial energy of the harmonic oscilator is
\begin{align}
    E_{\rm ini} = \frac{1}{2}A_0^2\omega_0^2
\end{align} 
and using Eq. 4 of \citet{Pontzen:2011ty}, the final energy in the adiabatic limit is:
\begin{align}
    E_{\rm fin} = \frac{1}{2}A_0^2\omega_1\omega_0.
\end{align}
If we thus lower the frequency of the harmonic oscillator by $\Delta \omega$ starting from $\omega_0$ and ending at $\omega_1$ during a time $\Delta t$, the condition for adiabaticity \ref{ad1} can then be written as:
\begin{align}
    &\left|\frac{\Delta \omega}{\omega_1 \Delta t}\right| \ll \frac{\omega_1}{2\pi}\label{con1}.
\end{align}
We will see that this condition will have to be fulfilled by any parametrization $\omega(t)$ in order for our ansatz \ref{hansatz} to provide a reasonable solution to the differential equation for the time-dependent harmonic oscilator (Eq.~\ref{harmonic}).
\begin{figure}
    \includegraphics[height=6.5cm,width=8.5cm,trim=0.5cm 0.75cm 0cm 0.0cm, clip=true]{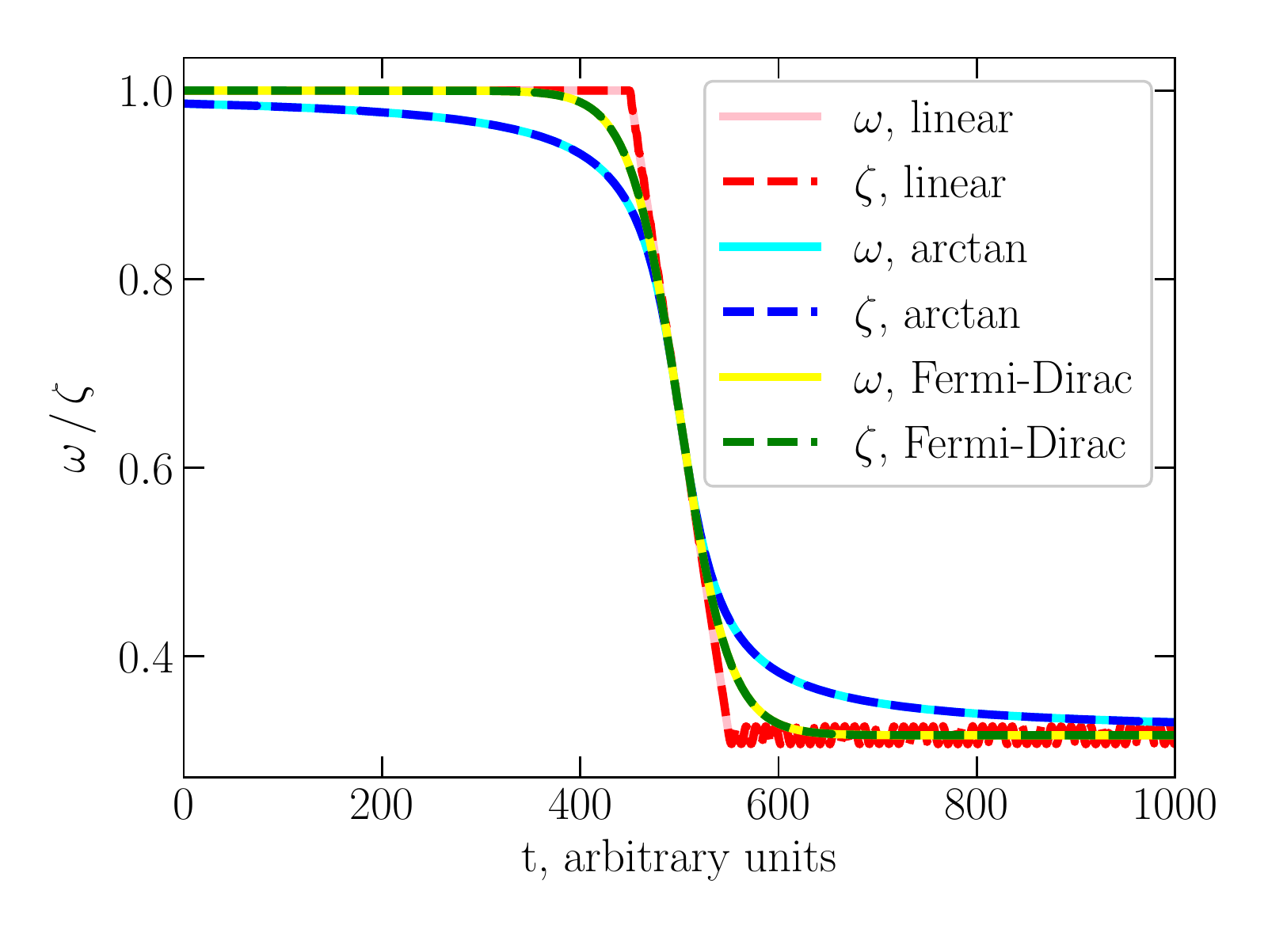}
	\caption{The solutions $\zeta(t)$ to equation \ref{deqz} are shown alongside $\omega(t)$ for three parametrizations of $\omega(t)$ (as given in the legend; see Eqs.~\ref{linear}-\ref{FD})  with $t_0 = 450$, $t_1=550$, $\omega_0 = 1$ and $\omega_1=\sqrt{0.1}$.
	Notice that the transition period between frequencies is strictly confined to the time interval $t_1-t_0$ only in the 'linear' case, but in this case, the solution oscillates around a constant value after the frequency transition is complete. In the 'arctan' case on the other hand, the solution is stable but the transition takes a long time to be complete. The 'Fermi-Dirac' case constitutes the best parametrization we found in terms of representing a sharp transition with a stable long term solution.}
	\label{solutions}
\end{figure}

\begin{figure*}
	\includegraphics[height=6.5cm,width=8.5cm,trim=0.5cm 0.5cm 0cm 0.0cm, clip=true]{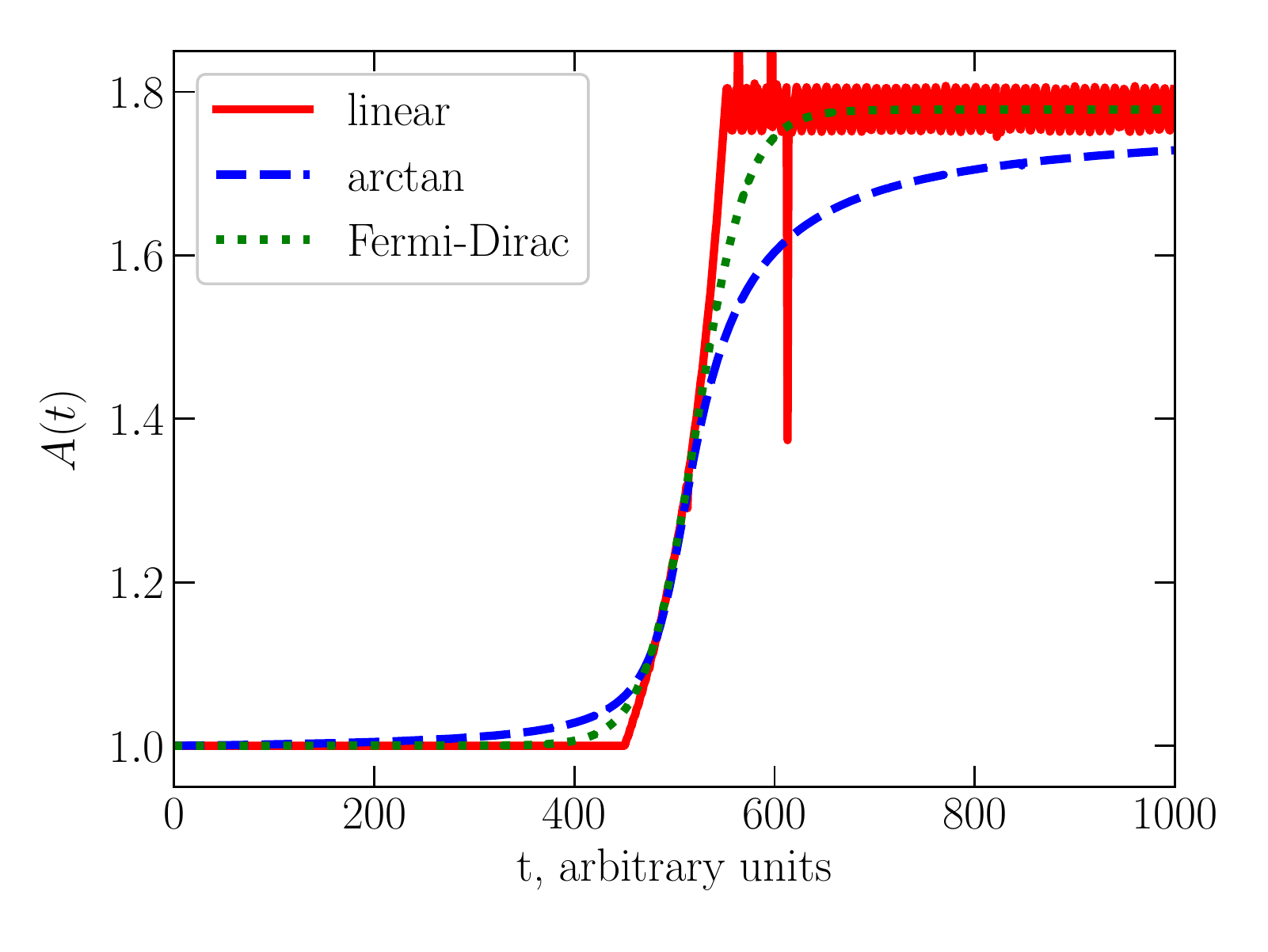}
	\includegraphics[height=6.5cm,width=8.5cm,trim=0.5cm 0.5cm 0cm 0.0cm, clip=true]{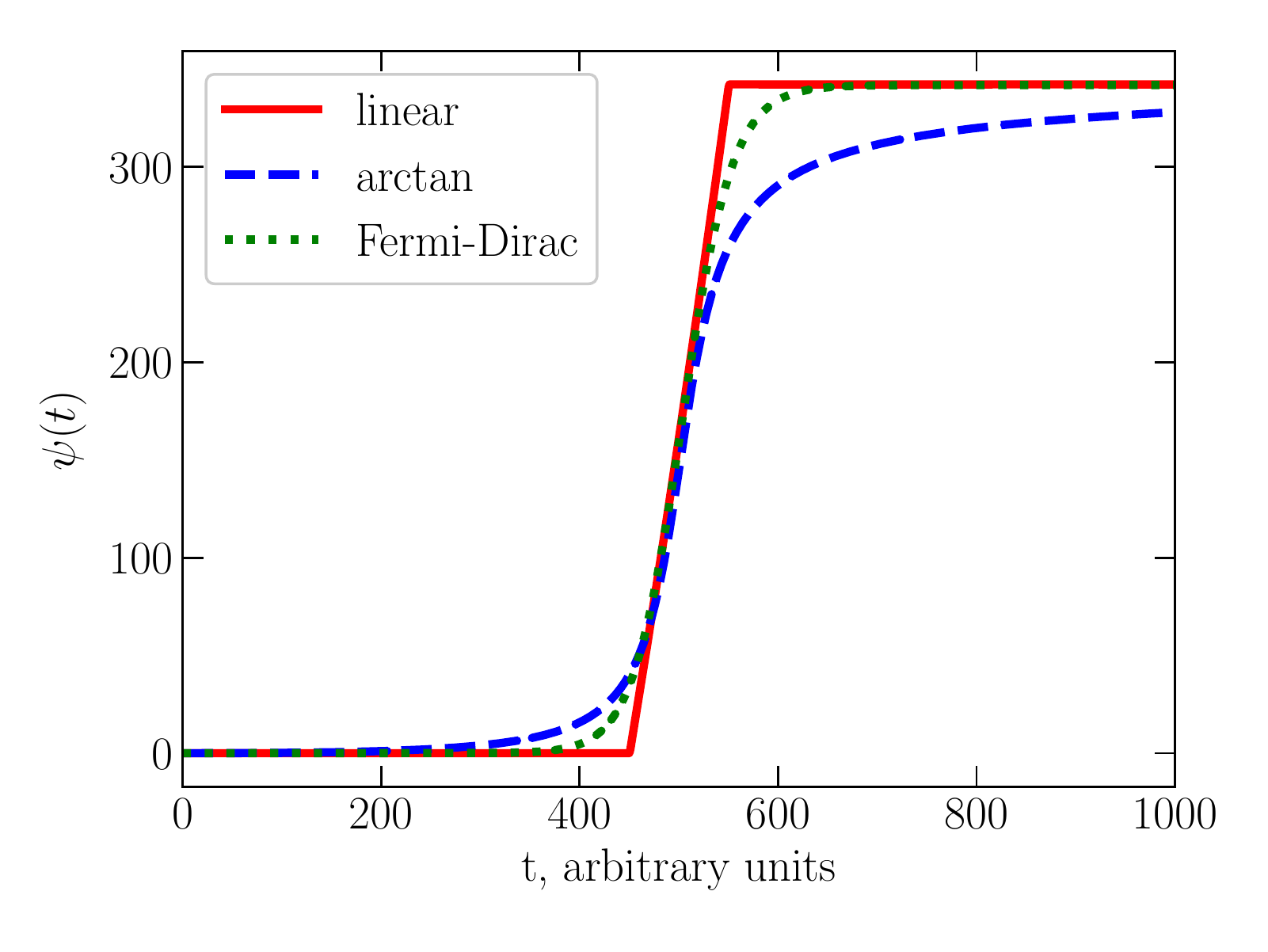}
	\caption{The solutions for the amplitude $A$ (left panel) and phase $\psi$ (right panel) of the harmonic oscillator under a change of frequency corresponding to the three different parametrizations shown in Fig.~\ref{solutions} (see Eqs.~\ref{linear}-\ref{FD}) for $A_0=1$ and $\psi_0=0$. For the phase, the transition is quickly reached in the 'linear' case, closely followed by the 'Fermi-Dirac' case, while the 'arctan' case takes much longer to complete the transition. When looking at the amplitude, it is clear that the 'linear' case is not appropriate since the amplitude shows numerical oscillatory features. Overall, the 'Fermi-Dirac' case is the most appropriate as the amplitude and phase reach their final values quickly and without numerical issues.}
	\label{results_adiabatic}
\end{figure*}

To calculate explicit solutions to Eq.~\ref{deqz}, we use a few specific functions to model the change in frequency. In particular, we consider 
three different parametrizations for $\omega(t)$. The first one is a linear transition: 
\begin{align}
\omega(t) = 
\begin{cases}
\omega_0 & t \le t_0 \\
\omega_0 + \frac{\omega_1-\omega_0}{t_1-t_0}(t-t_0) & t_0 < t \le t_1\\
\omega_1 & t_1 < t
\end{cases}
\label{linear}
\end{align}
which has the advantage that the frequency change is strictly confined to the transition period between $t_1$ and $t_0$. However, its obvious disadvantage is that at these two times the derivative of $\omega(t)$ is not continuous, which leads to numerical issues when solving the differential equation \ref{deqz} as we show below.

To circumvent this problem, we explore two different parametrizations which do not have sharp transitions at the boundaries but where the frequency change is still confined mostly to the time interval from $t_0$ to $t_1$. The other requirement we impose for these parametrizations is that at the
time $t = (t_0+t_1)/2$, the slope of the frequency function be $(\omega_1-\omega_0)/(t_1-t_0)$, just as in the linear case. 
Following these conditions, we explore these two functions:
\begin{align}
\omega(t) &= \frac{\omega_0+\omega_1}{2}+\frac{\omega_1-\omega_0}{\pi}\arctan\left(\frac{\pi}{t_1-t_0}\left[t-\frac{t_0+t_1}{2}\right]\right)\\
\omega(t) &= \omega_0 +\frac{\omega_1-\omega_0}{\exp\left(-\frac{4}{t_1-t_0}\left[t-\frac{t_1+t_0}{2}\right]\right)+1}\label{FD}
\end{align}
which we refer to as the 'arctan' and 'Fermi-Dirac' cases.
\begin{figure*}
	\includegraphics[height=6.5cm,width=8.5cm,trim=0.5cm 0.5cm 0cm 0.0cm, clip=true]{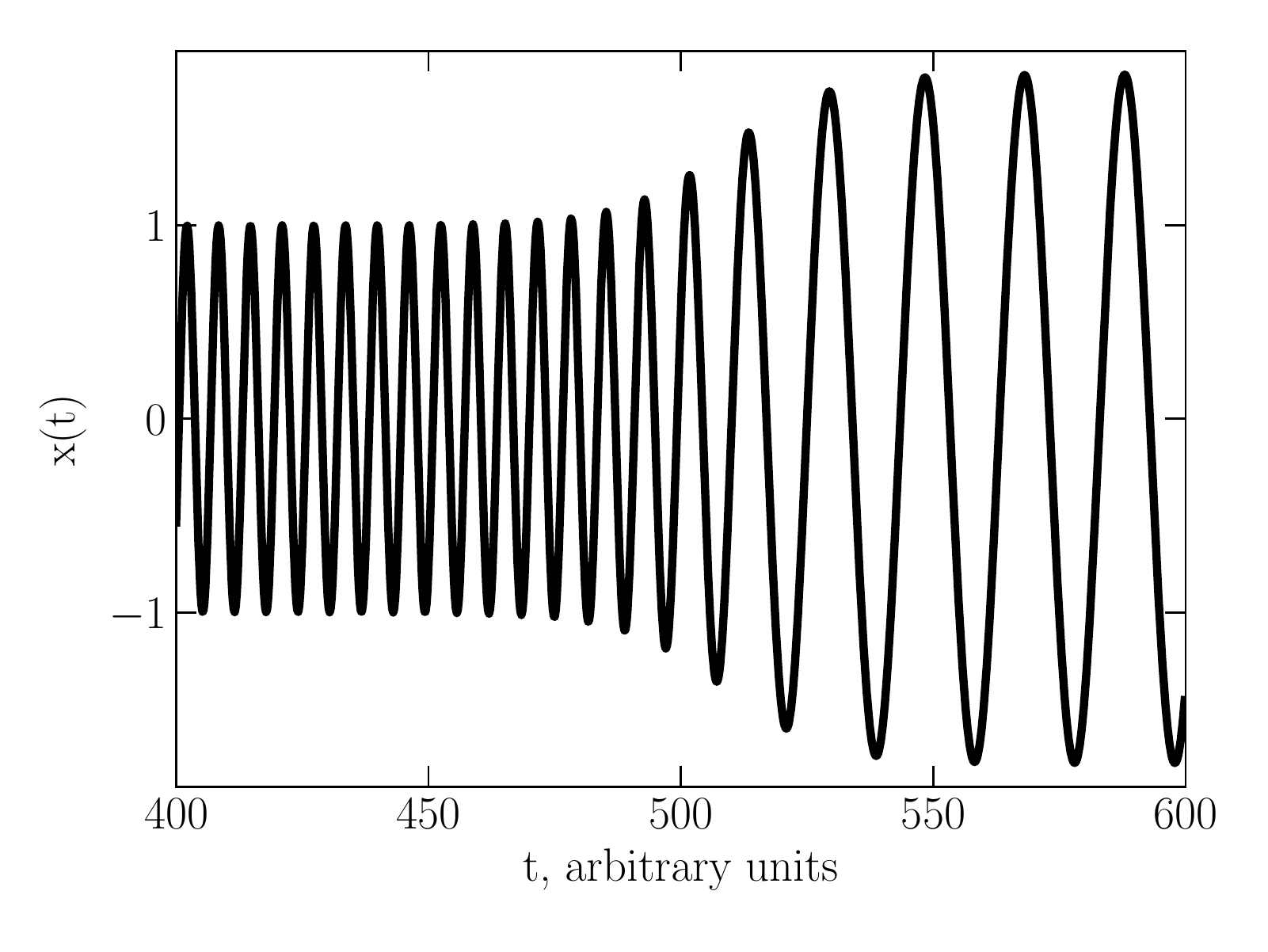}
	\includegraphics[height=6.5cm,width=8.5cm,trim=0.5cm 0.5cm 0cm 0.0cm, clip=true]{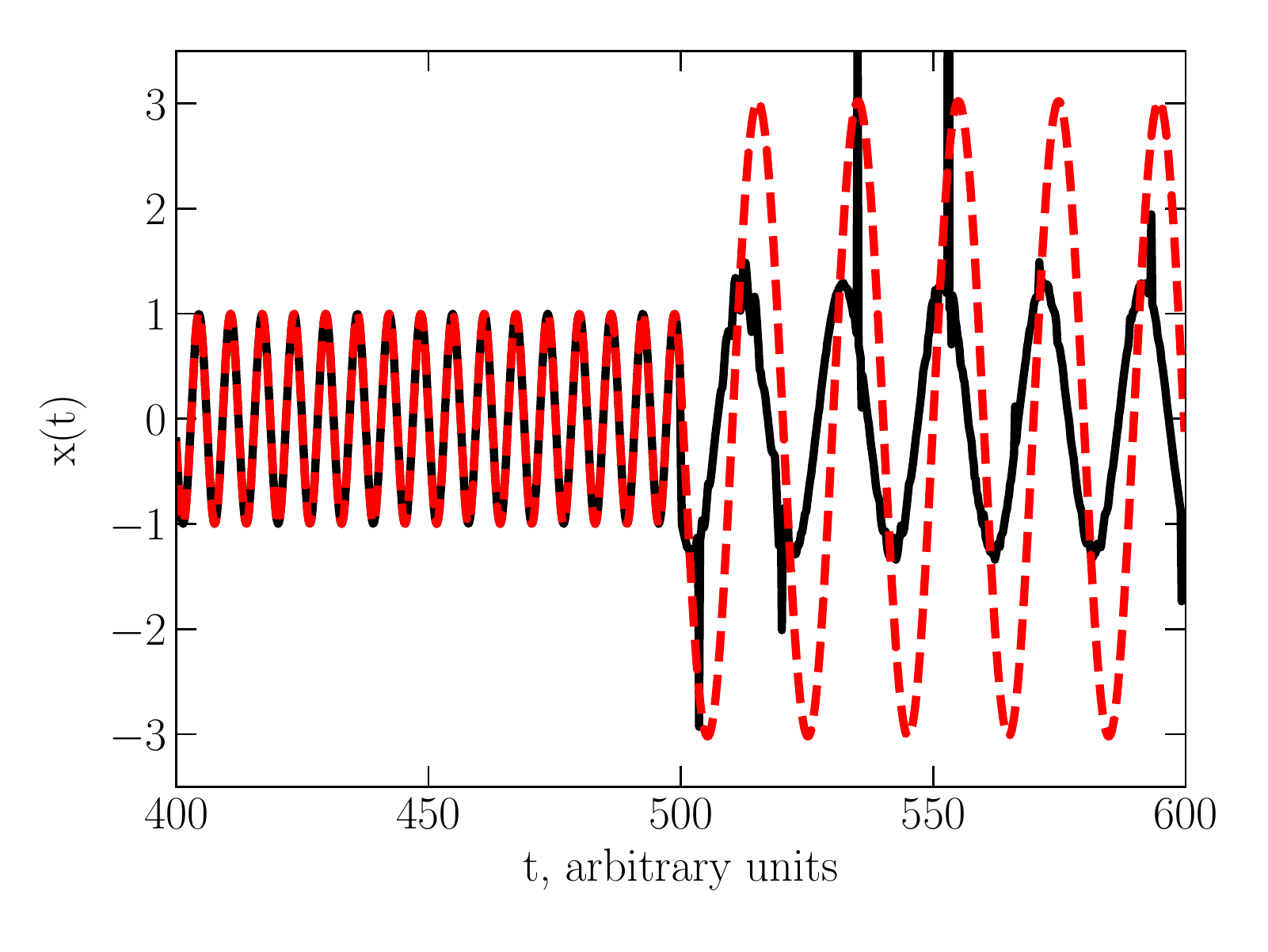}
	\caption{Solution to the harmonic oscillator equation 
	under a transition in frequency with a 'Fermi-Dirac' parametrization (see Eq. \ref{FD}). The left panel shows the case in which the frequency is altered over exactly two long orbital periods, whereas the right panel shows the case in which the transition period $t_1-t_0$ is too small for Eq. \ref{con1} to be fulfilled. In the left panel the solution is smooth and behaves just as expected from an adiabatic case (see Fig. 3 of \citealt{Pontzen:2011ty}). In the right panel, on the other hand, we observe random oscillations of the calculated solution and a very non-harmonic behavior, which implies that the original ansatz that led to equation \ref{deqz} is no longer applicable. Instead, the impulsive approximation is appropriate (see Eqs. \ref{amp} and \ref{phasetrafo}), which is shown as a red dotted line. 
	We caution the reader that in Eq.~\ref{amp}, the final amplitude of oscillation depends explicitly on the exact time of the transition.}
	\label{shorter}
\end{figure*}

The solutions to equation \ref{deqz} are shown in fig. \ref{solutions} (solid lines) along with the corresponding functions $\omega(t)$ (dashed lines) for each parameterization and for the parameters $t_0 = 450$, $t_1=550$, $\omega_0 = 1$ and $\omega_1=\sqrt{0.1}$. We find that $\omega$ = $\zeta$ just as predicted, and looking at the 'arctan' and 'Fermi-Dirac' cases, we find that precisely the condition \ref{con1} needs to be fulfilled in order for the left-hand-side of equation \ref{deqz} to be small and thus to have $\omega \sim \zeta$ at all times. 
In the 'linear' case, however, we find the predicted numerical effects occuring due to $\ddot{\omega}/\omega$ not being small at the times $t_0$ and $t_1$.
Since in the 'Fermi-Dirac' case the solution approaches the required asymptotic values significantly faster than in the 'arctan' case, we will focus on the 'Fermi-Dirac' case when conducting further analysis such as evaluating the solution to the original equation of motion.
Fig. \ref{results_adiabatic} shows the results for the amplitude (left) and phase (right) of the oscillations in all of the three cases. As anticipated, the discontinuity in the first derivative of the 'linear' parametrization of the frequency causes undesired numerical issues in the amplitude $A(t)$ of the oscillator. The phase shift, however, is not affected. Moreover, it is evident from fig. \ref{results_adiabatic} that the solution to the 'arctan' case displays a convergence behavior which is much slower than that of the 'Fermi-Dirac' case. This implies that out of all the three cases, the 'Fermi-Dirac' one turns out to be the best one to describe a smooth transition between two frequencies within a fixed interval of time. This is further confirmation that all reconstructions of $x(t)$ should be based on solutions of the 'Fermi-Dirac' case.

The final result for the solution $x(t)$ is a smooth function which
transitions continuously from a harmonic oscillator with a frequency $\omega_0$ to a final harmonic oscillator with a frequency $\omega_1$. The amplitudes and phases at the end of the transition are consistent with the approximations in Eqs.~\ref{AOFT} and \ref{psi_2}. In order to reproduce the results in Fig. 3 of \citet{Pontzen:2011ty}, we
need to choose the transition time interval to be equal to twice the longest orbital period possible: 
\begin{align}
	\Delta t =\frac{4\pi}{\omega_1}.
	\label{2p}
\end{align} 
We note that for our choice of parameters this transition time interval no longer strongly satisfies condition \ref{con1}, since the left-hand-side and the right-hand-side of the equation are of the same order. Nevertheless, the adiabatic approximation still holds as can be seen in the left panel of fig.~\ref{shorter}, where we show the solution to the harmonic oscillator equation in this case.

It is interesting to check at what point the adiabatic approximation breaks down in this simple toy model. 
Following Eq.~\ref{con1}, this should certainly happen when the period of time over which we change the potential is much shorter than the dynamical time-scale of the system:
\begin{align}
	\frac{\Delta t}{2\pi} \ll \frac{\Delta \omega}{\omega_1^2}.
	\label{lp}
\end{align}
To test this, we integrate Eq.~\ref{deqz} for $\Delta t =t_1-t_0 = 1$, and leave the rest of the parameters unchanged with respect to the fully adiabatic case shown in the left panel of Fig.~\ref{shorter}. 
The right panel of fig.~\ref{shorter} shows the numerical solution in this case (solid black line). 
It is apparent that when the change in potential is impulsive, the solution to equation \ref{deqz} that we obtain by numerical integration is no longer sensible, which confirms that we can no longer use the ansatz \ref{hansatz}. 
Instead we need to resort to the arguments summarized in Section~\ref{impulsive} and derived in \citet{Pontzen:2011ty}. A solution based on the latter is shown in the right panel of Fig.~\ref{shorter} as a red dashed line. We caution the reader that a specific choice of phase underlies this solution, in particular one that maximises the growth of the amplitude.

\section{Stability of haloes simulated in equilibrium}\label{app2}
In order to asses the intrinsic differences between the adiabatic and impulsive core formation scenarios, we need to ensure that numerical artefacts are well under control in the central regions of the halo where core formation takes place.
In particular we want to ensure that two effects are negligible in the timescales of our simulations within the region of interest for our work: (i) the growth of numerical errors coming from the initial setup of a DM halo in equilibrium, and (ii) the development of collision-less relaxation due to particle discreteness. In order to assess the magnitude of these two effects, we look at the stability of the halo set up in equilibrium in a self-consistent way in Section \ref{setup} (referred to as the equilibrium case throughout this work). Regarding (i) above, the method we have used \citep[developed in][]{Kazantzidis:2003im} has been used and tested extensively and we can be confident that numerical errors are under control as long as the amount of simulation particles sampled in the central region is large enough, which is connected to the second issue of particle discreteness in (ii).

\begin{figure*}
	 \includegraphics[height=6.5cm,width=8.5cm,trim=0.5cm 0.5cm 0cm 0.0cm, clip=true]{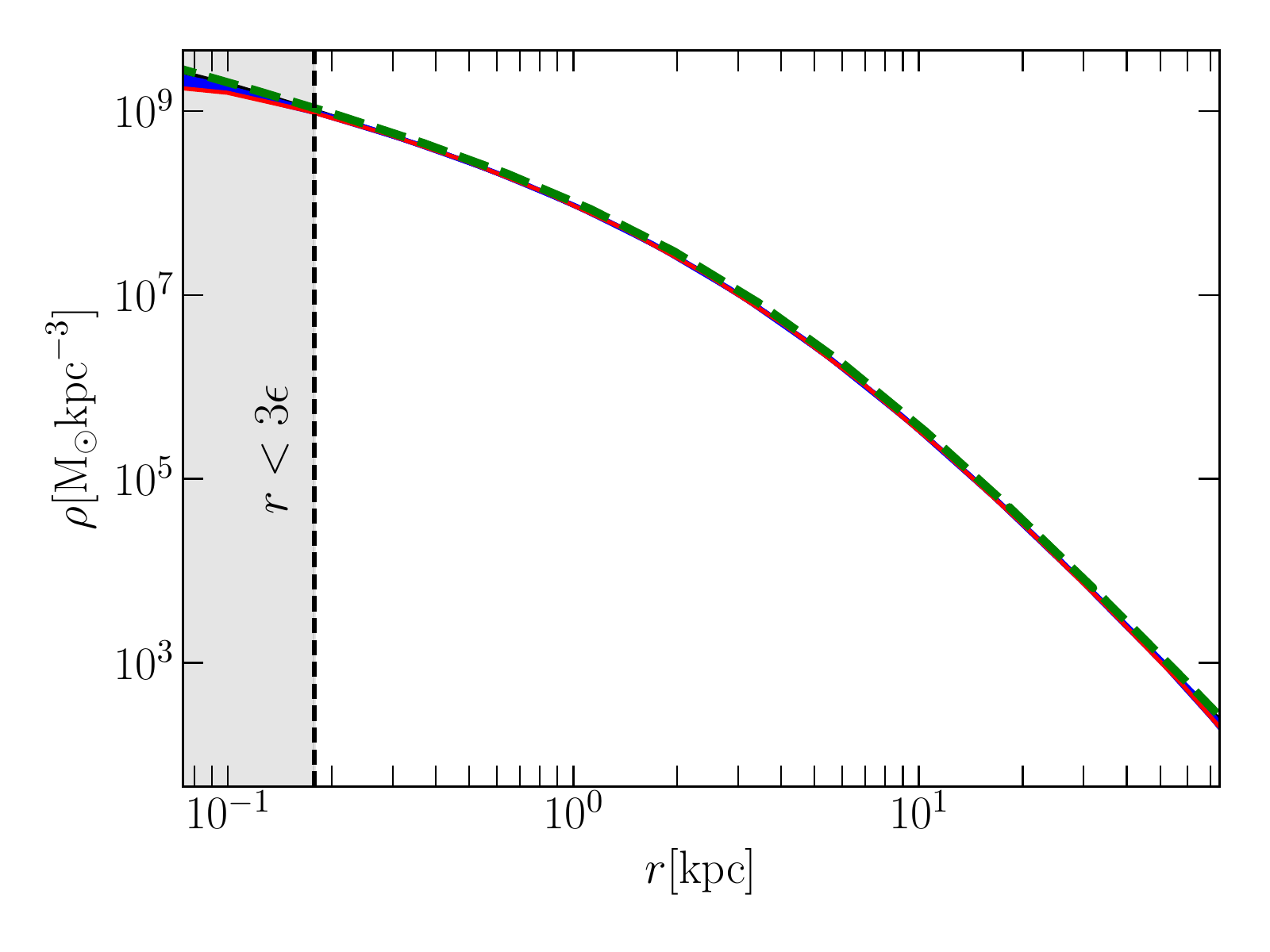}
	 \includegraphics[height=6.5cm,width=8.5cm,trim=0.5cm 0.5cm 0cm 0.0cm, clip=true]{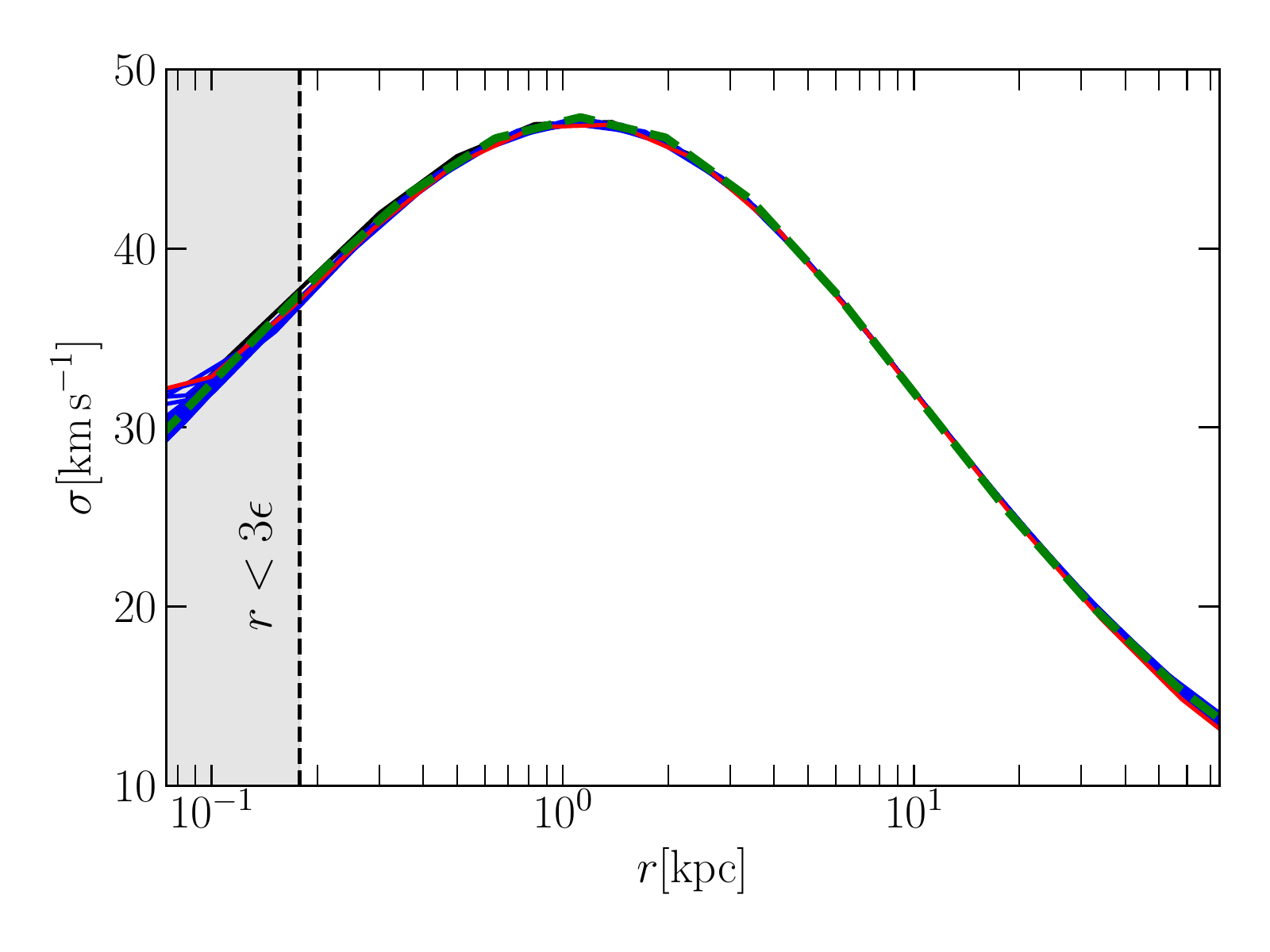}
	 \caption{Stability of the the equilibrium configuration for the (dwarf-size) Hernquist halo used in this work.
	 The evolution of the density (left) and velocity dispersion profiles (right) are shown over a simulation time of 4.2 Gyrs with outputs at every 0.3 Gyrs shown as blue solid lines.
	 The initial profile is highlighted in black, whereas the final profile is shown in red. 
	 The vertical dashed black line marks the radius beyond which we expect collision-less relaxation to be negligible (the softening length $\epsilon$ is given by Eq.~\ref{powerradius}). The region where collision-less relaxation effects may play a role is shaded in grey. A halo in perfect equilibrium would match the analytic formulae for the Hernquist profile shown as dashed green lines. The equilibrium configuration setup we have used is stable for $r>3\epsilon$.}
	 \label{profcomp}
\end{figure*}

Regarding collision-less relaxation, the relaxation time scale is given by
\citep{1987gady.book.....B}: 
\begin{align}
t_{\mathrm{relax}} \approx \frac{N}{8\ln N}t_{\mathrm{cross}}.
\end{align}
where $N$ is the number of collision-less particles and 
$t_{\rm cross}$ is the crossing time.
To ensure that an $N-$body representation of a DM halo does not suffer from gravitational two-body relaxation, the relaxation time needs to be longer than the simulation time. In order to achieve that, we need to avoid very large accelerations arising due to close encounters between simulation particles.
Choosing an appropriate softening length ensures this \citep{Power:2002sw}. In particular, by
demanding that the maximal gravitational force due to two-body encounters (equal to the maximum stochastic {\it softened} force) is smaller than the minimum mean field force in the simulation (which occurs at the virial radius), we get
\begin{align}
a_{\epsilon} \le a_{\mathrm{min}} \equiv \frac{Gm}{\epsilon^2}\le \frac{GM_{200}}{r^2_{200}}.
\end{align} 
From this it follows that a lower limit to the softening length required to avoid strong discreteness effects is:
\begin{align}
\epsilon \ge \frac{r_{200}}{\sqrt{N_{200}}}
\label{min_power}
\end{align}
where $M_{200} = N_{200}m$.

In CDM simulations, a safe choice is to set the softening length to 4 times the minimal one in Eq.~\ref{min_power}, which is the choice we use in our work (see Eq.~\ref{powerradius}). With this choice, it has been shown by many works in the past that the density profile of CDM haloes converges at $r\sim3\epsilon$, across a broad range of resolution levels (particle number). This implies that two-body relaxation becomes negligible for $r>3\epsilon$. 
In SIDM simulations, the physical collisional relaxation produced by self-scattering and the corresponding development of a central density core results in SIDM haloes with density profiles that converge at even smaller radii than in CDM \citep[e.g.][]{Vogelsberger2012}. We thus concentrate on the CDM case in its equilibrium configuration to explicitly check the stability of this configuration in the timescale of our simulations. Specifically, we investigate  
the stability of both the radial density and the velocity dispersion profile of the dwarf-size halo we have used as an initial condition for all our simulations when the core formation mechanisms are switched off (i.e. the equilibrium case). The properties of the halo are: ${\rm M}_{200} = 1.48\times10^{10}\, {\rm M}_{\odot}$, $r_{200}/r_{\rm s} = 15$, $N = 10^7$ and 3$\epsilon = 178$~pc. In this test, we
run the simulation for a total of 4.2 Gyrs, and take snapshots every 0.3 Gyrs. 
The density and velocity dispersion profiles for each snapshot are computed in the centre of mass frame of the halo at the corresponding time of the snapshot.
Fig. \ref{profcomp} shows the evolution of the radial profiles, density on the left panel and velocity dispersion on the right panel. Each simulation output is shown with blue solid lines, while the initial (final) times are shown with black (red) solid lines. We also show the analytical expectations for a Hernquist profile \citep{Hernquist:1990be} as green dashed lines.
The radial range in which the profiles deviate from each other is the range where the halo is no longer in equilibrium. In line with our expectations, only the innermost regions (well within $3\epsilon$, shown with a vertical dashed line and shaded in grey) are the ones that become visibly unstable within the simulation time. 
For $r>3\epsilon=178$~pc, we find essentially perfect stability in both the density and the velocity dispersion profiles. This is 5 times smaller than the $\sim1$~kpc size cores we study in this paper. Thus, we conclude that the halo equilibrium configuration we have set up is not affected by particle discreteness or numerical errors that could affect its stability at the relevant radii over the simulated time.

\bsp	
\label{lastpage}
\end{document}